\documentclass{aa}  

\usepackage{hyperref}
\usepackage{graphicx}
\usepackage{txfonts}
\linenumbers

\begin{document}
   \title{Ubiquitous yet forgotten: broad absorptions in the optical spectra of low-mass X-ray binaries}
   \titlerunning{Broad absorptions in the optical spectra of low-mass X-ray binaries}
   \author{D. Mata S\'anchez\inst{1,2}
          \and
          T. Mu\~noz-Darias\inst{1,2}          
          \and
          J. Casares\inst{1,2}
          \and          
          M. A. P. Torres\inst{1,2}
          \and
          M. Armas Padilla\inst{1,2} 
          }

   \institute{Instituto de Astrof\'isica de Canarias, E-38205 La Laguna, Tenerife, Spain\\
              \email{matasanchez.astronomy@gmail.com, dmata@iac.es}
         \and
             Departamento de Astrof\'isica, Univ. de La Laguna, E-38206 La Laguna, Tenerife, Spain\\
             }

   \date{Received 14/01/2026; accepted 11/02/2026}

  \abstract
  {Optical outburst spectra of low-mass X-ray binaries enable studies of extreme accretion and ejection phenomena. While some of their spectroscopic features have been analysed in detail, the appearance of broad absorptions in the optical regime has been traditionally neglected. In this work, we introduce the first population study dedicated to these features with the aim to understand their fundamental properties and discuss them in the context of their origin. We complement the study with a spectroscopic database of six low-mass X-ray binaries during outburst, in order to assess their evolution. We find that broad absorptions are ubiquitous, with the majority of black hole low-mass X-ray binaries exhibiting them in spite of a typically scarce outburst coverage. Their detection does not depend on the orbital inclination or the compact object nature, but they seem favoured in systems with orbital periods shorter than $< 11\, {\rm h}$. They predominantly occur in the hydrogen Balmer series, being stronger at shorter wavelengths, and they are detected across all X-ray states. Their profiles are best fitted with a Gaussian distribution of $\sigma_{\rm abs}=1400\pm 500\, {\rm km\, s^{-1}}$. They exhibit typical mean centroid velocities close to the systemic velocity, linking them to the accretion disc. Their equivalent widths, with typical values of $EW_{\rm abs}=4\pm 2\, \AA$, slowly evolve over weeks-to-months timescales, reaching up to $\sim 17\, \AA$ in the most extreme cases. We find that the normalised depth of these broad absorptions is anti-correlated with the system luminosity, and that they show constant line ratios over the whole sample. Based on these properties, we favour a scenario where BAs arise from a stable, optically thick layer of the accretion disc, below the hotter chromosphere-like region producing the emission line components. Our study is consistent with the continuous presence of broad absorptions during the whole outburst, with their visibility being conditioned by the emission lines filling the broad absorption profile and veiling by the X-ray reprocessed continuum.}

   \maketitle
%

\section{Introduction}

Accreting systems containing a compact object exhibit the most energetic events, triggered by mass accretion and/or ejection. These include cataclysmic variables (CVs) with a white dwarf accretor; low-mass X-ray binaries (LMXBs) containing a stellar mass black hole (BH) or a neutron star (NS); and active galactic nuclei (AGNs) with a supermassive BH. For LMXBs, analysis of the interplay between accretion and ejection is performed during their brightest active phases (known as outbursts), and relies on the identification of features associated with this phenomenon. In the optical regime, the canonical fingerprints of accretion are Doppler-broadened (sometimes double-peaked) emission lines (mainly of hydrogen, H, and helium, He). They arise from the outer (colder) rings of an accretion disc, built around the accretor due to mass transfer from a companion star as a result of Roche Lobe overflow and angular momentum conservation (see e.g., \citealt{Horne1986}). During the last decade, P-Cygni features (a straightforward signature of outflows) were also discovered during certain epochs of the outburst event \citep{Munoz-Darias2016}. The growing number of systems with outflow features both in LMXBs (see \citealt{MunozDarias2026} for a review) and CVs (e.g., \citealt{Cuneo2023}) alike shows that the role of mass ejection is widespread.

As a result of the renewed interest in analysing the optical outburst spectra of LMXBs, another spectroscopic feature was simultaneously found in an increasing number of systems: broad absorptions (hereafter BAs) where the emission lines are embedded. Even though reports of these features have been known for several decades (see e.g., \citealt{Joy1940} for CVs), they have usually been disregarded as contaminants hampering the study of other signatures of interest. However, the increasing number of detections during the last decade, as well as the pressing need to disentangle them from outflow features, requires that we now shift the focus towards them.

In this work, we first establish how frequent BAs are by exploring the known sample of LMXBs. We then analyse a spectroscopic database with six LMXBs exhibiting them (see Sec. \ref{sec:sample}). Through these samples, we inspect the conditions for the features to form, including their evolution over time (Sec. \ref{sec:analysis}). Altogether, they allow us to build a picture of the BAs properties to understand their origin (Sec. \ref{sec:discussion}).

\section{Sample description} \label{sec:sample}

\begin{figure*}
\centering
\includegraphics[keepaspectratio,trim=1cm 0cm 2.5cm 1cm, clip=true,width=0.95\textwidth]{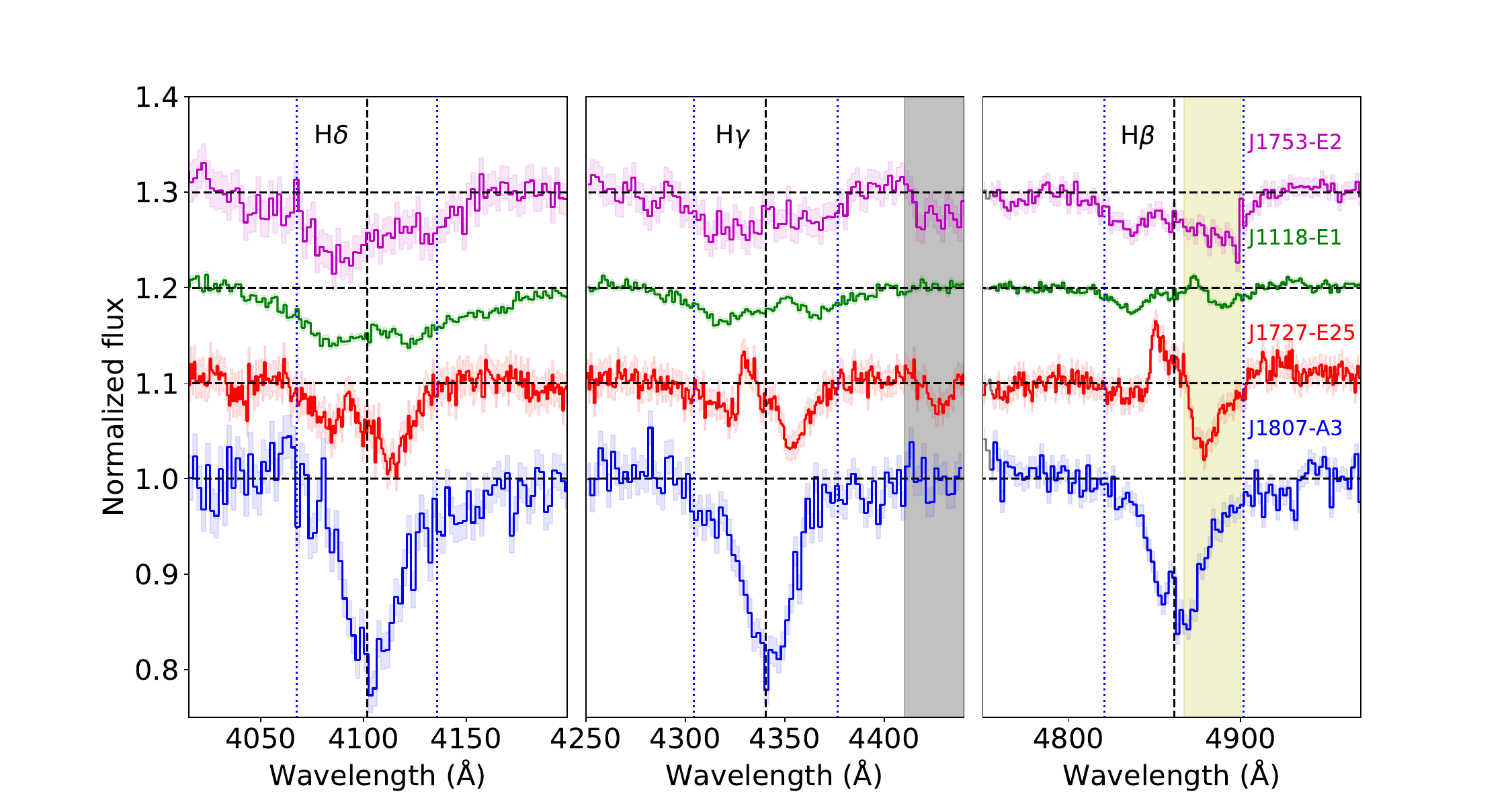}
\caption{Optical spectra from our database, zoomed into the Balmer $\rm H\delta$, $\rm H\gamma$ and $\rm H\beta$ transitions. They all show BAs with cores partially filled by the emission line. From bottom to top: J1807 (epoch 3 of dataset A, blue), J1727 (epoch 25, red), J1118 (epoch 1, green) and J1753 (epoch 2, magenta). Telluric bands and DIBs are shown as grey and yellow-shaded regions, respectively. The rest wavelength of the transition is marked with a black-dashed line, while blue-dotted lines mark velocity shifts of $\pm 2500\, \rm{km\,s^{-1}}$ as a visual reference.}
     \label{fig:specbest}
 \end{figure*}

We investigate BAs in the optical spectra of LMXBs following two complementary avenues. On one hand, we study the known population of LMXBs, including both BHs and NSs, and report on the detection of these features. On the other hand, we follow the temporal evolution of BAs within the outburst of six LMXBs with prolific spectroscopic datasets (see Fig. \ref{fig:specbest}). To fulfil both of these goals, we construct the following databases.

\begin{table}
\caption{Summary of the population study.}
\begin{tabular}{r|cccc}
CO &  BA & Non-BA  & Uncharted & Total \\
\hline \\
BH & 17(+2) &  5(+10)& 39 &   73  \\
NS & 7(+2) &  20(+18)& 238  &  285  \\
\hline 
\end{tabular}
\tablefoot{CO stands for compact object. BA and non-BA candidates are reported within parenthesis.}
\label{table:popsum}
\end{table}

\subsection{{Population database}}

We separate LMXBs according to the nature of their compact object (BH or NS) and compile their key physical properties.

\subsubsection{BHs}

We base our sample on the BlackCAT catalog from \citet{Corral-Santana2016}. At the time of publication, the updated online version\footnote{\url{https://www.astro.puc.cl/BlackCAT/}} consists of 73 sources. We individually searched the literature for studies on each system that reported optical spectroscopy during an outburst episode. We searched for BAs in the available spectra, both through visual inspection of the plots and previous reports by the original authors. This analysis led to the following classification of the population (see Table \ref{table:popsum}):

\begin{itemize}
    \item Uncharted (39): Optical spectra are not available for 39 systems, more than half of the known sample. The fact that most LMXBs are first discovered in the X-rays, together with their predominant presence across the Galactic plane, results in faint optical counterparts due to interstellar extinction (see e.g., GRS 1915+105, Av=19.7, \citealt{Chapuis2004}). Limitations from the observing facilities at the time of the X-ray discovery (leading to shallower limiting magnitudes) or a slow follow-up response might also lead to the absence of optical spectroscopy. We also include in this class those systems for which optical spectroscopy was obtained, but their counterparts were later discarded as interloper stars in the line of sight (e.g., MAXI J1813 in \citealt{Armas-Padilla2019}; MAXI J1631-479 in \citealt{Kong2019}). Finally, we decided to add to this subsample 4U 1543$-$475, whose relatively early-type companion dominates even during outburst \citep{SanchezSierras2023a}. Due to the broad and deep companion star stellar features, we cannot disentangle them from a potential BA, and therefore we cannot explore its presence. 
    \item BAs (17 confirmed + 2 candidates): A total of 19 systems exhibited BAs during at least one epoch of their outbursts (see \ref{table:bapop}). Systems where BAs are observed in at least one other transition than $\rm H\beta$ (e.g., $\rm H\alpha$, $\rm H\gamma$, $\rm H\delta$) are referred to as confirmed BA sources (17). Due to the presence of a diffuse interstellar band (DIB) contaminating the red wing of $\rm H\beta$, claiming the detection of BAs based on this sole line might be challenging. As a matter of fact, a blue-shifted absorption (such as those associated with outflows, see e.g., \citealt{MataSanchez2023b}) combined with this DIB would mimic a BA-like profile. For this reason, we refer to the two systems where the BA report is based only on $\rm H\beta$ as candidates, namely GX339$-$4 and GRS 1716$-$249. 
    \item Non-BAs (5 confirmed + 10 candidates): The remaining 15 sources have published optical spectra during their outbursts, but no obvious BAs were neither reported nor identified through visual inspection. In this regard, we note that most systems have only one (for 7 BHs) or at most two (for 3 BHs) epochs of optical spectroscopy of sufficient SNR to detect the emission lines from the accretion disc. We will refer to these 10 sources as candidate non-BAs. The remaining 5 targets are GS 2000+251 (11 epochs; \citealt{Borisov1989,Charles1991}), V404 Cygni (59 epochs across different outbursts, \citealt{MataSanchez2018}), XTE J1550$-$564 (7 epochs reported but no figures available, \citealt{Buxton1999}), SAX J1819.3$-$2525 (V4641 Sgr, with at least 23 outburst epochs when the early companion does not dominate; \citealt{Kiss2004,Munoz-Darias2018}) and MAXI J0637$-$430 (4 epochs during the same outburst, \citealt{Strader2019,Tetarenko2021}). Neither of them show BAs in spite of dedicated spectroscopic follow-ups, and they will be referred to as confirmed non-BAs. In spite of this classification, we remark that the non-detection of BAs on the available datasets does not forbid their presence at different epochs.  
\end{itemize}

\subsubsection{NSs}

The more prolific sample of LMXBs harbouring NSs allows us to expand our study independently of the compact object nature. We base our study on the most recent LMXBs catalogues published (\citealt{Avakyan2023} and \citealt{Fortin2024}), from which we remove BH and BH candidates previously considered.
As a result, our final sample of NS LMXBs consists of 285 sources, which we split among the following types using the same criteria defined in the previous section  (see Table \ref{table:popsum}):

\begin{itemize}
    \item Uncharted (238): There is no optical spectroscopy available during outburst for the majority of our sample. Similar arguments to those drawn for BHs allow us to explain this situation. Additionally, 17 sources are within globular clusters, preventing the obtention of optical spectroscopy for the counterpart due to crowding. Furthermore, 6 LMXBs had optical spectra obtained during outburst, but their early type companion star's contribution still dominated, preventing us from finding BA features. 
    \item BAs (7 confirmed + 2 candidates): A total of 9 targets show BAs. We label MXB 1837+05 and Aql X-1 as candidates, given that the only transition showing a BA is $\rm H\beta$. We also note that the subsample of confirmed BAs includes one ultracompact system (UW CrB), which has exhibited variable and broad absorption features (see e.g., \citealt{Kennedy2025,Fijma2025}). 
    \item Non-BAs (20 confirmed + 18 candidates): The remaining 38 systems correspond to those with published optical spectra during outburst, but without any trace of BA features. Candidates to non-BAs have either a single epoch (14 of them) or two epochs (4) of spectroscopy. On the remaining 20 targets, for which at least 3 epochs of spectroscopy were observed, BAs were not found in spite of the better coverage during the outburst event. Attending instead to the nature of the 38 non-BAs, we find 7 are candidates/confirmed transitional millisecond pulsars. These systems exhibit LMXB-like spectra during the so-called accretion powered state, but none has exhibited broad absorption features yet, maybe due to their low-luminosities compared to classic outbursting LMXBs. Furthermore, 11 systems are ultracompact candidates with hydrogen-poor companions, whose optical spectrum is either featureless or hydrogen-deficient (see e.g., \citealt{ArmasPadilla2023}). Therefore, only 20 non-BAs correspond to traditional LMXBs, out of which 9 are confirmed and 11 are candidates.

\end{itemize}

\subsubsection{Sample biases}

We identify biases during the sample compilation that might have an impact on the results discussed in the following sections. First and foremost, we have access to spectroscopic data of a limited fraction from the (already small) known population: barely 34 BHs ($47\%$ ) and 47 NSs ($16\%$) have at least 1 epoch of spectroscopy. 
The scarce number of spectroscopic epochs obtained for most of the outburst events (only 23 BHs and 29 NSs have more than 2 epochs of observations) plays against the detection of these features, as they are not persistently observed (see Sec. \ref{sec:analysis}). 
We are also biased towards the brightest systems in the optical range (whether due to their intrinsic brightness, their proximity to us or a lower extinction in their line of sight). Furthermore, our detection of BAs in the spectra depends on either literature reports or visual inspection of the published data. These features have traditionally not been the focus of their original studies, and as a result, we are naturally biased to mainly detect the deepest examples of BAs. 
Finally, BAs appear stronger at bluer wavelengths (see Sec. \ref{sec:analysis}), but most optical studies have traditionally focused on the $4500-7000\, {\rm \AA}$ range (covering \ion{He}{ii}$-4686$, H$\beta$ and H$\alpha$ transitions). Together with the contaminant DIB on the red wing of H$\beta$, which forbids us from unequivocally confirming a BA detection if it is only present in this transition, it further reduces the number of detections. All of these biases affect our sample in the same way: the number of systems exhibiting BAs compiled in this work should be treated as a conservative lower limit of the true sample.

\subsection{Spectroscopic database}
\label{sec:specdata}

In order to explore the evolution of BAs during outburst, dedicated spectroscopic datasets are required. To this aim, we construct a spectroscopic database which consists of six LMXBs where optical spectroscopy campaigns covered their outburst events and revealed BA features. This includes five BH transients either dynamically confirmed or established from H$\alpha$ scaling relations (Swift J1727.8$-$1613, GRO J0422$+$32, XTE J1118$+$480, Swift J1753.5$-$0127 and Swift J1357.2$-$0933; which we will refer to as J1727, J0422, J1118, J1753 and J1357, respectively), and one NS (MAXI J1807+132, hereafter J1807). A summary of the spectroscopic database is compiled in Table \ref{table:specdata}.

The dataset for J1727 consists of 25 spectroscopic epochs (50 individual spectra) obtained with the 10.4-m Gran Telescopio Canarias (GTC) at the Roque de los Muchachos Observatory (La Palma, Spain), equipped with the Optical System for Imaging and low-Intermediate-Resolution Integrated Spectroscopy (OSIRIS, \citealt{Cepa2000}). The first 20 epochs were already reduced and published in \citet{MataSanchez2024a}, dedicated to the discovery outburst of the source. The remaining 5 epochs correspond to later observations obtained during the soft state decay and transition to a low-hard state (see Table \ref{table:specdata}, also Sec. \ref{sec:hidsec}). We reduced them following the same prescription as in the original paper. We also flux-calibrated the spectra making use of spectrophotometric standard stars (Ross640, Feige 110 and GD191-B2B) observed at the end of the night for each epoch. This introduces uncertainties on the target flux calibration, as they were observed under different atmospheric conditions. We corrected for slit losses, considering wavelength dependent seeing and airmass correction. Once the spectra are calibrated in flux density units, we multiplied them by the transmission curves of the r- and g-band SDSS filters \citep{Abazajian2009}, and integrated the flux density over the wavelength dimension to obtain the optical brightness of the system on these photometric bands. J1727 data covered the full r-band regime in all 25 epochs with either the R1000B or R2500R grisms, while g-band was only fully covered with R1000B or R2000B in 20 epochs (excluding E5-E9).

We compiled two datasets for J1118. The first consists of 190 individual spectra distributed across 21 epochs during the outburst of 2000, previously published in \citet{Torres2002}. Another 4 more epochs (167 spectra) covering a later outburst in 2005 are also included in this study. Three of these epochs were previously presented in \citet{Elebert2006}, while the last of them remained unpublished to date. We obtained the reduced data for all epochs from the original authors, who followed the same prescription as in \citet{Torres2002}.

A number of observations have been reported for J0422 during its discovery outburst in 1992 and the following mini outburst a few months later. We identify three key datasets: \citet{Shrader1994}, \citet{Casares1995} and \citet{Callanan1995}. Given that 30-year old spectra are not available in public repositories, we digitized the plots in the original publications to extract the spectra. We use WebPlotDigitizer\footnote{\url{https://automeris.io/WebPlotDigitizer}} (v5.2) to this aim. A total of 8 spectra were obtained from figure 5 in \citet{Shrader1994}. They belong to the main 1992 outburst of the source and correspond to a period of 7 months, covering a wavelength range of $4000-7000\,\AA$. We also retrieved 10 spectra from Fig. 3 in \citet{Casares1995}. They correspond to the phase-binned average of 51 individual spectra obtained during four consecutive nights of the 1993 mini-outburst which followed the main event, covering $4250-5000\, \AA$. Finally, another 8 epochs proceed from Fig. 5 in \citet{Callanan1995}. They cover both the main outburst and the following echo outburst with different telescopes and instrumental setups. As a result, five spectra cover $4400-7000\, $\AA, two include $4050-7000\, \AA$, and a single spectrum extends down to $3900-7000\, \AA$ (see the original papers for further details).

We obtained 4 epochs of spectroscopy for J1753, a BH established from H$\alpha$ scaling \citep{Yanes-Rizo2025}, during the initial epochs of its 2023 outburst. They were observed with the instrumental set up detailed in Table \ref{table:specdata}, covering $3600-7700\, \AA$. We reduced this data following the same standard process as for J1727. Observations of spectrophotometric standard stars allowed us to flux-calibrate this dataset and obtain simultaneous photometry in r and g bands, following the same description as for J1727.

We compiled a dataset of observations for the high-inclination BH J1357 \citep{MataSanchez2015b,Casares2016} during its 2011 outburst event. It consists of a total of 4 epochs of spectra that were observed with different instrumental set-ups. The first epoch covers $6270-7000\, \AA$, while the remaining spectra extend down to $4500-7000\, \AA$. Inspection of the individual spectra shows they remain stable within each epoch, so we focus our analysis on the averaged spectra for each epoch to increase the SNR. Data was provided and reduced by the authors of the original publications.

The spectroscopic dataset for J1807, the only NS of our sample, covers two outburst events. We analysed the data presented by \citet{JimenezIbarra2019a}, consisting of 6 epochs (covered by 9 individual spectra) obtained during the 2017 outburst. Details on the data reduction are in the original paper. In addition, we performed a spectroscopic campaign during the latest outburst of the source in 2023, consisting of 10 epochs (28 individual spectra) detailed in Table \ref{table:specdata}. We reduced them following the same prescription as for J1727 datasets, which were obtained with the same telescope and instrument. Observations of spectrophotometric standard stars during J1807-B allowed us to flux-calibrate this dataset and obtain simultaneous photometry in r and g bands, following the same description as for J1727.

\section{Analysis and Results}
\label{sec:analysis}

\subsection{Population analysis}
\label{sec:analysispop}

\begin{figure*}
\centering
\includegraphics[keepaspectratio,  trim=0cm 0cm 0.5cm 0.5cm, clip=true,  width=0.5\textwidth]{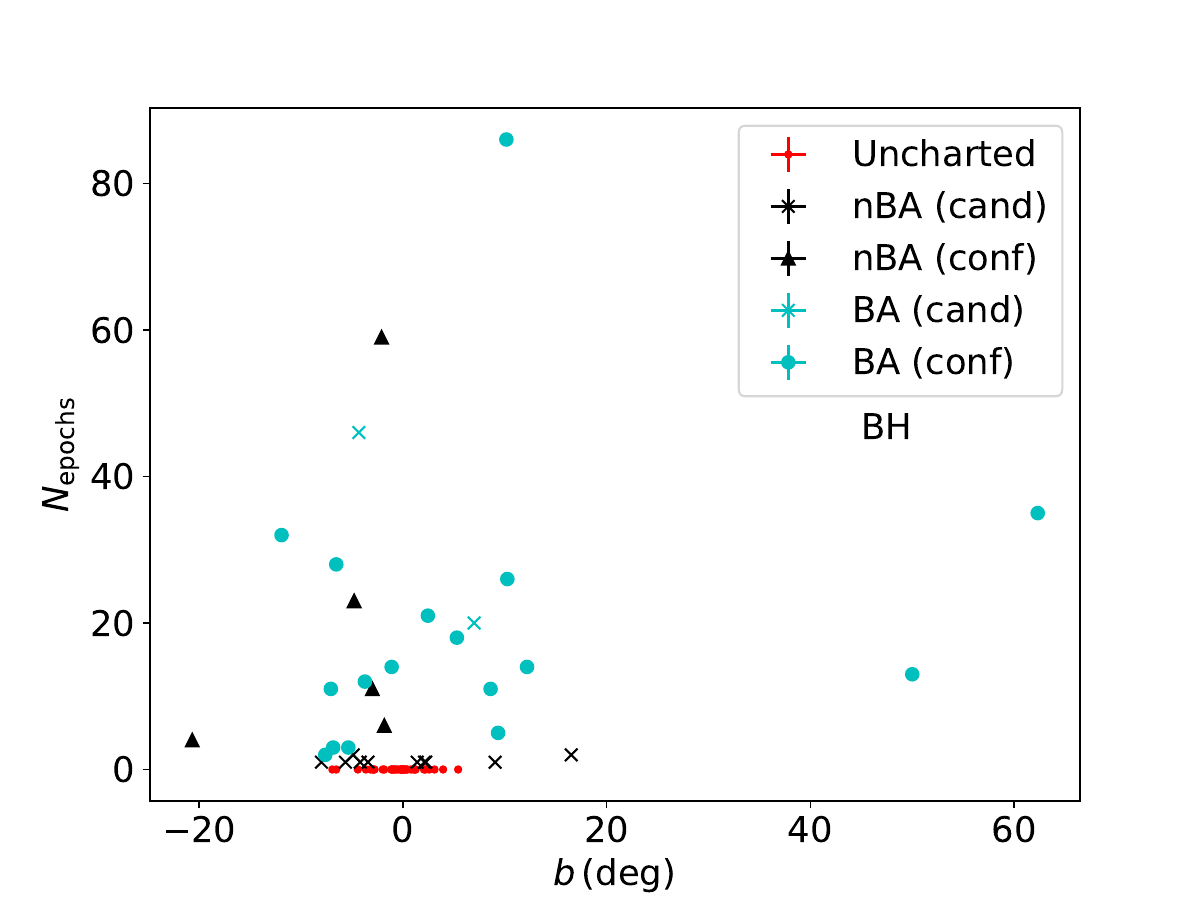}\includegraphics[keepaspectratio,  trim=0cm 0cm 0.5cm 0.5cm, clip=true,width=0.5\textwidth]{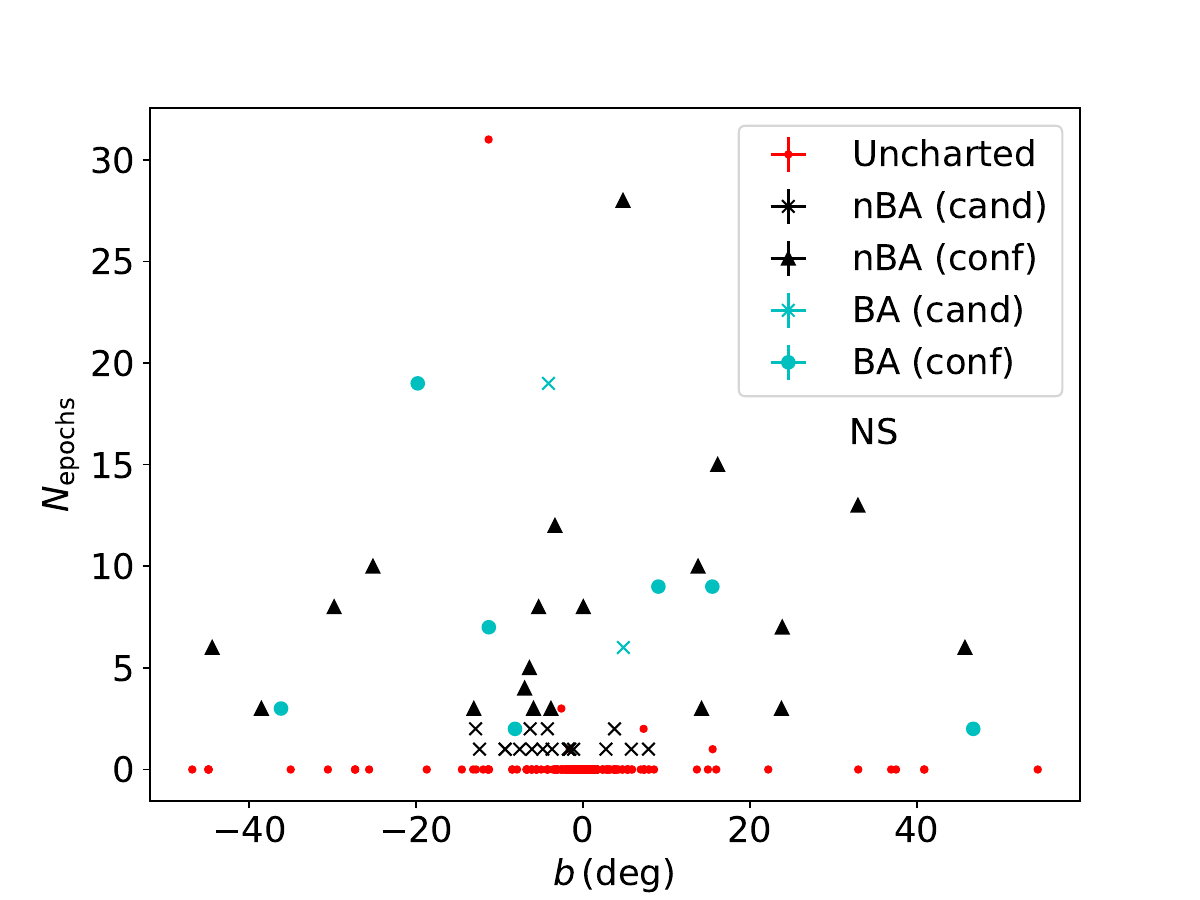}
   \caption{$b$ against $N_{\rm epoch}$ for the BH (left) and NS (right) population. Cyan symbols refer to systems with detected BA features (crosses for candidates, filled circles for confirmed systems), black symbols to non-BA systems (crosses for candidates, filled triangles for confirmed systems) and red dots to systems without available spectroscopic data (aka uncharted).}
     \label{fig:popepoch}%
 \end{figure*}

\begin{figure*}
\centering
\includegraphics[keepaspectratio,  trim=0cm 0cm 0.5cm 0.5cm, clip=true,  width=0.5\textwidth]{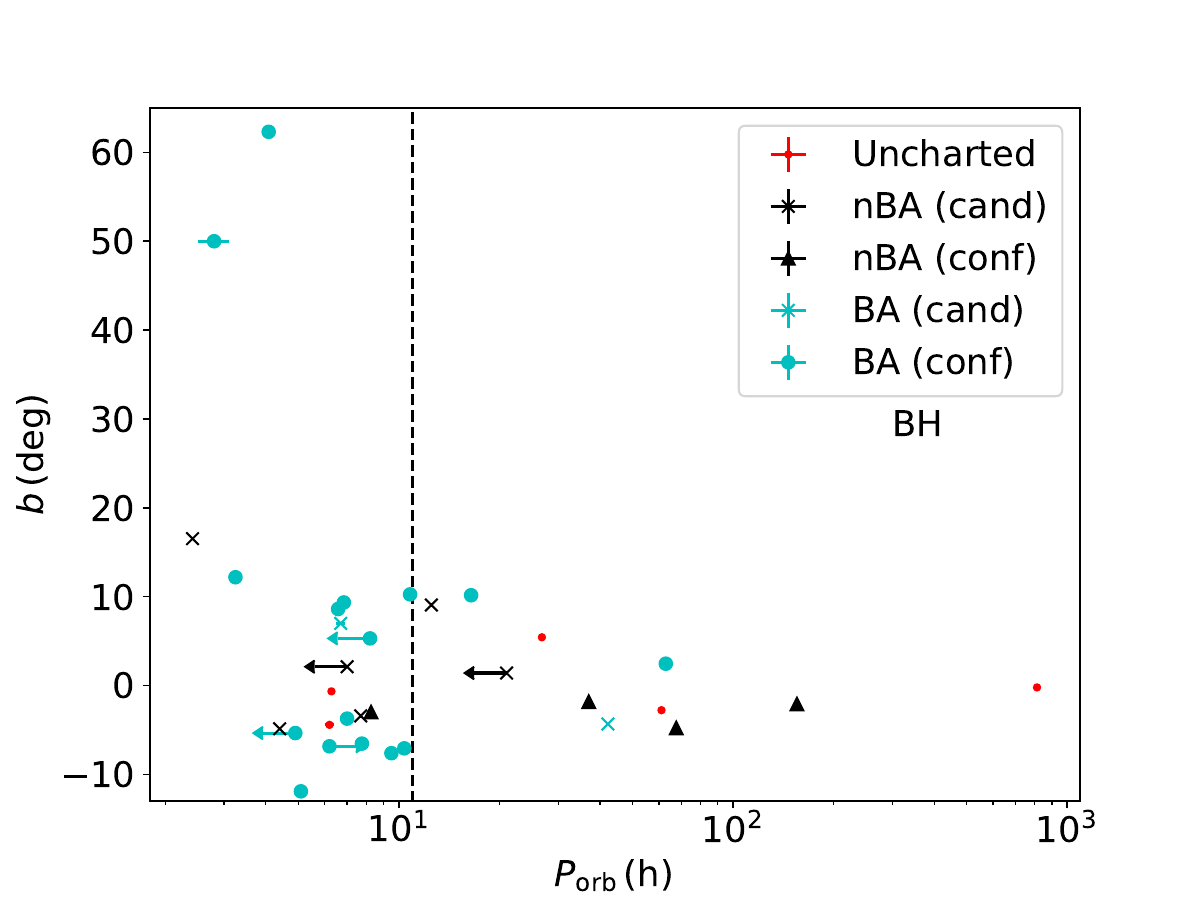}\includegraphics[keepaspectratio,  trim=0cm 0cm 0.5cm 0.5cm, clip=true,width=0.5\textwidth]{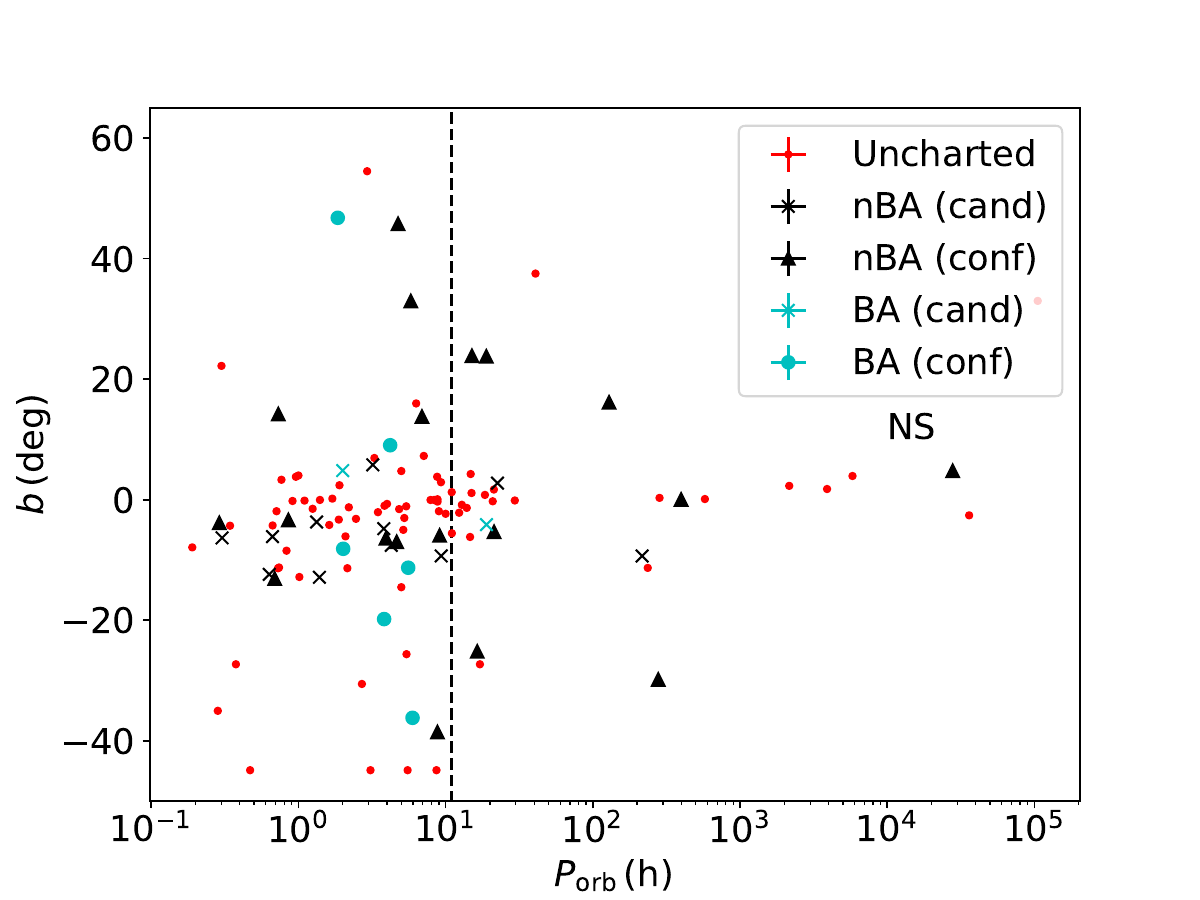}
   \caption{$P_{\rm orb}$ against $b$ for the BH (left) and NS (right) population. The colour and symbol description matches that of Fig. \ref{fig:popepoch}. The black-dashed line marks the $11\,{\rm h}$ threshold described in the text.}
     \label{fig:popb}%
 \end{figure*}

\begin{figure}
\centering
\includegraphics[keepaspectratio,  trim=0cm 0cm 0.5cm 0.5cm, clip=true,  width=0.5\textwidth]{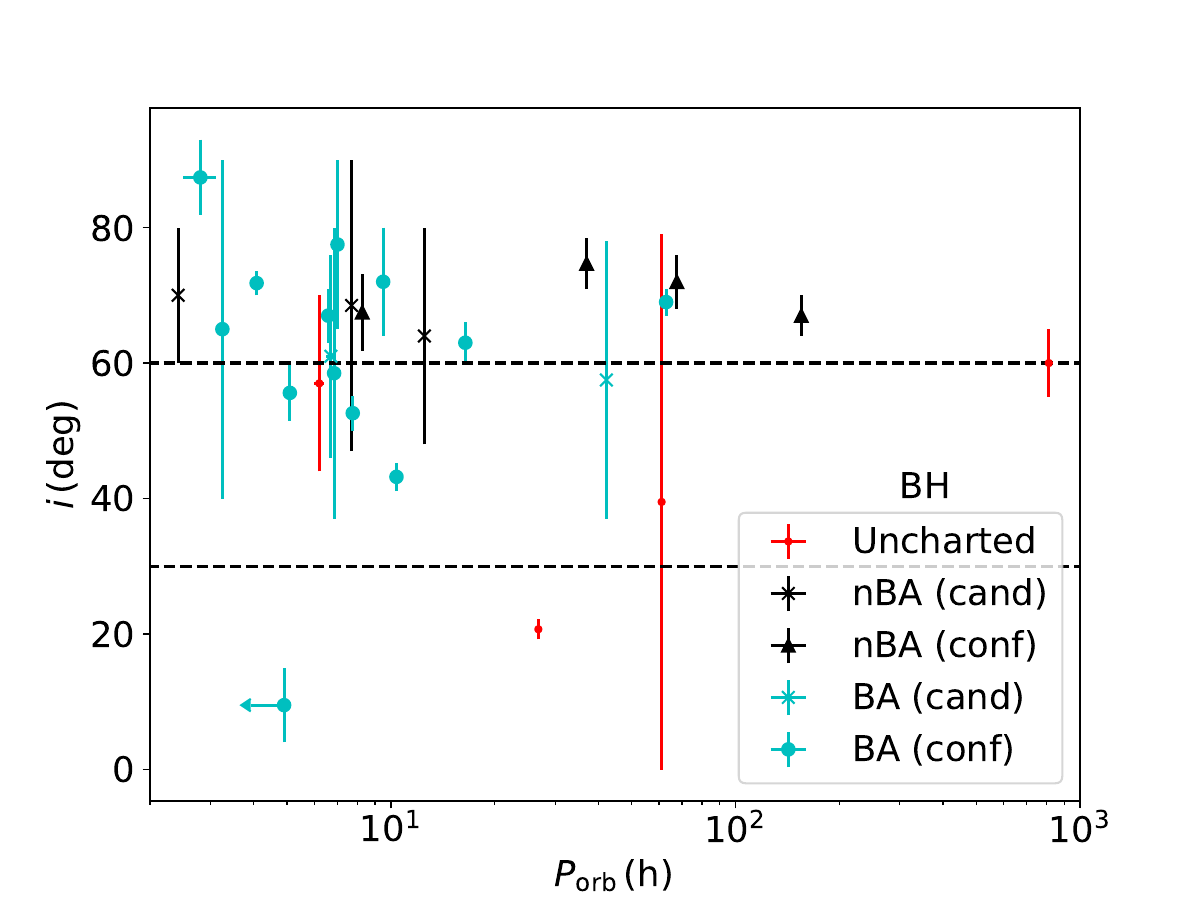}
   \caption{$P_{\rm orb}$ against $i$ for the BH population. The colour and symbol description matches that of Fig. \ref{fig:popepoch}. Dashed lines mark the thresholds at $30^ {\circ}$ and $60^ {\circ}$ defining the low, mid and high inclination regions. If only constraints to the $i$ parameter are available, we plot the central value and use the range as the associated uncertainty. }
     \label{fig:popinc}%
 \end{figure}

We analyse physical properties of the LMXB population, split into BHs and NSs, which might impact on the formation and/or detection of BAs in their optical spectra. Among them, we focus on those readily available for most members of the sample, namely the orbital period ($P_{\rm orb}$), the distance to the source ($d$) and the height over the Galactic plane ($|z|$). For BHs, the orbital inclination ($i$) is also reported. Symbols and colours allow us to separate systems exhibiting BA features at any point of their outburst evolution from those which do not (non-BAs). 
The full sample is shown in Fig. \ref{fig:popepoch}, where the number of epochs ($N_{\rm epochs}$) is plotted against the Galactic latitude ($b$). The remaining figures are limited to subsamples for which each pair of parameters is available. We have a systematically larger number of observed epochs among BHs than for NSs, as shown by a much larger fraction of uncharted targets for the latter (39/73, $53\%$ for BHs; 238/285, $84\%$ for NSs).

\subsubsection{Orbital period}

The distributions of $P_{\rm orb}$ within the known BH and NS populations are shown in Fig. \ref{fig:popb} ($b$ against $P_{\rm orb}$). Visual inspection of this figure leads us to set a threshold at $P_{\rm orb} \sim 11\,{\rm h}$ for both BH and NS alike. For BHs, this choice splits the parameter space into short (22/33 sources) and long (11/33 sources) $P_{\rm orb}$. Non-BAs are evenly split, being half (5/10 sources) below the threshold. Most BA systems ($83\%$, 15/18) are short $P_{\rm orb}$. Similar results are found for the NS population (Fig. \ref{fig:popb}, right panel), where $88\%$ BAs (7/8), $66\%$ non-BAs (20/30) and $68\%$ of the full NS population (100/146) lie below the threshold. Similar disc sizes in NSs and BHs require a longer $P_{\rm orb}$ (by a factor $\sim 2-3$) for the former, due to the lighter compact object mass. This might justify pushing the threshold for NSs to longer values, effectively removing the only BA candidate above the threshold. We decided to keep the same threshold for both populations in order to discuss a more conservative scenario. We explored the non-BA and BA period distributions to assess whether they belong to independent populations (analysing BHs and NSs separately). An Alexander-Govern test ($A=1.46$ and $p=0.23$ for BHs; $A=1.04$ and $p=0.31$ for NSs) did not allow us to reject the null hypothesis in either case.

Given our limited sample size, we are able to individually inspect the few BA outliers with $P_{\rm orb}$ above the threshold. Among NSs, the sole exception is Aquila X-1 ($P_{\rm orb}=18.95\,{\rm h}$, \citealt{Garcia1999}), a candidate BA where only H$\beta$ exhibited this feature at certain epochs (\citealt{Panizo-Espinar2021}). The remaining 10 sources above the period threshold for which optical spectra have been obtained are all non-BAs (most of them confirmed). 
Among the BH population, there are three long-period systems which exhibit BA features: a candidate BA (GX339$-$4; $P_{\rm orb}=42.21\,{\rm h}$, \citealt{Heida2017}) as well as two confirmed BAs, namely GRO J1655$-$40 ($P_{\rm orb}=62.92\,{\rm h}$, \citealt{Soria2000}) and MAXI J1820+070 ($P_{\rm orb}=16.45 \,{\rm h}$, \citealt{Torres2019}). The latter has been extensively observed (86 epochs over two different outbursts; see e.g., \citealt{Munoz-Darias2019} and \citealt{Sai2021}), but BAs have only been found in two epochs (\citealt{Garnavich2018,Yoshitake2024}). On the other hand, the detection of BAs in GRO J1655$-$40 is much more frequent \citep{Soria2000}. These observations suggest that, while BAs are not typically observed in long $P_{\rm orb}$, its detection is not forbidden. 

At this point, it is worth remarking that the larger accretion discs associated to a long $P_{\rm orb}$ can sustain longer outbursts. Given the transient detection of BAs, a more exhaustive coverage of the outburst might be required to properly cover the whole event, which could introduce an observational bias against their detection. However, we note that no BAs are found in systems with intensive coverage of their outburst (see in particular V404 Cygni, \citealt{MataSanchez2018} and SAX J1819.3-2525, \citealt{Munoz-Darias2018}). Together with the absence of definitive BA confirmation in any NS above the threshold, we conclude that BAs are preferentially shown in short $P_{\rm orb}$ systems.

\subsubsection{Orbital inclination}

Arguably the most elusive parameter to measure in LMXBs is $i$. As a result, we have measurements for a limited number of systems. There are 25 BHs for which $i$ has been measured or constrained, with uncertainties varying greatly from system to system. We collected them from the BlackCAT catalog \citep{Corral-Santana2016}, updated with the best $i$ measurements reported in \citealt{Casares2022}, and added the low inclination system MAXI J1836$-$194 ($i=4-15\, {\deg}$) as measured in \citet{Russell2014}.
To analyse the dependence with $i$, we define three different ranges: low inclination ($\lesssim 30^\circ$), medium inclination ($30-60^\circ$) and high inclination ($\gtrsim 60^\circ$). Plotting these separate ranges (see Fig. \ref{fig:popinc}) allows us to determine that $11\%$ of the BHs are low-inclination, $31\%$ are mid-range and $58\%$ are high inclination. This distribution is consistent with randomly oriented orbits (i.e., uniform in $\cos{i}$, favouring edge-on inclinations), which predicts $13.5\%$, $37.5\%$ and $50\%$ for the low, mid and high inclination ranges, respectively. This suggests a homogeneous sampling in spite of the limited number of systems. No clear connection between this physical parameter and the detection of BA features is observed. We nevertheless acknowledge that our sample is limited, specially among low inclination systems, precluding a detailed inspection. Indeed, low inclination LMXBs are intrinsically harder to characterise, due to both the uniform distribution in $\cos{i}$ and their weak ellipsoidal modulation in quiescence (from which most determinations of $i$ are obtained).

\subsection{Outburst-resolved spectroscopic analysis}

In order to explore the evolution of BAs during outburst, we performed a dedicated study on six LMXBs where optical spectroscopy campaigns covered their outburst events and revealed BA features (see Sec. \ref{sec:specdata}). We analysed the Balmer series (H$\alpha$, H$\beta$, H$\gamma$ and H$\delta$ transitions), with a particular focus on bluer transitions as they have exhibited the most prolific examples of BAs. To systematically assess the properties of BAs, we normalise each transition independently through the fit of a low order polynomial to the nearby continuum of each of the lines of interest, for each individual spectrum. The typical profiles of the Balmer series follow that of a canonical accretion disc (broad, double-peaked emission lines) embedded into a BA. To systemically capture the BA properties, we designed the following toy model. 

The emission component is fitted with a double Gaussian profile. The height of both Gaussians is controlled by the area of the red peak ($A_{\rm red}$) and the intensity ratio between the blue and red peaks ($I_{\rm ratio}=A_{\rm blue}/A_{\rm red}$), to allow for asymmetric profiles. For consistency, we define the area of the full double peak profile as $|EW_{\rm em_i}|=A_{\rm red_i}\cdot (1+I_{\rm ratio})$. Given that we are working with normalised spectra, this parameter is effectively the equivalent width ($EW$) of the emission component, where we follow the convention of positive values for absorption features and negative for emission features. The profile centroid is controlled by the offset ($\mu$) parameter (referred to the rest wavelength of each transition), the separation between the Gaussians' centroids is $DP$ and the width of each Gaussian is defined through $\sigma$. Other than $|EW_{\rm em_i}|$, which determines the area of the individual transitions, the remaining parameters are linked for all simultaneously fitted lines in each individual spectrum.

We tested three different profiles for the absorption component: Gaussian, Lorentzian and a Voigt function. Our tests showed that, while all three resulted in reasonable fits, the best-fitting $\chi^2$ for the Gaussian profile was slightly but systematically better than its Lorentzian counterpart, and that the Voigt profile fit always preferred a null contribution for the Lorentzian component. For this reason, we hereafter report only the results from a Gaussian fit to the absorption component, controlled by a common centroid ($\mu_{\rm abs}$), as well as independent width ($\sigma_{\rm abs_i}$) and depth (defined by the area of the absorption component, $EW_{\rm abs_i}$) for each transition. Both $\sigma$ and $\sigma_{\rm abs_i}$ parameters were corrected from the corresponding instrumental resolution on each measurement. To obtain the best fit to each spectrum we employ a python script based on \textsc{emcee}, an implementation of a Markov Chain Monte Carlo (MCMC) sampler \citep{Foreman-Mackey2013}. We define the likelihood function as the $\chi^2$ resulting from the comparison to our model, and report 16, 50 and 84$\%$ quantiles of the marginalised distributions for each fitted parameter (corresponding to $1\sigma$ uncertainties for a Gaussian-like distribution) in Table \ref{table:specfitparam}. It follows a dedicated study for each target.

\subsubsection{Swift J1727.8$-$1613}

J1727 spectroscopic sample was partly introduced (E1 to E20) in \citet{MataSanchez2024a}, where the authors mention the development of BAs in the Balmer series as the outburst evolves. The extended dataset we present in this work includes 5 further epochs (E21 to E25) where BAs become more prominent. We fitted all epochs with the model previously defined (see Fig. \ref{fig:J1727fitspec}). Both the nearby DIB ($\sim 4881\, \AA$) and the \ion{He}{i}-$4921.93\, \AA$ emission line appear close to the $\rm H\beta$ line, and they could bias the results from our fit. For this reason, we decided to mask out the affected region at the red wing of $\rm H\beta$, even though this diminishes our detection sensitivity to potential BAs on this line. The resulting best-fitting parameters are shown in Fig. \ref{fig:J1727fitparams}. The overall description of the fitting results follows. We focus exclusively on those fits resulting in a detection of BAs over the $3\sigma$ limit for the $EW_{\rm abs}$ parameter in at least one of the transitions.

The behaviour of the emission component parameters is typical from accretion discs in LMXBs. The double peak profile does not become particularly asymmetric, and the width of the Gaussians is remarkably stable over the whole outburst. We identify two regimes where the behaviour of $DP$ changes significantly (by a factor $\sim2$, see Table \ref{table:specfitparam}). We refer to them as the narrow-line regime (E1-E20), corresponding to the hard, hard-intermediate (HIMS), soft-intermediate (SIMS), and initial soft state decay; and the broad-line regime (E21-E25), limited to the low-luminosity soft state and the transition to the low-hard state. The regime swap occurred between two epochs separated by a 4-month gap (due to Sun constraints) during the soft state X-ray decay. 
$|EW_{\rm em}|$ evolves smoothly, being systematically weaker for shorter wavelengths across the full sample. The centroid of the emission ($\mu_{\rm{em}}=-210\pm 100\,\rm{km\,s^{-1}}$) is consistent with the systemic velocity ($\gamma= -181\pm 4\,\rm{km\,s^{-1}}$; \citealt{MataSanchez2025a}), supporting its association with the accretion disc of the binary. We nevertheless note its large standard deviation, which reveals significant variability on the parameter during the outburst, specially during the narrow-line regime.

On another vein, during the narrow-line regime, the absorption component remains either undetected (as for the majority of $\rm H\beta$ and $\rm H\gamma$, considering a $3\sigma$ threshold) or shallow ( $\rm H\delta$, $EW_{\rm abs}<2\,\AA$). The broad-line regime shows a smooth increase in the parameter for all transitions, reaching a maximum of $EW_{\rm abs}=5.3^{+0.4}_{-0.6}\,\AA$ for $\rm H\delta$. The width of the absorption is consistently broad for the $\rm H\delta$ transition across both regimes ($\sigma_{\rm abs}=1100\pm 200\,\rm{km\,s^{-1}}$). On the other hand, $\rm H\gamma$ and $\rm H\beta$ absorptions are narrow and variable during the initial regime but they become broad and consistent with $\rm H\delta$ after the transition to the broad-line regime. Finally, we note the centroid of the absorption shows large variability (similar to the emission component), still consistent with the systemic velocity. 

Given our intensive coverage during the outburst event, we also analysed a potential dependency of the measured parameters with the orbital phase. To this aim, we employed the ephemerides from \citet{MataSanchez2025a}. We found no orbital dependency for any of the parameters, suggesting the profile changes are rather linked to the outburst evolution.

\subsubsection{XTE J1118$+$480}

We have over 350 spectra from two outbursts of J1118, namely the 2000 and 2005 events. We first inspected the spectral evolution within each epoch. The most prolific example is E22, with 59 consecutive spectra covering over $5\,\rm{h}$. We find that the best fitting parameters remain stable during the whole dataset. This result holds for each of the epochs analysed, proving the stability of BAs features over periods of at least a few hours. Hereafter, we focus on the averaged spectrum for each epoch. We have access to 21 epochs covering the 2000 outburst and 4 further epochs during 2005. Both outburst events were hard-only outbursts, with low X-ray luminosity (see \citealt{Zurita2006}). The 2000 outburst was covered during 4 consecutive months, while the outburst remained stable in X-rays ($\sim 1.5-2.5\,\rm{counts\, s^{-1}}$ flux variability, see \citealt{Torres2002}). The 2005 outburst coverage is scarcer, but includes the outburst peak and decay ($R\sim 13.5$ to $R\sim 14.5$ during E21-24, and dropping to $R\sim 16.5$ on E25, see \citealt{Zurita2006}). We fitted the spectra following the general prescription (see Fig. \ref{fig:J1118fitspec} and \ref{fig:J1118fitparams}). We find no clear differences between the 2000 and 2005 outbursts in any of the fitted parameters, so we will analyse the dataset as a whole.

The emission component centroid velocity $\mu_{\rm{em}}$ is highly variable, with a mean and standard deviation of $\mu_{\rm{em}}=50\pm 100\,\rm{km\,s^{-1}}$ (consistent with the systemic velocity of $16\pm 6\, \rm{km\, s^{-1}}$; \citealt{Torres2004}). The peak-to-peak ratio is similarly variable, with epochs of particularly asymmetric profiles (e.g., E9). The emission component width and double peak separation are stable across both outbursts. 
The $|EW_{\rm em}|$ evolves smoothly and remains below $|EW_{\rm em}|<4\,\AA$ during the entire dataset.

The absorption component is consistently broad over the whole outburst event, particularly for $\rm H\delta$ ($\sigma_{\rm abs}=2200\pm 400\,\rm{km\,s^{-1}}$). $EW_{\rm abs}$ follows a similar behaviour, where $\rm H\delta$ exhibits the deepest BA examples, peaking at E25 with $EW_{\rm abs}=14.3\pm 1.2\, \AA$, corresponding to the faintest epoch recorded. The centroid of the absorption component is remarkably variable as well. We performed a Lomb-Scargle periodogram on $\mu_{\rm abs}$ and found maximum power at a period of $\sim 53\, {\rm d}$ for the 2000 outburst event, consistent with the disc precession period (see Sec. \ref{sec:orbitalperiod}). A sinusoidal fit to our limited dataset resulted in a radial velocity semi-amplitude of $\sim 240 \pm 30\, \rm{km\,s^{-1}}$, oscillating around a central velocity of $\sim 260 \pm 30\, \rm{km\,s^{-1}}$, notably offset from the systemic velocity.

\subsubsection{GRO J0422$+$32}

The sample for J0422 has limited and uneven spectral coverage. It did not exhibit the X-ray characteristic features indicating a transition to the soft state, neither during the main outburst in 1992, nor during the subsequent minioutburst in 1993 (see e.g., \citealt{Shrader1997}). For this reason, we will consider them all as hard-only outbursts (see e.g., \citealt{Alabarta2021}). We analysed the three datasets separately (A, B and C; see Table \ref{table:specdata}). The systemic velocity is $11\pm 8\, \rm{km\,s^{-1}}$ \citep{Harlaftis1999}.

Dataset A shows 8 spectra obtained from August 1992 to March 1993, covering the main 1992 outburst with a monthly cadence. They cover $\rm H\delta$ to $\rm H\beta$. The emission component shows highly variable radial velocity and $I_{\rm ratio}$, while stable $\sigma_{\rm em}$ and $DP$ (after removing an outlier in A4). The absorption component seems consistently blue-shifted during most of the outburst A1-A7 ($\mu_{\rm em}=-130\pm 40\, \rm{km\,s^{-1}}$), but redshifted by the end of it (A8, $\mu_{\rm em}=200\pm 50\, \rm{km\,s^{-1}}$). The width and depth of the BA are variable, but overall consistent across the explored transitions, showing deeper profiles in bluer wavelengths.

Dataset B spectra consistently covers $\rm H\gamma$ and $\rm H\beta$. It corresponds to a phase-folded set of spectra obtained during the initial decline of the 1993 December mini-outburst. The short period when these observations were obtained (4 consecutive nights) allows us to inspect the evolution of the BAs on short timescales. Most of the emission and absorption component parameters are stable at typical values (except for an outlier on epoch B3). 
The original paper \citep{Casares1995} reported that the emission component of Balmer lines experienced Doppler-shifting at the orbital period of the system with semi-amplitudes of the order of $\mu_{\rm em}\sim 100\, \rm{km\,s^{-1}}$. We confirm this trend for the emission component in $\rm H\gamma$ and $\rm H\beta$, varying around a mean velocity of $\mu_{\rm em}=20\pm 40\, \rm{km\,s^{-1}}$, consistent with the systemic velocity. 
A similar analysis reveals periodic variability for the absorption component centroid $\mu_{\rm abs}$. We performed a sinusoidal fit to this parameter and obtained a radial velocity semi-amplitude of $\sim 48\pm 10\, \rm{km\,s^{-1}}$, slightly off-phase compared to the known ephemerides ($\phi_0=0.10\pm 0.03$) and oscillating around a central velocity of $\sim 146\pm 7\, \rm{km\,s^{-1}}$, significantly larger than the systemic velocity. 

For dataset C, the uneven spectral coverage led us to mainly focus on $\rm H\beta$ for the analysis, and include  $\rm H\gamma$ when present. The observations span over a year, including a short period at the beginning of the 1992 outburst (C1-5) as well as the 1993 mini outburst (C6-8). The 1992 outburst starts by exhibiting relatively deep BAs in $\rm H\beta$ (up to $EW_{\rm{abs}}=3.7\pm 0.7\, \AA$ in C1), becoming shallower as the outburst evolves ($EW_{\rm{abs}}=0.73\pm 0.04\, \AA$ in C5). During 1993 outburst, the BAs are less variable, and exhibit values comparable to those of the final epochs of 1992 outburst. The width of $\rm H\beta$ BAs remains stable across all epochs. The centroid velocity is highly variable during 1992 outburst, from $\sim -500\,\rm{km\,s^{-1}}$ in epoch C1 to $\sim 50\,\rm{km\,s^{-1}}$ in epoch C5. During the 1993 minioutburst, the radial velocity is stable and consistent with C5. 

\subsubsection{Swift J1753.5$-$0127}

Observations were performed during the outburst of the source which started on September 28th 2023 and lasted $\sim 5$ months \citep{Alabarta2024}
We obtained 4 epochs which allow us to explore the $\rm H\beta$, $\rm H\gamma$ and $\rm H\delta$ transitions. These epochs were obtained during the initial rise of the 2023 outburst (days 5 to 15, referred to the outburst trigger on HJD 2460215.5), and before the source reached a plateau in optical brightness ($r \sim 17$, \citealt{Alabarta2024}). The few references regarding X-ray observations at the time of publication suggest the source was in the hard state during our 4 epochs \citep{Alabarta2023b}. We focus on the Balmer series to compare with the rest of the population, but we note that a BA appeared on \ion{He}{ii}--4686 (E1, see Sec. \ref{sec:heii}).

The emission component is stable over the duration of our observations (Fig. \ref{fig:J753fitspec}), with typical values comparable to the rest of the sample and a centroid velocity consistent with the systemic value ($131\pm 55\, {\rm km\, s^{-1}}$, \citealt{Yanes-Rizo2025}). The centroid velocity of the absorption component is also consistent with the above. However, contrary to other systems where the BAs become deeper at later epochs of the outburst, J1753 exhibits the deepest example on the initial observation ($EW_{\rm{abs}}=17\pm 4\, \AA$ on $\rm H\delta$ during E1). In this regard, it is worth remarking that our initial epoch is the faintest of our sample, obtained during the outburst rise. Table \ref{table:specfitparam} compiles the mean values for each fitted parameter, but given the low number of spectra (only E3 and E4 show $>3\sigma$ BA detections), individual measurements dominate. 

\subsubsection{Swift J1357.2$-$0933}

The limited wavelength coverage of the 4 epochs observed for J1357 only makes $\rm H\alpha$ and $\rm H\beta$ available for our study. We fitted the averaged spectra corresponding to four epochs (days 27, 47, 72 and 101 referred to the outburst trigger). The outburst peaked soon after its discovery, and remained in the hard state as the luminosity decayed. BAs clearly developed in the later, lower-luminosity epochs (see Fig. \ref{fig:J1357fitspec}). A systemic velocity of $\gamma = -130 \pm 17\,\rm{km\,s^{-1}}$ was proposed by \citet{JimenezIbarra2019b} based on the emission line analysis in outburst, as the companion star features remain undetected even in quiescence (see e.g., \citealt{Torres2015,MataSanchez2015b,Anitra2023}).

We fit the line profiles and reported the averaged values in table \ref{table:specfitparam}, though the scarce number of spectra makes them dominated by individual measurements. The emission component is similar to that observed in other members of our sample, being the most remarkable feature its large $DP= 1707\pm 13\, {\rm km\,s^{-1}}$, consistent with its optically-dipping nature. BAs show quite variable centroids, with a minimum of $\mu_{\rm abs}=-380\pm 30\,\rm{km\,s^{-1}}$ (E3) and a maximum of $\mu_{\rm abs}=40\pm 30\,\rm{km\,s^{-1}}$ (E4). The maximum depth of $EW_{\rm{abs}}=7.9\pm 0.2\, \AA$ is observed for $\rm H\beta$ on E4, being also the broadest BA of our sample ($\sigma_{\rm abs}=2980\pm 30\, {\rm km\, s^{-1}}$).

\subsubsection{MAXI J1807$+$132}

The dataset for J1807 covers the 2017 (6 epochs) and 2023 (10 epochs) outbursts. \citet{JimenezIbarra2019a} proposed a systemic velocity of $\gamma = -145 \pm 13\,\rm{km\,s^{-1}}$, based on a line fitting model (similar to that employed in our study) to dataset A spectra, but including a larger set of lines (e.g., $\rm H\alpha$, \ion{He}{ii}--4686). 

Only three spectra from the 2017 dataset have enough SNR to perform meaningful fits (A1-3, \citealt{JimenezIbarra2019a}). They correspond to a short period spanning 2 weeks. Comparison of the photoindex observed in \citet{JimenezIbarra2019a} with the later 2023 outburst \citep{Rout2025} suggests the system was in the hard state during 2017 observations. The best fitting model reveals consistent parameters for the emission component, including a centroid velocity matching the systemic velocity. Regarding the absorption component, $\sigma_{\rm abs}$ is quite stable, while $EW_{\rm abs}$ increases from A1 to A3, reaching a maximum in  $\rm H\delta$ of $EW_{\rm abs}=7.9\pm 0.9\,\AA$. The centroid of the absorption retains positive velocities, again consistent with the systemic velocity.

During its 2023 outburst, J1807 exhibited the canonical X-ray behaviour of an LMXB \citep{Rout2025}. The 10 epochs we observed are associated with their corresponding X-ray state following the aforementioned paper (see Fig. \ref{fig:J1807fitspec} and \ref{fig:J1807fitparams}). Both the emission and absorption components followed the same behaviour as observed in 2017, with a maximum $EW_{\rm abs}=8.0\pm 0.3\,\AA$ being achieved by $\rm H\delta$ during B6. J1807 study shows how similar BAs resurge from the same source in two different outburst episodes 6 years apart, developing across all X-ray states.

\subsubsection{Combined analysis of the spectroscopic sample}
\label{sec:datacomp}

\begin{figure*}
\centering
\includegraphics[keepaspectratio, trim=0cm 0cm 0cm 0cm, clip=true, width=0.52\textwidth]{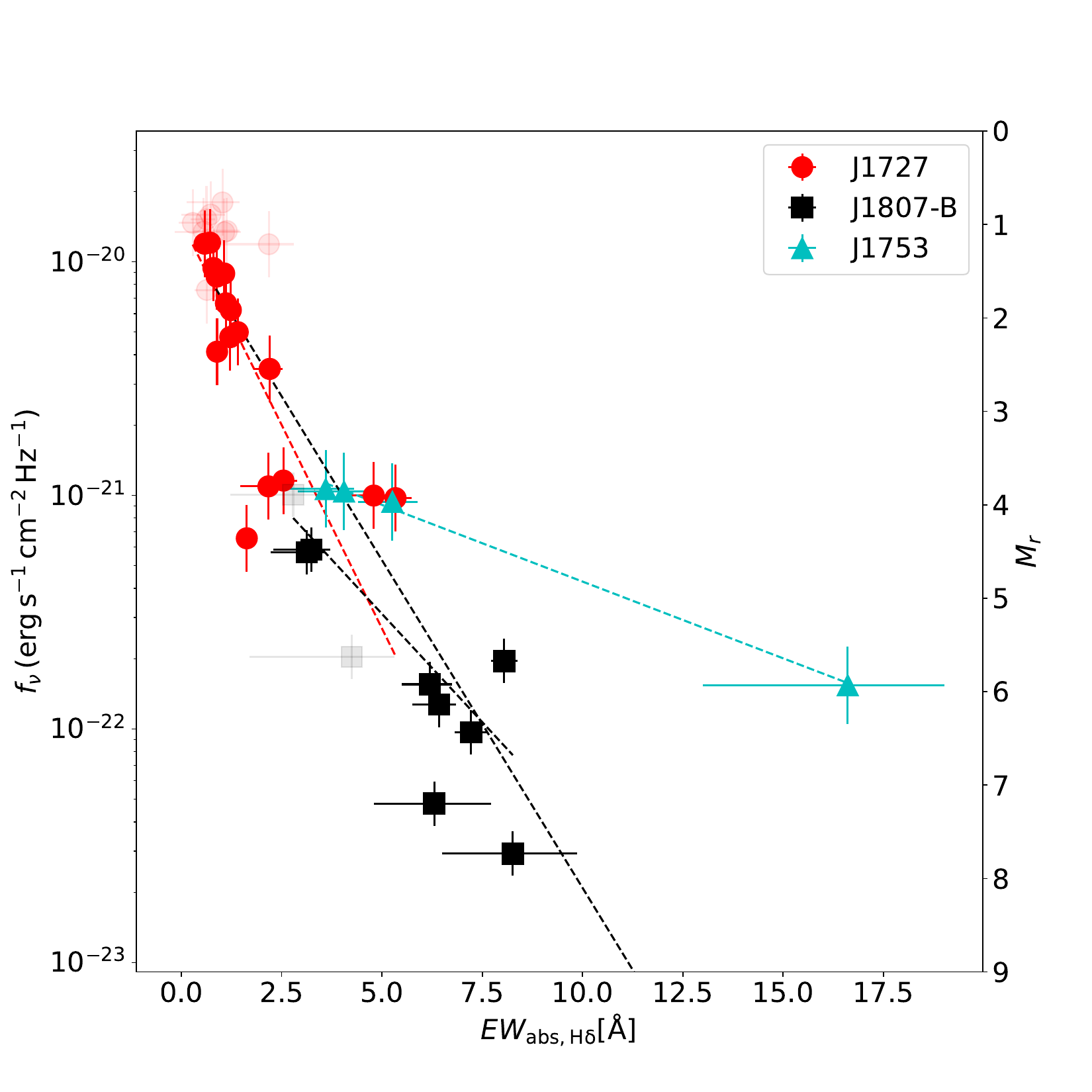}\includegraphics[keepaspectratio, trim=1cm 0cm 1cm 2cm, clip=true,  width=0.48\textwidth]{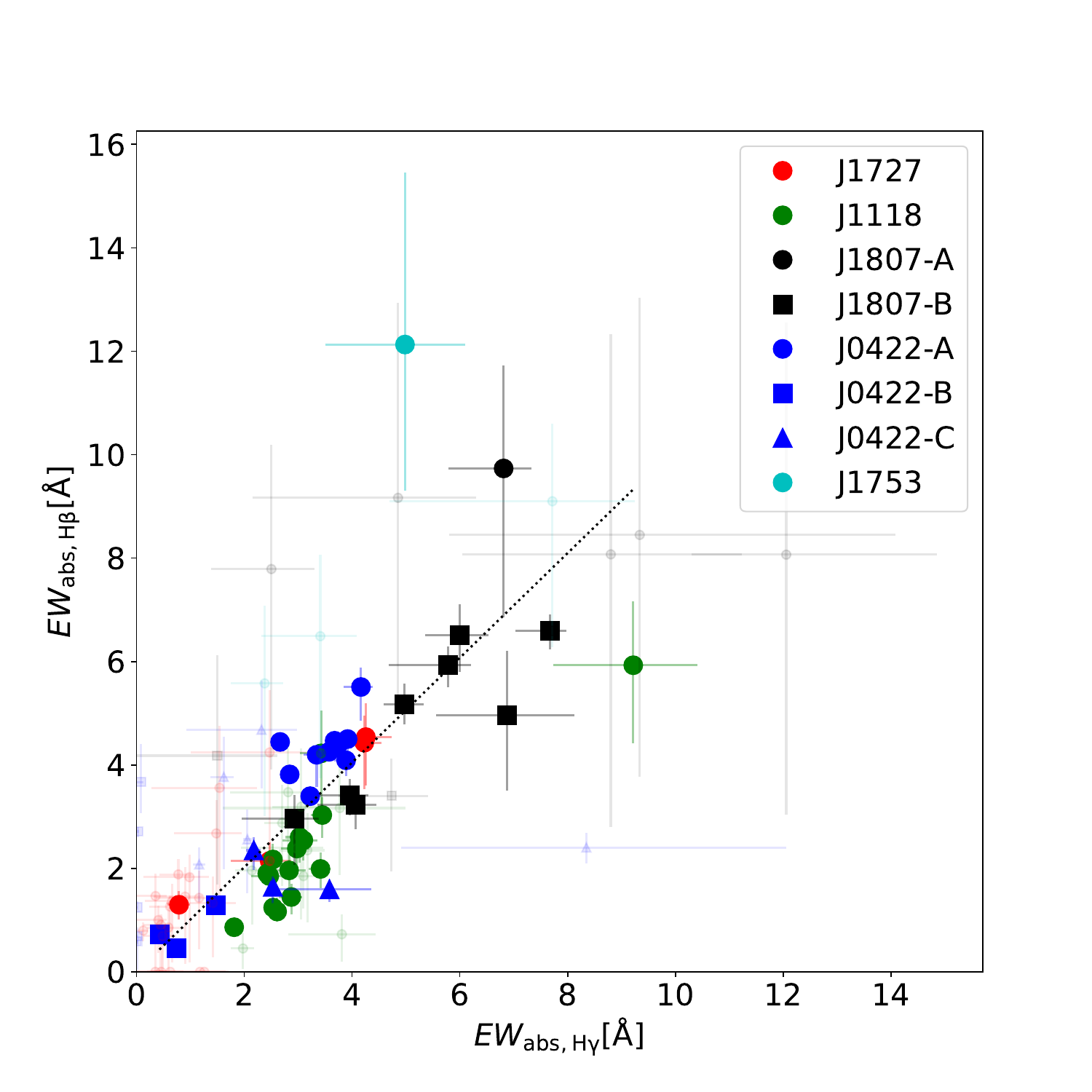}
\caption{Left panel: $EW_{\rm abs,H\delta}$ against r-band brightness (left axis for flux density units, $f_{\rm \nu}$; right axis for absolute magnitudes, $M_{\rm r}$) from the best fitting results of the subsample with flux-calibrated spectra. The colour code mark the systems associated with each data point. A black, dashed line depicts the best linear fit to the correlation, being cyan-dashed, red-dashed and black-dashed lines the equivalent for individual datasets. Right panel: $EW_{\rm abs,H\beta}$ against $EW_{\rm abs,H\gamma}$ from the best fitting results of the whole spectroscopic sample. The colour code marks the systems associated with each data point, while symbols differentiate between subsets as described in the legend. A black, dashed line depicts the best linear fit to the correlation, consistent with a 1:1 ratio. Transparent symbols correspond to fitting parameters consistent with null BA within $3\sigma$. }
     \label{fig:brigthnessvsEWabs}%
 \end{figure*}

\begin{figure*}
\centering
\includegraphics[keepaspectratio,  trim=0cm 0cm 0.5cm 0.5cm, clip=true,width=0.33\textwidth]{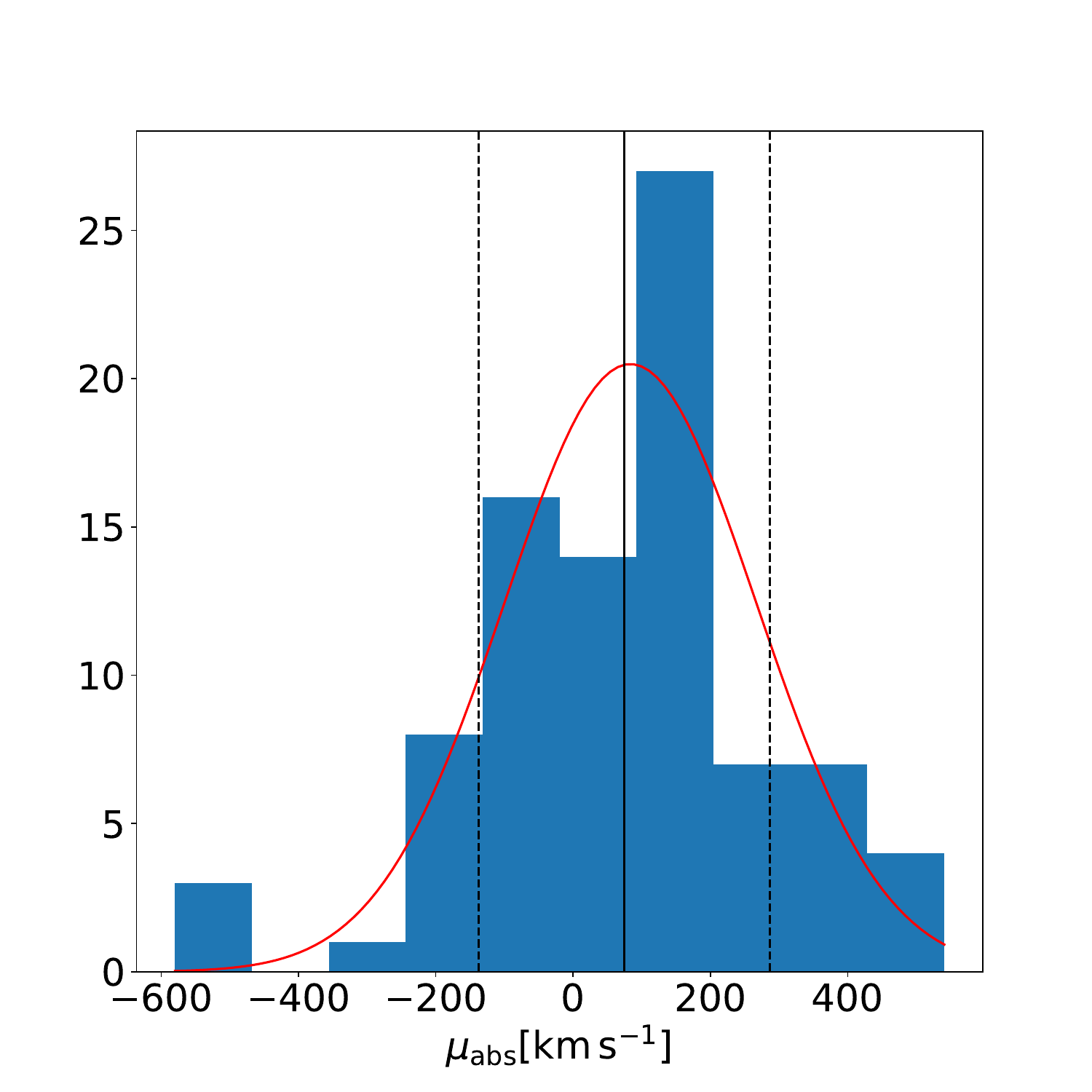}\includegraphics[keepaspectratio,  trim=0cm 0cm 0.5cm 0.5cm, clip=true,width=0.33\textwidth]{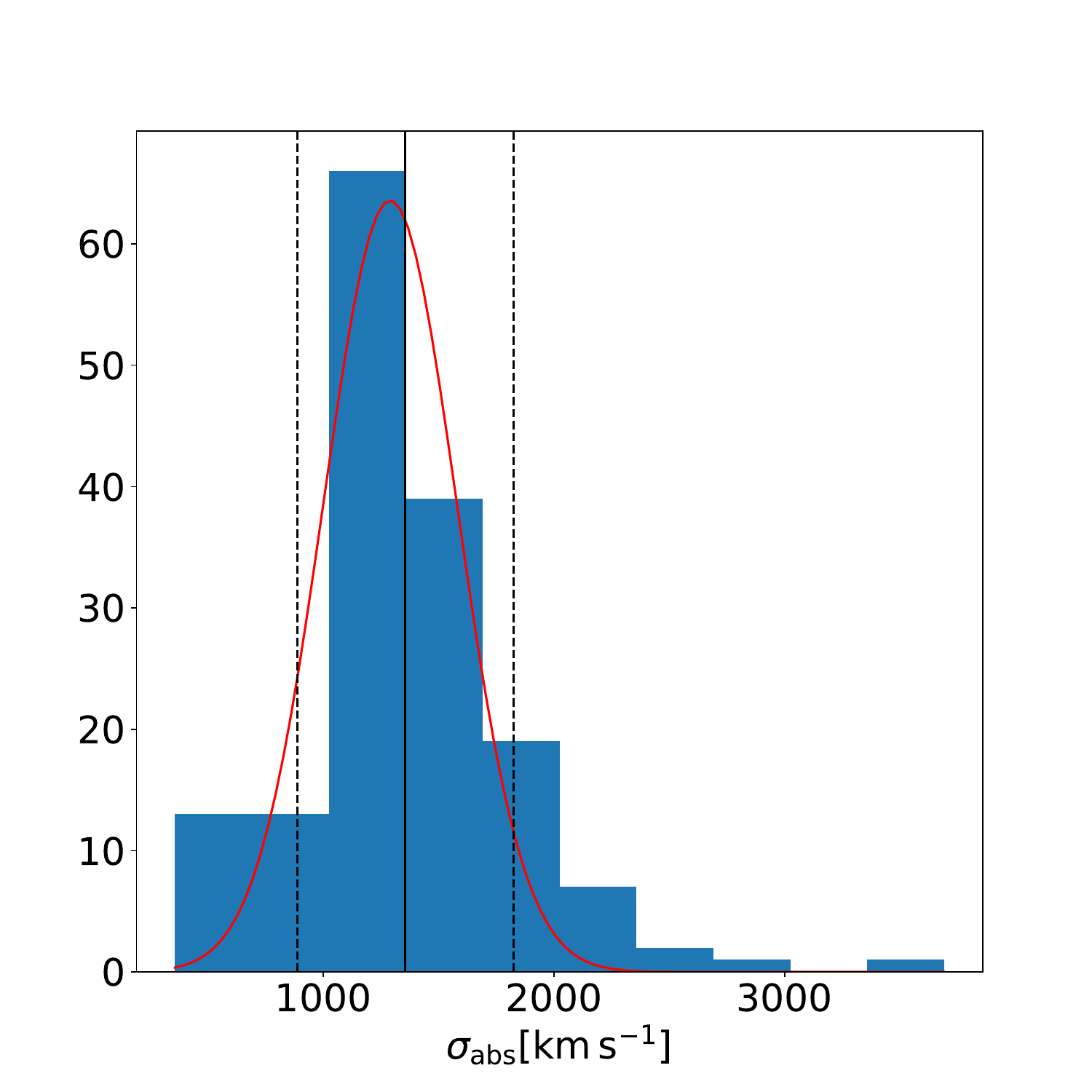}\includegraphics[keepaspectratio,  trim=0cm 0cm 0.5cm 0.5cm, clip=true,  width=0.33\textwidth]{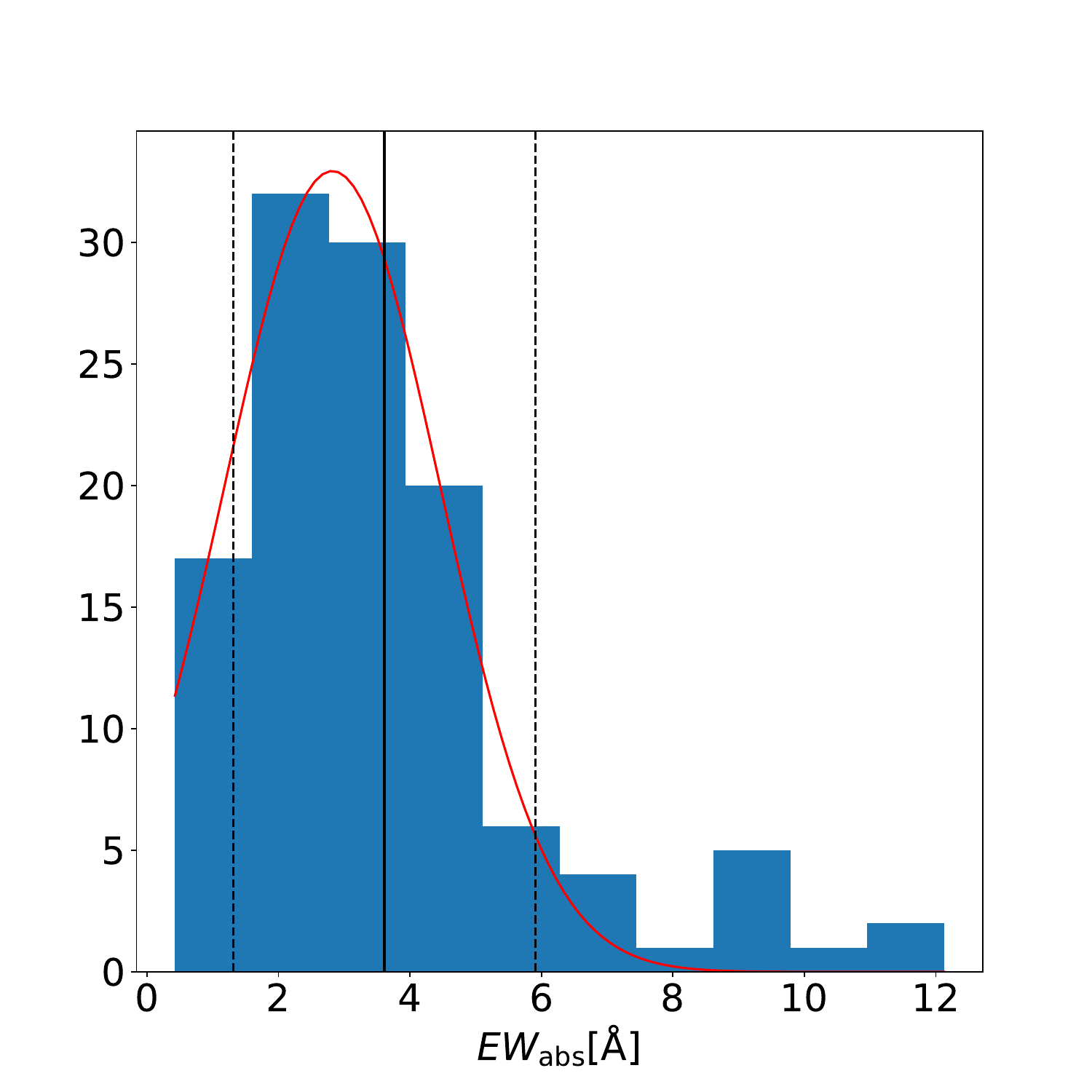}
   \caption{Histograms of the spectroscopic database for $\mu_{\rm abs}$, $\sigma_{\rm abs}$ and $EW_{\rm abs}$ (combining the results for $\rm H\beta$ and $\rm H\gamma$, which produced consistent distributions). The black continuous and dashed lines mark the mean and $1\sigma$ of the distribution, respectively. A Gaussian fit to the histogram is also shown as a red line.}
     \label{fig:histograms}%
 \end{figure*}

We searched for correlations between the fitted parameters in the combined spectroscopic database. To this aim, we analysed all possible pair combinations, employing the Spearman’s correlation coefficient ($r_s$) to assess their strength.
The strongest correlation (the only above $r_s \sim 0.7$) is found for $EW_{\rm abs,H\beta}$ against $EW_{\rm abs,H\gamma}$ (Fig. \ref{fig:brigthnessvsEWabs}, right panel; $r_s=0.89$). A linear fit to the full sample reveals a line ratio of $EW_{\rm abs,H\beta}/EW_{\rm abs,H\gamma}=1.05\pm 0.05$. To further investigate this, we also analysed the weaker correlation ($r_s=0.63$) between $EW_{\rm abs,H\gamma}$ and $EW_{\rm abs,H\delta}$. A linear fit reveals $EW_{\rm abs,H\gamma}/EW_{\rm abs,H\delta}=0.67\pm 0.05$, suggesting that $\rm H\delta$ is systematically deeper than redder transitions.
Flux ratios of absorption lines trace the physical conditions of the line forming region, being particularly sensitive to density and temperature. The inspected transitions cover a narrow range of wavelengths ($4300-4900\, {\AA}$), so under the assumption of a flat continuum contribution at this range, $EW$ ratios might be a proxy for the flux ratios. Although we do not have access to enough transitions to extract reliable physical parameters (see e.g., \citealt{Ilic2012}), the stability of the line ratios suggests physical conditions of the BA forming region remain stable. This occurs both during the outburst evolution and for all systems analysed, in spite of the extreme variability of the optical continuum, supporting a scenario where BAs are persistent.

We also produced histograms combining fits to all members of the sample, focusing on parameters describing the BA. We study the joint results for the $\rm H\beta$ and $\rm H\gamma$ line profiles, consistent with each other (see Fig. \ref{fig:histograms}). $\rm H\delta$ was not included in the histogram because it suffers from systematically lower SNR and overall scarcer sampling. We find that the distribution of centroid velocities (corrected from their respective systemic velocity) has a mean and standard deviation of $\mu_{\rm abs}=80\pm 200\, {\rm km\, s^{-1}}$. The spread is large, but it remains fully consistent with that of the emission component ($\mu=100\pm 180\, {\rm km\, s^{-1}}$). This shows that the BAs in our sample are not systematically blue-shifted. The width of the BA $\sigma_{\rm abs}$ is quite stable, showing a Gaussian distribution with mean and standard deviation of $\sigma_{\rm abs}=1400\pm 500\, {\rm km\, s^{-1}}$. Finally, the typical depth of the features are $EW_{\rm abs}=4\pm 2\, \AA$. In this regard, we note that the deepest examples occur at low luminosities, reaching $EW_{\rm abs}\sim 12\, \AA$ (and even deeper in H$\delta$, $EW_{\rm abs}= 17\pm 4\, \AA$ in J1753).

Certainly, the observation of deeper BAs at fainter epochs appears a general property of our sample (see e.g., E25 in J1118). To test this, we compare the fitted parameters of the spectroscopic sample against the brightness of the source at the time of observations. We only have simultaneous optical photometry for J1727, J1807-B and J1753 datasets (see Sec. \ref{sec:specdata}).
In order to compare all systems, we decided to convert their apparent magnitudes to absolute magnitudes. To this aim, we employed the best-fit extinction and distance reported for each source, namely: $d=3.4 \pm 0.3\, {\rm kpc}$, $E(B-V)=0.6\pm 0.3$ for J1727 \citep{MataSanchez2025a}; $d=6.3 \pm 0.7\, {\rm kpc}$, $E(B-V)=0.099\pm 0.008$ for J1807 \citep{Saavedra2025}; and $d=3.9 \pm 0.7\, {\rm kpc}$, $E(B-V)=0.45\pm 0.09$ for J1753 \citep{Yanes-Rizo2025}. The resulting absolute magnitudes were plotted against all the available parameters from the spectral line profile fits. We find an inverse correlation ($r_s = 0.9$) between the brightness of the system and the $EW_{\rm abs, H\delta}$ (see Fig. \ref{fig:brigthnessvsEWabs}, left panel). 
A linear fit to this correlation results in $M_r=(1.1\pm 0.5)  + (0.70\pm 0.11) EW_{\rm abs, H\delta}$ (and similarly, $M_g= (1.0\pm 0.7)  + (0.75\pm 0.15) EW_{\rm abs, H\delta}$). There is a clear outlier to the linear correlation, corresponding to the deepest BA observed in J1753 sample. We nevertheless note this was obtained at the faintest epoch during the rise of the outburst, and that it remains consistent within $3\sigma$ (due to the larger uncertainties).
The fact that the datasets of three different systems (two BHs and a NS) follow a common correlation implies that the compact object nature is not key to develop BAs. Furthermore, it shows that the optical luminosity range where BAs are visible is quite extended, spanning 8 magnitudes.

\section{Discussion} \label{sec:discussion}

The above analysis revealed a number of common features for BAs, which allows us to systematically study these features for the first time. From the population analysis, we find the following key characteristics:

\begin{itemize}
    \item Ubiquity: They have been observed in $78\%$ of BHs with at least 3 epochs of optical spectroscopy, as well as in $29\%$ of NSs; a firm lower limit due to BAs transient visibility.
    \item Frequent at short periods: They are more frequently observed below $P_{\rm orb}< 11\, \rm{h}$ for both BH and NS systems alike.
    \item Non-dependent of $i$ or compact object nature: BAs occur at all orbital inclinations, as well as for BHs and NSs alike.
    \item Only seen in outburst: No BA features have been reported in the quiescent state of neither CVs nor LMXBs.
    \item  Predominantly shown in the H Balmer series: Its presence in other atomic species is limited to a handful of examples. These reports mainly refer to the \ion{He}{ii}$-4686$ line; in particular, GRO J0422+32 (originally reported in \citealt{Shrader1992}, but not mentioned/shown in \citealt{Shrader1994}); Swift J1753.5$-$0127 (observed in E1, during the initial rise of 2023 outburst); and Swift J1357.2$-$0933 (appearing only during the optical dips, \citealt{JimenezIbarra2019b}). Even fewer examples exist for \ion{He}{i} transitions: 3A 0620 shows a BA in \ion{He}{i}$-4471$; \citealt{Ciatti1977}. 
\end{itemize}

The study of our spectroscopic database reveals:

\begin{itemize}
    \item Deeper features at shorter wavelengths: If $\rm H\delta$ transition is available with sufficient SNR, it systematically exhibits deeper examples of BAs. 
    \item Slow profile evolution: Hydrogen Balmer emission lines from the accretion disc or the disc wind have been seen to evolve drastically during outburst, within timescales as short as just a few minutes (see e.g., \citealt{MataSanchez2018}, V404 Cygni). On the other hand, BAs evolve at a slower pace, and once they appear, they are sustained over long periods (e.g., AT2019wey shows deep broad absorptions in the Balmer series through several months; \citealt{Yao2021}). The only example with fast cadence observations of our database (J1118) showed stability over periods of at least a few hours.
    \item They are deeper in fainter epochs, and otherwise independent of the X-ray state: There is an inverse correlation between $EW_{\rm abs}$ and optical brightness, common among different systems and outburst epochs. BAs are equally observed in the hard and soft states.
    \item Their centroid velocities are variable but their mean is consistent with the systemic velocity: They oscillate around the central value, without preference towards red-shifted or blue-shifted velocities and detached from the emission component. 
   \item Their width is stable: They follow a Gaussian distribution both during outburst and across systems ($\sigma_{\rm abs}=1400\pm 500\, {\rm km\, s^{-1}}$).
    \item The normalised Balmer line ratios for the BAs remain constant: This suggests similar conditions of temperature and density at the BA forming region, independently of the system and/or outburst state.
\end{itemize}

\subsection{Precedents of BAs in CVs}
\label{sec:CVs}

CVs (see \citealt{Warner1995} for general review) are a natural benchmark to test accreting compact systems phenomenology due to their more prolific population (over $10^4$, see e.g., \citealt{Jackim2020}). There are reports of BAs in their spectra during active outburst phases since the onset of the field \citep{Joy1940}. They have been found in both persistent (i.e. nova-like, disc-dominated; e.g., UX UMa; \citealt{Neustroev2011}) and transient (i.e. dwarf novae; e.g., GW Lib, SS Cyg and V455 Andromedae; see \citealt{vanSpaandonk2010}, \citealt{Hessman1984} and \citealt{Tovmassian2022}, respectively) CVs. Studies report BAs in H and \ion{He}{i} with very broad full-width at zero intensity (FWZI$\sim 50-100\,\AA$), and cores filled by a narrower emission component (e.g., \citealt{Szkody1990}). BAs in CVs are comparable to those compiled in this work for LMXBs (see e.g., H$\beta$ in Fig. \ref{fig:J1727fitspec} and \ref{fig:J1807fitspec}, with wings reaching the mark of $\pm 2500\, \rm{km\, s^{-1}}$, i.e. FWZI$\sim 77\,\AA$). \citet{Morales-Rueda2002} report on 40 systems where BAs are observed, whose H$\gamma$ BA properties are consistent with (though slightly narrower than) our LMXB sample.

Models considering the absorption originates from an optically thick accretion disc were developed to explain these features (e.g., \citealt{Herter1979,Cheng1989}). Despite their limited success reproducing observational profiles, they all agree that BAs should be more prominent in low $i$ systems due to limb darkening. \citet{Herter1979} model considers the broadening of the line is, for $i>15\, {\deg}$, dominated by Doppler broadening. However, they expected a flat-bottom core for such a broadening profile, which is not observed. Instead, Gaussian-like profiles are found (as we do for LMXBs), which led them to suggest pressure broadening must be at play as well. As a result, they state BAs are formed in a region of the disc somewhat denser than the line-forming regions. When the outburst triggers, the disc begins to transition from an optically thin (quiescence) state to an optically thick state. As the instability wave propagates, ionizing the gas, the former emission lines gradually develop an absorption component, leading to mixed profiles. There are several reports of BAs becoming deeper as the CV brightens during the outburst and the core emission diminishes (see \citealt{Hessman1984,Szkody1990,Clarke1984,Tovmassian2022} and references therein). Their EW increases during the outburst rise, only to remain stable during the rest of the event until they vanish during the decay to quiescence (e.g., \citealt{Clarke1984}). These studies claim that BAs might always be present, and their visibility would be conditioned by the filling of the emission core component.

\subsection{The current BA view on LMXBs}
\label{sec:lmxbdubus}

\begin{figure}
\centering
\includegraphics[keepaspectratio, trim=0cm 0cm 0cm 0cm, clip=true,width=0.5\textwidth]{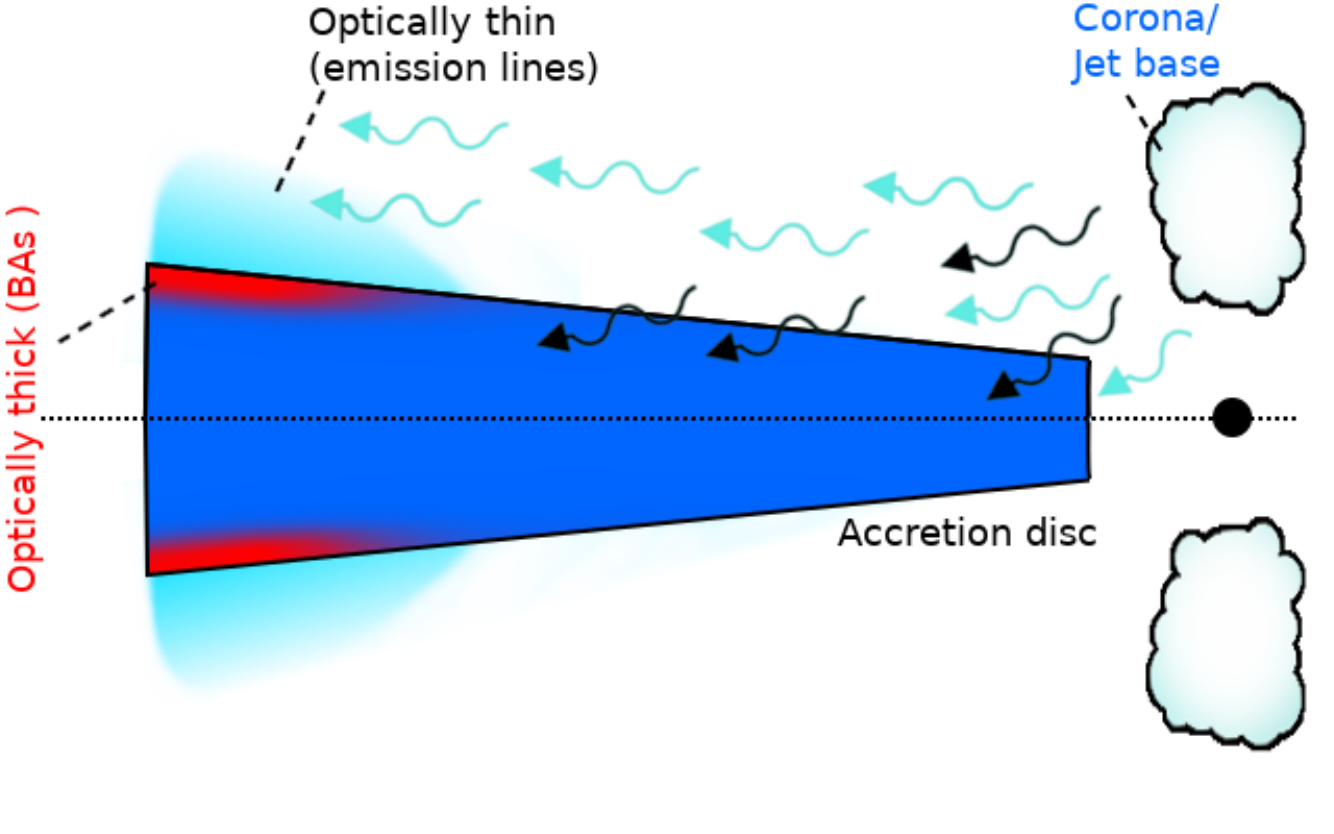}
\caption{Schematic of the model for BA formation. The white cloud on top of the compact object is the illuminating X-ray source (e.g., the base of the jet or the corona). It emits both soft (cyan) and hard (black) X-ray photons. The former heats up the disc chromospheric region giving rise to emission lines, while the later contributes to heating up deeper layers of the accretion disc. This produces a stellar-like temperature profile, with a relatively cold region on the surface producing the observed BAs. Note that the accretion disc also exhibits a radial gradient of temperature, so only the outermost rings develop such a region.}
     \label{fig:Dubus}%
 \end{figure}

In spite of the frequent appearance of BAs in LMXB spectra, they have been systematically set aside, as they hampered the study of other phenomena of interest. Among the few works mentioning BAs, \citet{Dubus2001} discussed them in a paper dedicated to J1118 outburst. They briefly compared their findings with previous detections in a handful of systems from the literature, as well as described a qualitative and phenomenological picture inspired by previous CV studies (e.g., \citealt{Sakhibullin1998}). According to this scenario, emission lines (which usually dominate LMXB outburst spectra) are produced in an optically thin (cromosphere-like) region above the accretion disc as a result of photoionisation by soft X-rays. On the contrary, BAs are formed in a lower, optically-thick region of the accretion disc, where the temperature stratification with height over the orbital plane resembles that of a stellar atmosphere (see Fig. \ref{fig:Dubus}). Following this picture, a number of predictions were stated: 

\begin{itemize}
    \item[i)] The formation of BAs depends on the spectral energy distribution of the irradiating X-ray flux. Soft X-rays are responsible for heating up the optically thin region responsible for the emission lines. On the other hand, hard X-rays penetrate deeper into the accretion disc, enhancing the temperature at the base of the optically thick atmosphere. For this reason, harder X-ray spectra might favour the formation of BAs. 
    \item[ii)] High values of overall X-ray luminosity might lead to the photoionisation of deeper layers, ultimately reaching the region of the accretion disc producing the BAs, and inverting the profile temperature. Therefore, BAs might disappear at high luminosities. If the optically thin atmospheres in LMXBs are more extended than those observed in CVs (an expected feature due to the larger irradiation), BAs should be less frequently observed in LMXBs than in CVs.
    \item[iii)] Limb-darkening has a larger impact on systems with edge-on orbital inclination, as photons do not penetrate as deep into the accretion disc atmosphere as in a face-on case configuration. As a result, the formation of BAs should be inclination dependent, being shallower for edge-on systems.
\end{itemize}

The authors analysed these predictions on a handful of systems compiled from existing literature, which supported their conclusions. However, our work involving the whole BH population shows otherwise, as described in the following sections.

\subsection{Testing the model predictions}
\subsubsection{On the brightness and X-ray hardness}
\label{sec:hidsec}
BAs have never been detected in quiescence even for systems which exhibited them during outburst. 
The optically thin, low-viscosity accretion discs found in quiescence favour the formation of emission lines. Combined with veiling effects (as the relative companion star contribution increases) and the broader emission line wings (as the accretion disc contracts), the non-detection of BAs in quiescence is expected. During outburst, an optically thick photosphere develops as a result of internal viscous heating, enabling the formation of absorption profiles. Following \citet{Dubus2001} model described in the previous section, the BA formation would then depend on both the overall luminosity of the system and its X-ray spectral state.

Our spectroscopic analysis shows that BAs become deeper at fainter epochs (see Fig. \ref{fig:brigthnessvsEWabs}, left panel). This is the opposite behaviour to that reported for CVs (see Sec. \ref{sec:CVs}), which develop deeper BAs as their luminosity increases (see e.g., \citealt{Hessman1984}). The brightest CVs ($\sim 10^{29-33}\, \rm{erg\,s^{-1}}$) are comparable to the faintest LMXBs ($\sim 10^{33-36}\, \rm{erg\,s^{-1}}$; see e.g., \citealt{Armas-Padilla2013}), especially if we rescale them by their Eddington luminosity. This might suggest that the luminosity dependence of the BA formation mechanism saturates at this intersection ($\sim 10^{33}\, \rm{erg\,s^{-1}}$). CVs continuum during outburst is  dominated by internal viscous heating of the accretion disc. Therefore, brighter CVs overall correspond to hotter and deeper layers of the accretion disc, favouring the formation of stronger BAs. On the contrary, the continuum in LMXBs is dominated by soft X-ray reprocessing of the central source irradiation at the accretion disc atmosphere. As such, brighter LMXBs do not necessarily translate into heating at the base of the accretion disc. At high luminosities, the disc atmosphere expands due to Compton heating from the X-ray irradiation, making the observability of BAs a balance between the relative depth of the absorption and emission components. As a result, BAs become weaker in bright LMXBs due to being i) filled by a stronger emission component; and ii) veiled by the extremely variable X-ray reprocessed continuum (given that $EW$ is a normalised measure relative to the continuum).

\begin{figure}
\centering
\includegraphics[keepaspectratio,  trim=0cm 0cm 0cm 0cm, clip=true,  width=0.5\textwidth]{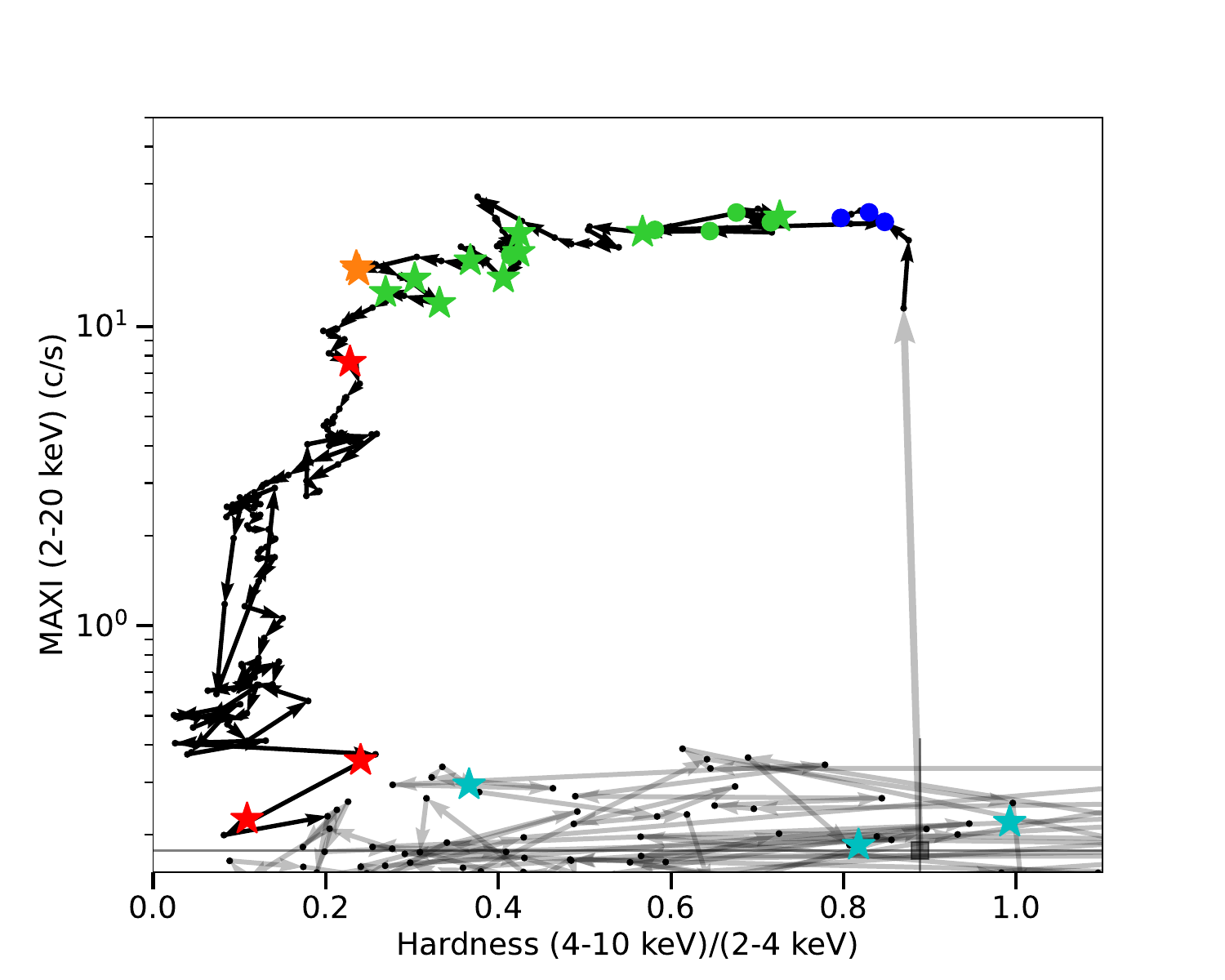}
   \caption{Hardness-Intensity diagram of J1727 (black, arrow-connected dots; grey for low X-ray fluxes with colours possibly affected by background subtraction). Coloured symbols mark epochs where our optical spectroscopy was obtained. Filled circles mark spectra where no BAs were detected, while filled stars correspond to BA detection in at least one of the inspected transitions (see Section \ref{sec:analysis}). The colour code separates the initial rise in the hard state (blue), hard-intermediate state (HIMS, green) and soft-intermediate state (SIMS, orange), as well as the soft state decay (red). The last five epochs, published for the first time in this work, correspond to the final decay during the soft state and the subsequent transition to the low-luminosity hard state (cyan). }
     \label{fig:hidJ1727}
 \end{figure}

Regarding the X-ray state dependence, our analysis shows BAs are detected during both hard and soft states alike for the two systems exhibiting full transitions (J1727 and J1807). In particular, J1727 (see Fig. \ref{fig:hidJ1727}) allows us to study the BA evolution over a full outburst event, showing that it is mainly driven by the continuum brightness rather than being linked to X-ray state transitions. The remaining systems exhibit hard-only outbursts, forbidding us from performing a more detailed study.

We therefore conclude that BAs are present and remarkably stable over a wide range of luminosities, and only truly vanish when they return to quiescence.

\subsubsection{On the orbital inclination}

The population analysis already showed that the distribution of $i$ for LMXBs with BAs is perfectly consistent with randomly oriented orbits (see Sec. \ref{sec:analysispop}). There are no hints of shallower BAs for edge-on systems, contrary to the model prediction. Our spectroscopic sample allows for a closer inspection: J0422 is in the low-intermediate region ($\sim 10-50 \, \deg$, \citealt{Casares1995c,Beekman1997MNRAS,Webb2000}), J1727 inclination is only loosely constrained by an upper limit of $74 \, \deg$ (\citealt{Wood2024}, \citealt{MataSanchez2025a}), and the remaining targets are in the high inclination regime: J1118 ($68-79 \, \deg$, \citealt{Khargharia2013}), J1807 ($i=72\pm 5\, \deg$, see \citealt{Saavedra2025}), J1753 ($79\pm 5 \, \deg$, \citealt{Yanes-Rizo2025}), and the most edge-on system known J1357 ($87.4^{+2.6}_{-5.6} \, \deg$, \citealt{Casares2022}; see also \citealt{Corral-Santana2013,MataSanchez2015b}).

The largest $EW_{\rm abs}$ is found for J1753 in the $\rm H\delta$ transition, one of the most edge-on systems of our BH sample. The following systems in terms of $EW_{\rm abs}$ are J1357, J1118 and J1807 (again at the higher end of $i$), which shows consistently large values at all inspected transitions. The lowest $EW_{\rm abs}$ is found for J1727, for which we only have an upper limit to the inclination suggesting a mid-range value at most. Finally, the most face-on system of our sample (J0422) shows $EW_{\rm abs}$ measurements between those of J1807 and J1727. In this regard, we note that J0422 inclination has been recently proposed to be as high as $i=56\pm 4\, \deg$ \citep{Casares2022}. This opens the possibility of J1727 being closer to face-on than the former. We also explored the width of the BAs ($\sigma_{\rm abs}$, see Fig. \ref{fig:histograms}), and found that only J1727 ($\sigma_{\rm abs} =700\pm 300 \,\rm{km\,s^{-1}}$) is consistently below the typical values of our sample ($\sigma_{\rm abs}=1400\pm 500\, {\rm km\, s^{-1}}$, which updates to $\sigma_{\rm abs} =1500\pm 300 \,\rm{km\,s^{-1}}$ if J1727 is removed). In order to further explore the inclination dependence, we turned to the most extreme example for a low-inclination system among the full BH population exhibiting BAs: MAXI J1836$-$194 ($4-15 \, \deg$, \citealt{Russell2014}). The authors performed a Gaussian fit to the absorption component and found, respectively for $\rm H\beta$ and $\rm H\gamma$, $EW_{\rm abs}=2.54\, \AA$ and $EW_{\rm abs}=2.5\, \AA$, as well as $\sigma_{\rm abs} =920\,\rm{km\,s^{-1}}$ and $\sigma_{\rm abs} =680\,\rm{km\,s^{-1}}$. Compared with the sample presented in our work, the features in MAXI J1836$-$194 are on the lower end of our spectroscopic sample (see Fig. \ref{fig:histograms}). 

We conclude that lower $i$ does not produce stronger BAs, contrary to the model prediction. Instead, based on the few available examples of low-inclination LMXBs, their BAs actually appear narrower, further supporting a Doppler broadening mechanism for BAs.

\subsubsection{On the orbital period and the system size}
\label{sec:orbitalperiod}
The orbital period is a proxy for the size of the system, and in particular to that of the accretion disc (given the late type companions and low mass ratio of LMXBs with BHs). A similar statement holds for NSs, even though they have smaller discs than BHs for similar $P_{\rm orb}$ (by a factor $\sim 2-3$) and larger mass ratios. The population analysis shows that BAs form in the majority of short $P_{\rm orb}<11\, \rm{h}$ systems. This does not prevent the formation of BAs in larger systems (e.g., GRO J1655$-$40, MAXI J1820$+$070), but there are clear examples of BA non-detection in spite of an intensive coverage during the outburst.

To explore the possibility of larger discs preventing the formation or visibility of BAs, we focus on V404 Cygni ($P_{\rm orb}=155.31 \,{\rm h}$, \citealt{Casares2019}) during its 2015 outburst \citep{Munoz-Darias2016}, arguably the best covered outburst among LMXBs. Conversion of the apparent magnitudes exhibited during the outburst (Fig. 1 in \citealt{MataSanchez2018}) to absolute magnitudes (using $d=2.39\pm 0.14 \, \rm{kpc}$, \citealt{MillerJones2009}; and extinction $E(B - V ) =
1.3$, \citealt{Chen1997}) results in a range of $M_r=0$ to $-5$. These values are above the brightest values of Fig. \ref{fig:brigthnessvsEWabs} (left panel). This might imply that large accretion discs are simply too luminous for BAs to be detected, as they are completely veiled by the continuum. Another possibility is that they are hidden by other features developing in the spectra during outburst (e.g., nebular phases, an early type donor star, or P-Cygni profiles).

We studied the orbital phase dependence of BAs. Analysis of the individual spectra of the phase-folded J0422-B dataset showed sinusoidal variability on the $\mu_{\rm abs}$ parameter, with a radial velocity semi-amplitude of $K=48\pm 10\, {\rm km\, s^{-1}}$. This result is remarkably aligned with the expected compact object radial velocity semi-amplitude ($K_1\sim 40\, {\rm km\, s^{-1}}$, derived from the companion star $K_2$ and the mass ratio $q$ reported in \citealt{Webb2000}). The orbital ephemerides employed for the phase folding were derived in the original work from Doppler maps of a narrow emission component in the \ion{He}{ii}-4686 line \citep{Casares1995}. The periodicity was later confirmed to be the orbital period of J0422, and its origin was traced to a disc hotspot \citep{Casares1995c}. The modulation we find on the BA radial velocity is consistent with this interpretation. 

The analysis of this same parameter on the epoch-averaged spectra in J1118 revealed a periodic modulation at $\sim 53\, \rm{d}$, much longer than its orbital period ($P_{\rm orb}=4.08 \,\rm{h}$; \citealt{Torres2004,GonzalezHernandez2014}). There are previous reports on long period variability ($\sim 52\, {\rm d}$) for the emission lines centroids in J1118, based on near and in quiescence observations (see \citealt{Torres2004}, \citealt{Calvelo2009}). These are associated with the precession period of an eccentric accretion disc. Emission lines in our dataset, obtained during more active phases of the outburst, did not exhibit such modulation even though it was explicitly searched for (see \citealt{Torres2002}). We now find that BAs exhibit variability at the precession period, with fully consistent radial velocity semi-amplitude ($\sim 240 \pm 30\, \rm{km\,s^{-1}}$) to the near and in quiescence observations.

Both of the above results further support a disc origin for the BA component. Furthermore, the observed BA evolution with the disc precession period locates these features at the outer radius. However, it is worth noting that the central velocity of the BAs in both datasets ($140\pm 40\, {\rm km\,s^{-1}}$ and $260 \pm 30\, \rm{km\,s^{-1}}$) are well over the corresponding systemic velocity ($11\pm 8\, {\rm km\,s^{-1}}$, \citealt{Harlaftis1999}; $16\pm 6\, {\rm km\,s^{-1}}$, \citealt{Torres2004}). Related to this, we find that $\mu_{\rm abs}$ parameter is highly variable in all examples of our spectroscopic sample (see e.g., Fig. \ref{fig:J1727fitparams} and \ref{fig:J1807fitparams}), which suggests that it might evolve over even larger timescales. While such a variability is usually associated with precessing accretion discs, our limited sample precludes further inspection. The remaining parameters of the absorption remained stable and did not display any periodic modulation.

\subsection{The origin of the BAs}

The results compiled in this work enable us to draw a number of common properties for BAs, which we now use to explore their origin. Our study has been greatly influenced by previous models developed originally for CVs in order to explain the BAs, which assigned these features to the accretion disc. Certainly, such an origin is further favoured in LMXBs by a number of reasons. First and foremost, the accretion disc dominates the optical spectrum during the outburst. The emission line profiles, whose broadening naturally arises from Doppler movement of the gas in the disc, show full-widths-at-half-maximum ($ \rm{FWHM}\sim 1000\, {\rm km\, s^{-1}}$) comparable to those of BAs, suggesting they both arise from a similar outer radius of the disc. Comparison of the broadness of BAs in LMXBs ($ \rm{\sigma_{\rm abs}}= 1400 \pm 500\, {\rm km\, s^{-1}}$) with those in CVs ($ \rm{\sigma_{\rm abs}}= 1200 \pm 300\, {\rm km\, s^{-1}}$; combining the sample for H$\gamma$ in Table 3 from \citealt{Morales-Rueda2002}) shows the absorptions are slightly larger in the former. This is consistent with larger accretion disc velocities around more massive accretors at the same radii. Additionally, we present two examples of BA periodic evolution, further strengthening the connection with the disc outer region. However, it is worth taking a step back to analyse other possibilities.

Episodes of mass outflows are commonly observed in X-ray outbursts (see e.g., \citealt{Ponti2012}). During the past decade, they have been systematically discovered in the optical regime as well (see \citealt{MunozDarias2026} for a review). They are mostly found as both P-Cygni profiles and broad emission lines (nebular phase, see \citealt{Munoz-Darias2016}). The exact geometry of the outflow is still under debate, but \citet{Munoz-Darias2022} suggests a dynamical, multi-phase wind. A broader context is provided by the comparison of the outburst (and outflow) phenomenology in LMXBs and that of AGNs \citep{FernandezOntiveros2021}. In this regard, the AGN BAL subclass is known to show blue-shifted broad absorptions, which are traditionally associated with outflows (e.g., \citealt{Weymann1991,Proga2000}). While they usually appear at high-ionisation transitions (e.g., \ion{N}{V} and \ion{C}{IV}), examples on the hydrogen series exist, possessing comparable widths ($FWZI=2000\pm 200\, {\rm km\,s^{-1}}$, \citealt{Hall2007}) to those compiled in this paper for BAs. Our analysis of the BAs in LMXBs does not reveal a systematic blue-shift of these features, but instead a highly variable centroid velocity around the systemic velocity (see Fig. \ref{fig:histograms}). Based on these results alone, we cannot favour an outflow interpretation of BAs, neither can we fully disregard this possibility. 

On this regard, the high-inclination, optical-dipping system J1357 is known to exhibit broad ($\sigma_{\rm abs}\sim 2000\, {\rm km\, s^{-1}}$), blue-shifted ($\mu_{\rm abs} \sim -800\, {\rm km\,s^{-1}}$) absorptions exclusively during the optical dips (hereafter, dip-features, aka DFs). These were observed during a time-resolved ($\sim30\,\rm{s}$ cadence) spectroscopic campaign on its 2017 outburst and interpreted as outflows \citep{JimenezIbarra2019b}. We now reveal BAs in a longer exposure (1-h) averaged spectra during the previous 2011 outburst (E4, $\sigma_{\rm abs}=2980\pm 30\, {\rm km\, s^{-1}}$). We investigate if BAs and the blue-shifted DFs correspond to the same, albeit diluted, phenomena. The veiling of a DF in an averaged spectrum depends on the continuum flux variation and the time spent in and out of the dips during the exposure. We focus on the deepest BA found for J1357 in H$\beta$ (E4, 2011). The closest epoch with optical photometry \citep{Corral-Santana2013} shows a duty cycle of $50\%$ (4-minute long dips with a recurrence time of 8-minutes) and a typical brightness variability of $\Delta R\sim 0.6$. Nevertheless, only 2017 temporal resolution allows \citet{JimenezIbarra2019b} to build inside-of-dip and out-of-dip spectra separately. For this reason, we use their photometric observations from June 21st 2017, as similar recurrence times for the dips to those of our E4 2011 observations are reported. The closest spectroscopic dataset to 2017 epoch (June 16th) shows a duty cycle of $33\%$, brightness variability of $\sim 0.5$, as well as absorption depth during the dips of $EW_{\rm dip}=4.4\pm 0.4\, \AA$ in the H$\beta$ line. In order to compare this DF with our BA, we note that it would show as a weak BA of $EW_{\rm abs}\sim 1.7\, \AA$ in a 1-h long-exposure spectrum. Alternatively, our observed BA in 2011 E4 ($EW_{\rm abs}= 7.9\pm 0.2\, \AA$) would imply a DF stronger than $EW_{\rm dip}\sim 20\, \AA$, i.e., $\sim 5$ times deeper than those observed at any time during the 2017 outburst.
To further explore this, we analyse the six individual 10 minute exposures of E4 and found that the BA still remained with similar depths in H$\beta$. Additionally, E4 BA shows a centroid of $\mu_{\rm abs}=50\pm 40\,\rm{km\,s^{-1}}$, far from the extreme absorptions observed during the dips ($\mu_{\rm abs} \sim -800\, {\rm km\,s^{-1}}$, \citealt{JimenezIbarra2019b}). Not even the minimum of $\mu_{\rm abs}=-380\pm 30\,\rm{km\,s^{-1}}$ of E1 reaches such values, and rather resembles the BAs behaviour in other systems. Finally, E4 BA shows the broadest example of our sample ($\sigma_{\rm abs}\sim 2900\, {\rm km\, s^{-1}}$), while typical DFs are $\sigma_{\rm abs}\sim 2000\, {\rm km\, s^{-1}}$. 
Together, these results suggest that the fast-evolving dipping absorptions and the long-lasting BAs observed outside the dips might have a different origin.

\subsection{The case of \ion{He}{ii}-4686}
\label{sec:heii}
While BAs are predominantly observed in the H Balmer series, there are a handful of reports in other species. Of particular interest is the \ion{He}{ii}-4686 line, which requires higher temperatures to ionise. Precedents of BAs on this transition exist in at least two dwarf novae (\citealt{Morales-Rueda2002}). They exhibit slightly broader BAs than their hydrogen counterparts (though still consistent within uncertainties), in line with a formation region in an inner (i.e. hotter) radius. There have been two previous reports of \ion{He}{ii}-4686 BAs in LMXBs: J1357, limited to the optical dips and always blue-shifted (associated with outflows, \citealt{JimenezIbarra2019b, Charles2019}); and J0422, where they just report on their presence \citep{Shrader1992,Shrader1994}. We now report one more case during the initial rise of J1753 outburst (E1). A fit to the \ion{He}{ii}-4686 line using the same model introduced in Sec. \ref{sec:analysis} reveals BA properties of $\sigma_{\rm abs, HeII}=800\pm 180\, {\rm km\, s^{-1}}$ and $EW_{\rm abs, HeII}=3\pm 2 \, {\rm \AA}$. 
The BA in J1753 is shallower than the contemporaneous features in hydrogen, consistent with a forming region temperature comparable to an early-type star photosphere. However, the narrower BA for \ion{He}{ii} (compared to $\sigma_{\rm abs}=1500\pm 200\, {\rm km\, s^{-1}}$ in H$\gamma$ and H$\beta$) is puzzling. It would suggest that a line with a higher ionisation potential forms at an outer radius of the accretion disc. Assuming a disc origin, it would imply an inversion of the temperature gradient at some critical radius, increasing with the distance to the central source instead. In order to attain such configuration, we propose shielding by a vertical structure might be at play. If large enough, it would fully shield the outer accretion disc beyond the structure. If smaller, it would only shield a certain range of radii, leaving the outermost regions again exposed to irradiation. This is supported by the constant $EW_{\rm abs}$ line ratios observed (Fig. \ref{fig:brigthnessvsEWabs}, right panel) observed, which suggest similar physical conditions both during the whole outburst event and between systems. In order to survive the extreme heating corresponding to the luminosity range displayed in Fig. \ref{fig:brigthnessvsEWabs} (left panel, $\Delta r\sim 8$ optical magnitudes), shielding of the BA forming region is an appealing solution (see e.g., \citealt{Motta2017}, Armas-Padilla et al 2025). The specific conditions required to reproduce this scenario are in line with the scarce number of BA detections in \ion{He}{ii}. Further studies in other systems are required to properly test this possibility.

\section{Conclusions}

We present a systematic study of the LMXB population focused on the detection of BAs. We confirm they are ubiquitous during outburst, and more frequently observed in short orbital period systems. No dependence on the orbital inclination or the compact object nature is found. They are mostly seen in the Balmer series, with deeper examples at shorter wavelengths. Their profiles evolve slowly, as well as appear indistinctly across all X-ray states. More importantly, we observe an inverse correlation between the depth of the BAs and the system luminosity. We propose this arises as a result of veiling from the X-ray reprocessed continuum, while BAs profiles themselves remain stable and are present over the whole outburst duration. Their visibility appears conditioned at high luminosities by such a veiling, while the low luminosity cut-off depends on the SNR of the observations. The fact that no quiescent spectra has ever shown BAs suggests there is a true luminosity cut-off where BAs do disappear from the spectra. Even though no data is available to observe this transition in LMXBs, comparison with the CV population, which shows an increasing depth for the BA with luminosity, provides further support. We note that these properties do not fit all the predictions made in previous works, based on a qualitative model inherited from CVs. We conclude that, while the proposed origin of BAs in a thick accretion disc still stands, other explanations (e.g., outflows) remain as viable alternatives, as long as they fit the diverse properties compiled in this work.

\begin{acknowledgements}
DMS and M.A.P. acknowledge support through the Ram\'on y Cajal grants RYC2023-044941 and RYC2022-035388-I, funded by MCIU/AEI/10.13039/501100011033 and FSE+. DMS, TMD, JC, MAP and MAPT acknowledge support by the Spanish Ministry of Science via the Plan de Generacion de conocimiento PID2021-124879NB-I00, PID2022-143331NB-I00 and PID2024-161863NB-I00. We are deeply grateful to Tom Marsh for developing the \textsc{molly} software, one of his many contributions to advancing the field of compact objects. We are thankful to the GTC staff for their prompt and efficient response at triggering the time-of-opportunity program at the source of part of the spectroscopy presented in this work. Based on observations collected at the European Southern Observatory under ESO programme 105.20LK.002. This work has made use of data from the European Space Agency (ESA) mission {\it Gaia} (\url{https://www.cosmos.esa.int/gaia}), processed by the {\it Gaia} Data Processing and Analysis Consortium (DPAC, \url{https://www.cosmos.esa.int/web/gaia/dpac/consortium}). Funding for the DPAC has been provided by national institutions, in particular the institutions participating in the {\it Gaia} Multilateral Agreement. \textsc{pyraf} is the python implementation of \textsc{iraf} maintained by the community. This work has made use of the LMXB catalogue (\url{https://binary-revolution.github.io/LMXBwebcat/}) maintained by the Binary rEvolution team (\url{https://github.com/Binary-rEvolution}). The spectra and associated calibration files
are available upon reasonable request to the corresponding author.

\end{acknowledgements}

%
%

\bibliographystyle{aa} 
\bibliography{bibliography} 

\onecolumn

\begin{appendix}
\section{Database description.}

\begin{table*}
\caption{List of BHs and NSs with confirmed and candidate BA detections.}
\begin{tabular}{l|ccc}
Name &  CO &  BA lines &  References \\
 &  & &     \\
\hline \\
GRO J0422+32 (V518 Per) & BH &   Balmer, \ion{He}{ii}$-4686$  &  \citealt{Shrader1992,Shrader1994,Shrader1997}, \citealt{Bartolini1994}, \\
& & &  \citealt{Casares1995}, \citealt{Callanan1995}\\
3A 0620$-$003 & BH &   Balmer, \ion{He}{i}$-4471$   &  \citealt{Whelan1977}, \citealt{Ciatti1977} \\
GRS 1009$-$45 (N Vel 1993) & BH &   Balmer  &  \citealt{Bailyn1995}, \citealt{Masetti1997} \\
XTE J1118+480 &  BH &  Balmer  &  \citealt{Torres2002}, \citealt{Dubus2001} \\
GRS 1124$-$684 (N Mus 1991) &  BH &  Balmer  &  \citealt{dellaValle1991,dellaValle1998}\\
MAXI J1305$-$704 &  BH &  Balmer  & \citealt{Charles2012}, \citealt{Miceli2024}   \\
Swift J1357.2$-$0933 &  BH &  Balmer, \ion{He}{ii}$-4686$  &  \citealt{JimenezIbarra2019b}  \\
GRO J1655$-$40 &  BH &  Balmer  & \citealt{Hynes1998},\citealt{Soria2000} \\
Swift J1727.8$-$16 & BH &  Balmer  & \citealt{MataSanchez2024a}\\
Swift J1753.5$-$0127 &  BH &  Balmer, \ion{He}{ii}$-4686$ & \citealt{Torres2005}, \citealt{Rahoui2015}, \\
& & & \citealt{Neustroev2023} \\  
MAXI J1820+070 &   BH & Balmer  & \citealt{Yoshitake2024}, \citealt{Garnavich2018}  \\
XTE J1859+226 &  BH &  Balmer  &  \citealt{Zurita2002}, \citealt{Wagner1999}, \citealt{Welsh2002}  \\
Swift J1910.2-0546 &  BH &  Balmer  & \citealt{Casares2012}, \citealt{CorralSantana2025} \\
AT 2019wey &  BHC & Balmer  & \citealt{Yao2021}, \citealt{Mereminskiy2022} \\
MAXI J1348$-$630 &  BHC &  Balmer  & \citealt{Panizo-Espinar2022} \\
MAXI J1803$-$298 &  BHC & Balmer  & \citealt{MataSanchez2022}\\
MAXI J1836$-$194 &   BHC & Balmer  & \citealt{Russell2014}   \\
\hline \\
1H 1659$-$487 (GX339-4) &  BH & H$\beta$ (candidate) & \citealt{Rahoui2014} \\
GRS 1716$-$249 &  BH & H$\beta$ (candidate) & \citealt{dellaValle1994}, \citealt{Cuneo2020b} \\
\hline \hline \\
EXO 0748$-$676 &  NS &   Balmer   & \citealt{Crampton1986}, \citealt{Mikles2012},\citealt{Pearson2006} \\
 1E 1603.6+2600 (UW CrB) &  NS &   Balmer   & \citealt{Kennedy2025}, \citealt{Fijma2025} \\
 3U 1728$-$16 &  NS &   Balmer   &  \citealt{Shahbaz1996}, \citealt{Cornelisse2007}  \\
 SAX J1808.4$-$3658 &  NS &   Balmer   & \citealt{Elebert2009}\\
 MAXI J1807+132 &  NS &   Balmer   & \citealt{JimenezIbarra2019a} \\
3A 1822$-$371 &  NS &   Balmer   & \citealt{Casares2003} \\
XTE J2123$-$058  &  NS &   Balmer   &  \citealt{Hynes2001}\\
\hline \\
Aql X$-$1 &  NS & H$\beta$ (candidate) & \citealt{Cornelisse2007}, \citealt{Panizo-Espinar2021} \\
MXB 1837+05 &  NS & H$\beta$ (candidate) & \citealt{Cornelisse2013}, 
\end{tabular}
\tablefoot{Description of acronyms in the table: CO (Compact object), BHC (BH candidate, not confirmed neither dynamically nor through H$\alpha$ scaling relations yet).}
\label{table:bapop}
\end{table*}

\begin{table*}
\caption{Observation log for the spectroscopic database of 5 BHs and 1 NS.}
\begin{tabular}{c|cccccc}
Name &  Epoch & Date   & Setup & $N_{\rm spec}\times T_{\rm exp}$ & Coverage& References \\
 & & & & (s) & \rm{\AA} &      \\
\hline \\
J1727 & E1-E20 & 26/08/23 - 24/10/23 & GTC+OSIRIS & & $3930-7650$& \citet{MataSanchez2024a} \\
 &  E21 & 20/02/24 & GTC+OSIRIS (R2000B)& 2x450  &  $3930-5630$  & This work \\
 &   &  & GTC+OSIRIS (R2500R)& 2x450  & $5550-7650$ &  \\
  &  E22 & 26/02/24 & GTC+OSIRIS (R2000B)& 2x400  &  $3930-5630$& This work \\
 &   &  & GTC+OSIRIS (R2500R)& 2x400  &$5550-7650$  &  \\
  &  E23 & 19/03/24 & GTC+OSIRIS (R2000B)& 2x400  & $3930-5630$ & This work \\
 &   &  & GTC+OSIRIS (R2500R)& 2x400 & $5550-7650$ & \\
   &  E24 & 31/03/24 & GTC+OSIRIS (R2000B)& 2x400 & $3930-5630$ & This work \\
 &  &   & GTC+OSIRIS (R2500R)& 2x400 & $5550-7650$ &  \\
    & E25 & 07/04/24 & GTC+OSIRIS (R2000B)& 2x400 & $3930-5630$ & This work \\
 &   &  & GTC+OSIRIS (R2500R)& 2x400 &$5550-7650$  &  \\
\hline \\
J1118 & E1-E21 & 31/03/00 - 10/06/00 & T-FLWO 1.5-m+ FAST  & &  $3650-7500$ & \citet{Torres2002} \\
 & E22 & 13/01/05  & T-FLWO 1.5-m+ FAST  & 59x300 & $3600-7800$ & \citet{Elebert2006} \\
 & E23 & 18/01/05  & T-FLWO 1.5-m+ FAST  & 52x300 & $3600-7800$ & \citet{Elebert2006} \\
 & E24 & 01/02/05  & T-FLWO 1.5-m+ FAST  & 27x600 & $3600-7800$ & \citet{Elebert2006} \\
 & E25 & 14/02/05  & T-FLWO 1.5-m+ FAST  & 6x300 & $3600-7800$ & This work \\
 & &   & T-FLWO 1.5-m+ FAST  & 7x450 & $3600-7800$ &  \\
 &  &   & T-FLWO 1.5-m+ FAST  & 16x600 & $3600-7800$ &  \\
\hline \\
J0422 & A1-A8 & 18/08/92 - 16/03/93 & Perkins 1.8-m+ CCDS  & & $4000-7000$ &  \citet{Shrader1994} \\
& B1-B10 & 16/10/93 - 19/10/93 & WHT+ISIS  
&  &$4250-5000$   &\citet{Casares1995} \\
 & C1-C8 & 19/11/92 - 10/12/93 & KPNO 4-m & 3 & $4400-7000$ &\citet{Callanan1995} \\
  & & & MDM 1.3-m + MIII& 2 & $4400-7000$   &\\
  & & & MDM 1.3-m + MIII& 2 & $4050-5000$   &\\
  & & &MDM 1.3-m + MIII& 1 & $3900-5000$   &\\
\hline \\
J1753 & E1 & 03/10/23 & GTC+OSIRIS (R2000B)  & 2x400 & $3930-5630$ &  This work \\
 &  &  & GTC+OSIRIS (R2500R)  &  2x400 & $5550-7650$ &  This work \\
 & E2 & 06/10/23 & GTC+OSIRIS (R1000B)  & 2x400 & $3600 - 7700$ &  This work \\
 & E3 & 09/10/23 & GTC+OSIRIS (R1000B)  & 2x400 & $3600 - 7700$ &  This work \\
 & E4 & 13/10/23 & GTC+OSIRIS (R1000B)  & 2x400 & $3600 - 7700$ &  This work \\
\hline \\
J1357 & E1 & 24/02/11 & INT+IDS  & 1800x2 & $6270-7000$ &  \citet{Casares2011} \\
 &  & 25/02/11-26/02/11 &  & 2000x4 & &   \\
&E2 & 19/03/11 & WHT+ACAM  & 61x250 & $4400-9000$ &  \citet{Corral-Santana2013} \\
&E3 & 13/04/11  & WHT+ACAM  & 20x600 &$4400-9000$   &  \citet{Corral-Santana2013}  \\
&E4 & 12/05/11  & MAG+LDSS  &  6x600 &$4500-7500$   &This work \\
\hline 
\hline \\
J1807 & A1-A6 & 28/03/17 - 18/08/17 & GTC+OSIRIS  & &  $3600-7700$ & \citet{JimenezIbarra2019b} \\
 & B1 & 18/07/23 & GTC+OSIRIS (R1000B)  & 1x300  &  $3600-7700$ & This work \\
 & B2 & 20/07/23 & GTC+OSIRIS (R1000B)  & 3x200  & $3600-7700$  & This work \\
 & B3 & 23/09/23 & GTC+OSIRIS (R1000B)  & 3x200  & $3600-7700$  & This work \\
 & B4 & 25/07/23 & GTC+OSIRIS (R1000B)  & 3x200  & $3600-7700$  & This work \\
 & B5 & 29/07/23 & GTC+OSIRIS (R1000B)  & 3x400  & $3600-7700$  & This work \\
 & B6 & 02/08/23 & GTC+OSIRIS (R1000B)  & 3x800  & $3600-7700$  & This work \\
 & B7 & 11/08/23 & GTC+OSIRIS (R1000B)  & 3x800  & $3600-7700$  & This work \\
 & B8 & 14/08/23 & GTC+OSIRIS (R1000B)  & 3x800  & $3600-7700$  & This work \\
 & B9 & 19/08/23 & GTC+OSIRIS (R1000B)  & 3x800  & $3600-7700$  & This work \\
 & B10 & 21/08/23 & GTC+OSIRIS (R1000B)  & 3x800 & $3600-7700$   & This work \\
\end{tabular}
\tablefoot{Description of acronyms in the table: GTC+OSIRIS (Gran Telescopio Canarias equipped with the OSIRIS spectrograph); T-FLWO 1.5-m+ FAST (Tillinghast telescope of the Fred Lawrence Whipple Observatory 1.5-m telescope, equipped with the FAST spectrograph); Perkins 1.8-m+ CCDS (Perkins 1.8-m telescope at the Lowel Observatory, equipped with the Ohio State University CCD Spectrograph); WHT+ISIS (William Herschell Telescope at the Roque de los Muchachos Observatory, equiped with the Intermediate-dispersion Spectrograph and Imaging System); WHT+ACAM (William Herschell Telescope at the Roque de los Muchachos Observatory, equiped with the Auxiliary-port CAMera); KPNO 4-m + RCS  (Kitt Peak National Observatory 4-m telescope, equipped with the Ritchey-Chr\'etien spectrograph); MDM 1.3-m + MIII (Michigan-Dartmouth-MIT Observatory 1.3-m telescoped, equipped with the Mark III spectrograph); INT+IDS (Isaac Newton Telescope equipped with the Intermediate Dispersion Spectrograph). For further details, we refer the reader to the original works.}
\label{table:specdata}
\end{table*}

\section{Spectroscopic fitting}

\begin{table*}
\caption{Best fit parameters from our modelling.}
\begin{tabular}{l|c|ccccc|ccc}
 & Line & $\mu_{\rm em}$ & $DP$ & $I_{\rm ratio}$ & $\sigma_{\rm em}$ & $|EW_{\rm em}|$ & $\mu_{\rm abs}$ & $\sigma_{\rm abs}$ & $|EW_{\rm abs}|$   \\
 & & $(\rm{km\, s^{-1}})$ &  $(\rm{km\, s^{-1}})$& & $(\rm{km\, s^{-1}})$ & $(\rm{\AA})$ & $(\rm{km\, s^{-1}})$ &  $(\rm{km\, s^{-1}})$ &  $(\rm{\AA})$  \\
 \hline & & & & & & & &\\
J1727 &  $\rm H\beta$ & $-210\pm 100$ &  $500\pm 200$ & $1.0\pm 0.3$ & $250\pm 70$ & $0.6\pm 0.9$ &  $-130\pm 190$& $700\pm 400$& $1.7\pm 1.3$ \\
 &  $\rm H\gamma$ & & &  & & $0.5\pm 0.3$ &  & $800\pm 400$ & $1.3\pm 1.2$ \\
 &  $\rm H\delta$ & & &  & & $0.3\pm 0.2$ & & $1100\pm 200$& $1.8\pm 1.3$ \\
J1727-N
&  $\rm H\beta$ & $-190\pm 50$& $420\pm 60$ & $1.0\pm 0.3$& $220\pm 60$ & $0.4\pm 1.0$ & $-100\pm 190$ & $500\pm 300$ & $1.4\pm 1.0$ \\
 &  $\rm H\gamma$ & & & & & $0.42\pm 0.18$ & & $700\pm 400$& $0.7\pm 0.4$ \\
 &  $\rm H\delta$ & & & & & $0.22\pm 0.08$ & & $1200\pm 300$& $1.1\pm 0.4$ \\
J1727-B
&  $\rm H\beta$ & $-250\pm 160$& $840\pm 40$& $0.94\pm 0.2$ & $330\pm 40$ & $1.1\pm 0.4$ & $-200\pm 140$  & $1100\pm 200$ & $2.7\pm 1.5$  \\
 &  $\rm H\gamma$ & & &  & & $0.8\pm 0.3$ & & $1080\pm 150$ & $2.6\pm 1.4$  \\
 &  $\rm H\delta$ & & &  & & $0.66\pm 0.18$ & & $1130\pm 130$  &  $3.3\pm 1.5$ \\
\hline & & & & & & & & \\
J1118 &  $\rm H\beta$ &  $50\pm 100$&  $1230\pm 160$ & $0.9\pm 0.4^{a}$ & $520\pm 90$ & $0.9\pm 0.5$ &  $130\pm 170$ &  $1400\pm 200$&  $2.4\pm 1.1$ \\
 &  $\rm H\gamma$ & & & & & $0.6\pm 0.4$ & &  $1800\pm 300$&  $3.2\pm 1.4$ \\
 &  $\rm H\delta$ & & & & & $0.5\pm 0.5$ & &  $2200\pm 400$ &  $5\pm 2$ \\
\hline & & & & & & & & \\
J0422-A &  $\rm H\beta$ & $-60\pm 280$ &  $800\pm 150^{b}$ & $1.4\pm 0.6$& $650\pm 30$ &  $0.5\pm 0.33$& $-90\pm 120$& $1400\pm 200$ &  $2.6\pm 1.0$\\
 &  $\rm H\gamma$ & & & & & $0.5\pm 0.8$ & & $1600\pm 600$ & $3\pm 2$ \\
 &  $\rm H\delta$ & & & & & $0.22\pm 0.14$ & & $1600\pm 300$ & $2.9\pm 0.6$ \\
J0422-B$^{c}$&  $\rm H\beta$ & $20\pm 40$ & $800\pm 40$ & $1.01\pm 0.12$&  $320\pm 50$& $1.0\pm 0.2$ & $140\pm 40$ &  $1190\pm 70$& $4.4\pm 0.5$   \\
 &  $\rm H\gamma$ & & & & & $0.47\pm 0.12$ & & $1220\pm 40$  & $3.5\pm 0.5$  \\
 J0422-C &  $\rm H\beta$ & $0\pm 300$ & $800\pm 500$ &$1.0\pm 0.7$ & $520\pm 150$ & $0.21\pm 0.19$ & $100\pm 300$  & $1400\pm 150$  & $1.5\pm 1.1$  \\
 &  $\rm H\gamma$ & $-100\pm 200$ &  $700\pm 300$ & $0.9\pm 0.3$ & $440\pm 190$ &  $0.01\pm 0.06$  & $100\pm 200$  & $2000\pm 600$  & $0.9\pm 0.4$  \\
\hline & & & & & & & & \\
J1753 &  $\rm H\beta$ &  $-20\pm 140$&  $1100\pm 90$ & $0.89\pm 0.9$ & $840\pm 50$ & $2.3\pm 1.3$ &  $80\pm 70$ &  $1320\pm 60$&  $8\pm 3$ \\
 &  $\rm H\gamma$ & & & & & $0.7\pm 0.4$ & &  $1540\pm 40$&  $5\pm 2$ \\
 &  $\rm H\delta$ & & & & & $1.1\pm 1.0$ & &  $1660\pm 170$ &  $7\pm 5$ \\
\hline & & & & & & & & \\
J1357 &  $\rm H\alpha$ &  $-200\pm 200$&  $1707\pm 13$ & $0.53\pm 0.13$ & $620\pm 40$ & $2.9\pm 0.6$ &  $-300\pm 400$ &  $1643\pm 14$&  $6\pm 3$ \\
 &  $\rm H\beta$ & & & & & $0.27\pm 0.19$ & &  $2700\pm 1000$&  $6\pm 4$ \\
\hline\hline & & & & & & & & \\
J1807 - A$^{d}$ &  $\rm H\beta$ & $-240\pm 120$ & $500\pm 300$ &  $1.2\pm 0.17$ &  $410\pm 110$ & $0.40\pm 0.08$ & $140\pm 100$ & $1250\pm 180$ & $4.5\pm 1.6$ \\
 &  $\rm H\gamma$ & & & & &  $0.22\pm 0.10$& & $1300\pm 300$ &  $4.1\pm 1.8$ \\
&  $\rm H\delta$ & & & & & $0.6\pm 0.2$ & & $1080\pm 50$ &  $5\pm 2$  \\
J1807 - B &  $\rm H\beta$ & $-70\pm 160$ & $700\pm 300$ & $0.82\pm 0.17$ & $450\pm 150$ & $0.6\pm 0.6$& $30\pm 80$ & $1510\pm 190$ & $4.6\pm 1.3$ \\
 &  $\rm H\gamma$ & & & & & $0.4\pm 0.3$ & & $1400\pm 300$& $4.8\pm 1.7$ \\
&  $\rm H\delta$ & & & & & $0.36\pm 0.17$ & & $1400\pm 400$ &  $6\pm 2$ \\
\hline
\end{tabular}
\tablefoot{We report, for each transition, the average and standard deviation of the best fitting results. For those systems where different regimes are identified, we also report the separate means and standard deviations. a - J1118 data marked with this note are the mean and standard deviation after removing the outlier from epoch 9. b - J0422-A data marked with this note are the mean and standard deviation after removing the outlier from epoch A4. c- J0422-B data marked with this note are the mean and standard deviation after removing the outlier from epoch B3. d - J1807-A includes the mean and standard deviation of the first three epochs, as the SNR of later observations was too low.}
\label{table:specfitparam}
\end{table*}

\section{Spectroscopic fitting results.}

\begin{figure*}
\centering
\includegraphics[keepaspectratio, trim=1cm 0cm 2.5cm 3cm, clip=true,  width=0.5\textwidth]{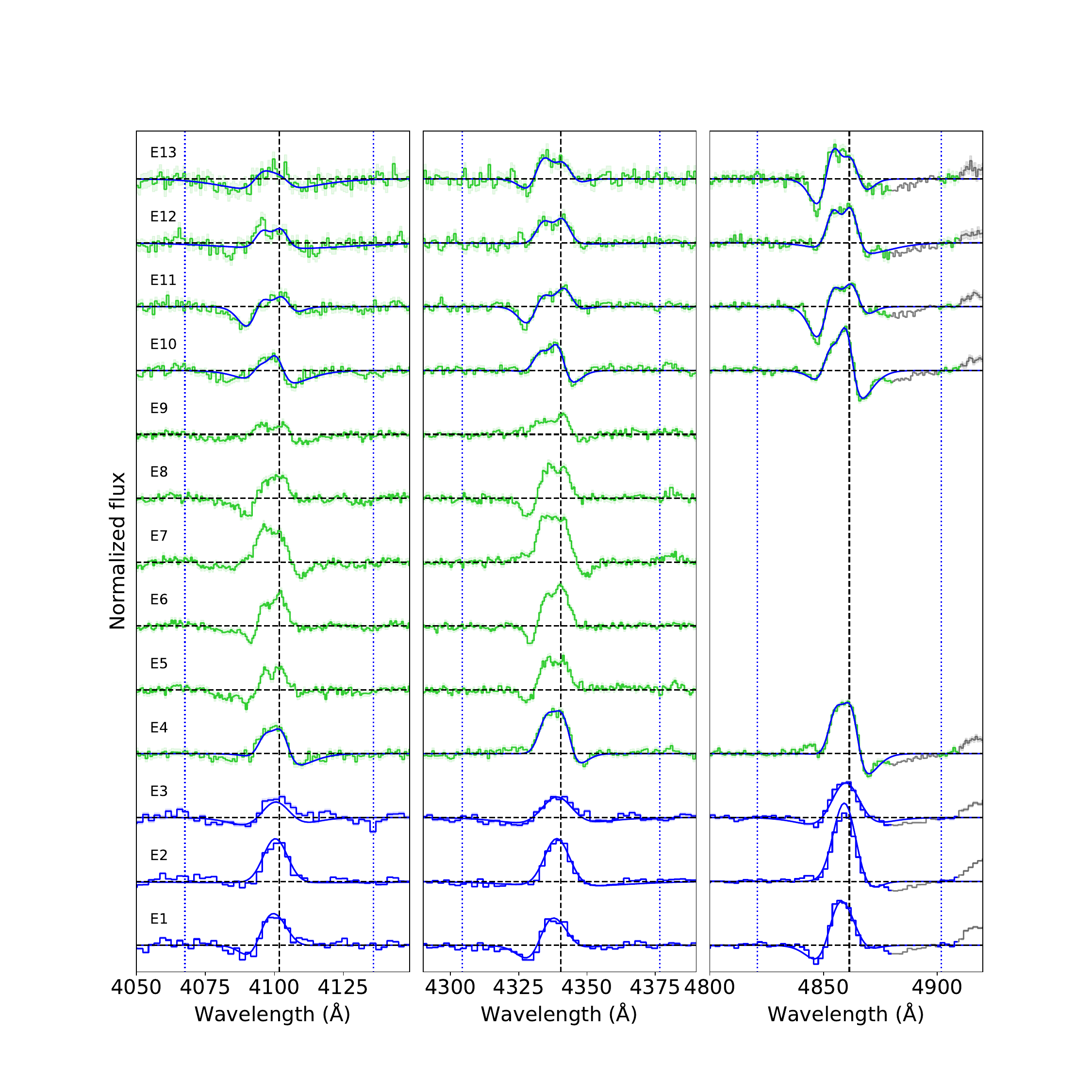}\includegraphics[keepaspectratio,trim=1cm 0cm 2.5cm 3cm, clip=true,width=0.5\textwidth]{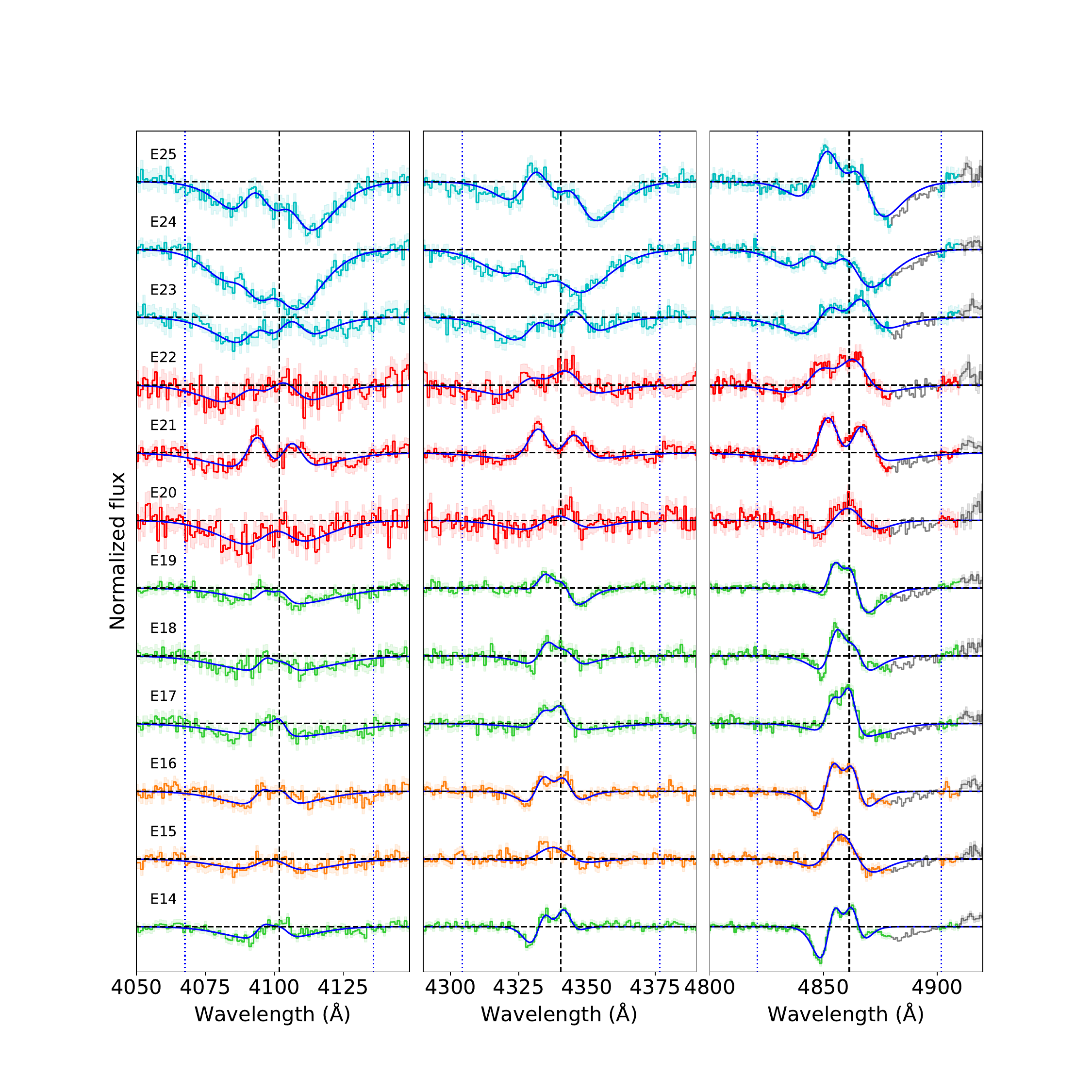}
\caption{J1727 spectroscopic database showing the BAs evolution during the outburst, zoomed into the Balmer $\rm H\delta$, $\rm H\gamma$ and $\rm H\beta$ transitions. The colour of the spectra corresponds to the X-ray spectral state, following the same convention as in \citet{MataSanchez2024a}: hard-intermediate state (HIMS, green), soft-intermediate state (SIMS, orange) and soft state (red). In addition, the low-luminosity transition between the soft and hard state is also marked for the new data (cyan). The best fit corresponding with the parameters shown in Fig. \ref{fig:J1727fitparams} is shown as a solid blue line. The DIB next to $\rm H\beta$, as well as a nearby He I transition are shown in grey. The rest wavelength of the transition is marked with a black-dashed line, while blue-dotted lines mark velocity shifts of $\pm 2500\, \rm{km\,s^{-1}}$ as a visual reference.}
     \label{fig:J1727fitspec}%
 \end{figure*}

\begin{figure*}
\centering
\includegraphics[keepaspectratio, trim=0cm 0cm 1cm 1cm, clip=true,width=0.5\textwidth]{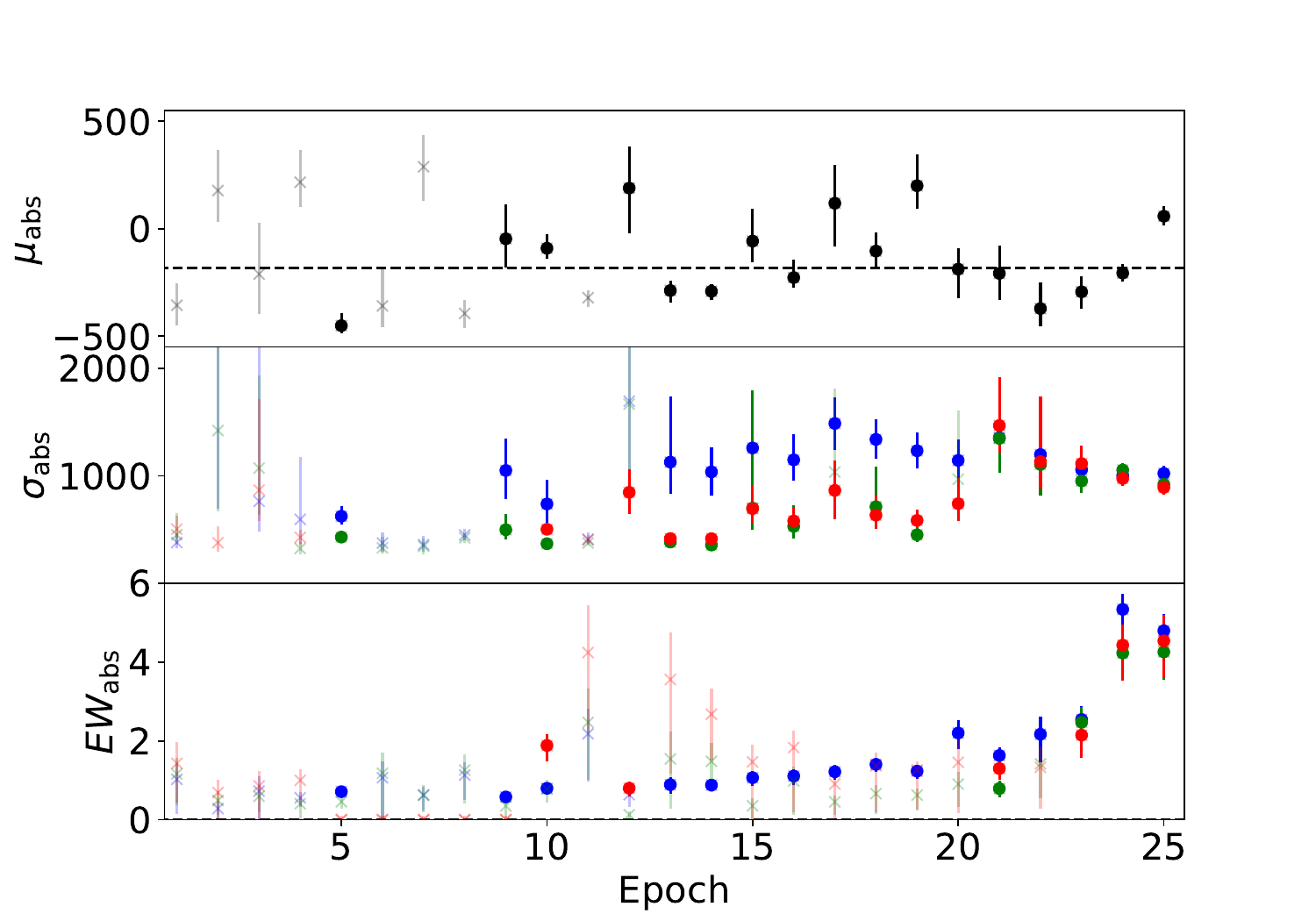}\includegraphics[keepaspectratio, trim=0cm 0cm 1cm 0cm,width=0.5\textwidth]{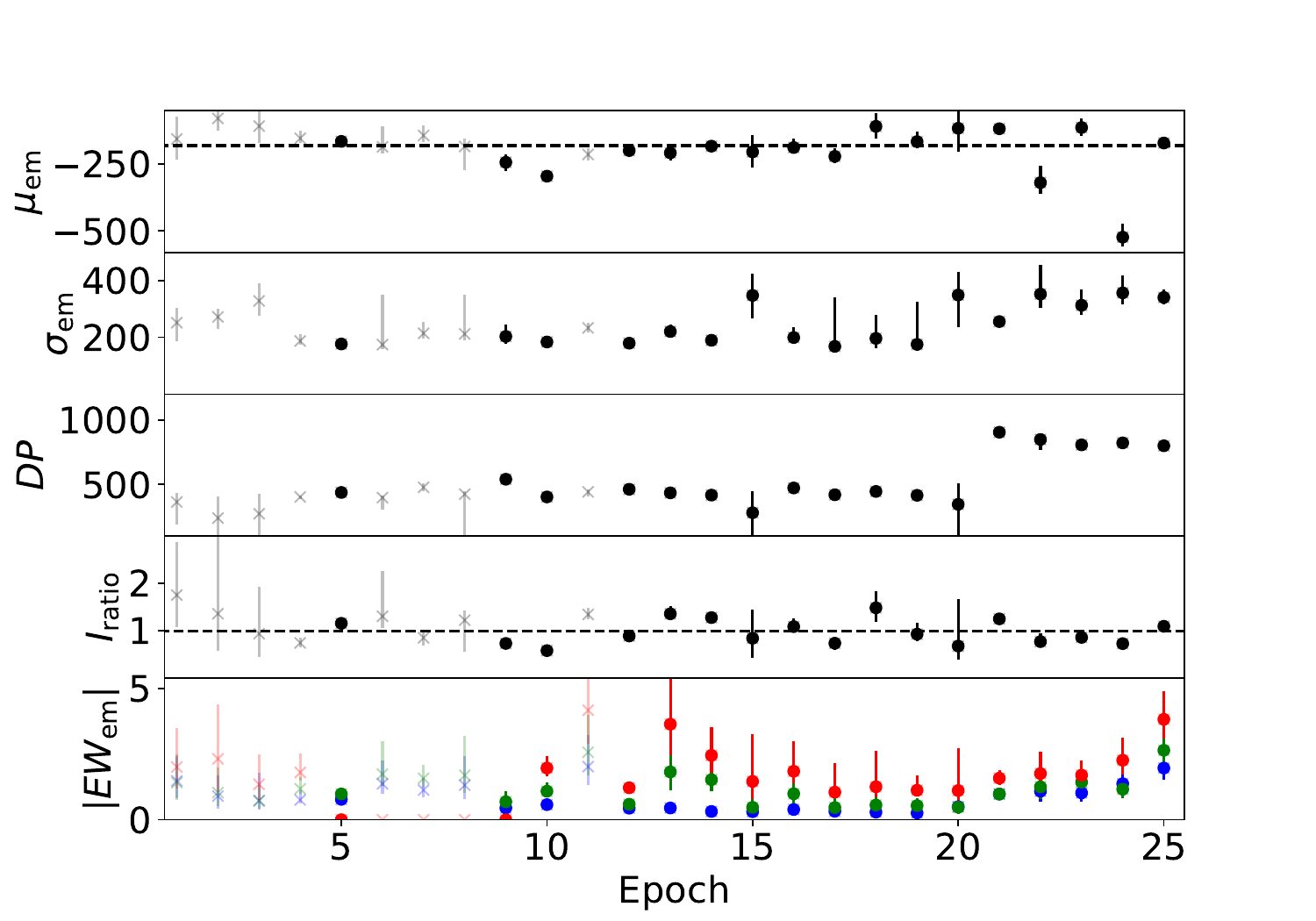}
\caption{Best fitting parameters for the complete spectroscopic sample of J1727 (see Fig. \ref{fig:J1727fitspec}). The left figure contains parameters associated with the absorption component, while the right figure shows the emission component parameters. For those parameters which are independent for each of the fitted lines, the following colour code applies: $\rm H\delta$ (blue), $\rm H\gamma$ (green) and $\rm H\beta$ (red). The black-dashed line in the top panel marks the systemic velocity reported in the text. Transparent points mark best-fit solutions where $EW_{\rm abs}$ is consistent with 0 within $3\sigma$, and therefore consistent with a non-detection of BA. A similar line in the $I_{\rm ratio}$ panel marks the symmetric profile ($I_{\rm ratio}=1$). Units are [$\rm km\,s{-1}$] for $\mu_{\rm abs}$, $\mu_{\rm em}$, $\sigma_{\rm abs}$, $\mu_{\rm em}$, $DP$; and [$\rm\AA$] for $EW_{\rm abs}$ and $|EW_{\rm em}|$.}
     \label{fig:J1727fitparams}%
 \end{figure*}

\begin{figure*}
\centering
\includegraphics[keepaspectratio, trim=1cm 0cm 2.5cm 3cm, clip=true,  width=0.5\textwidth]{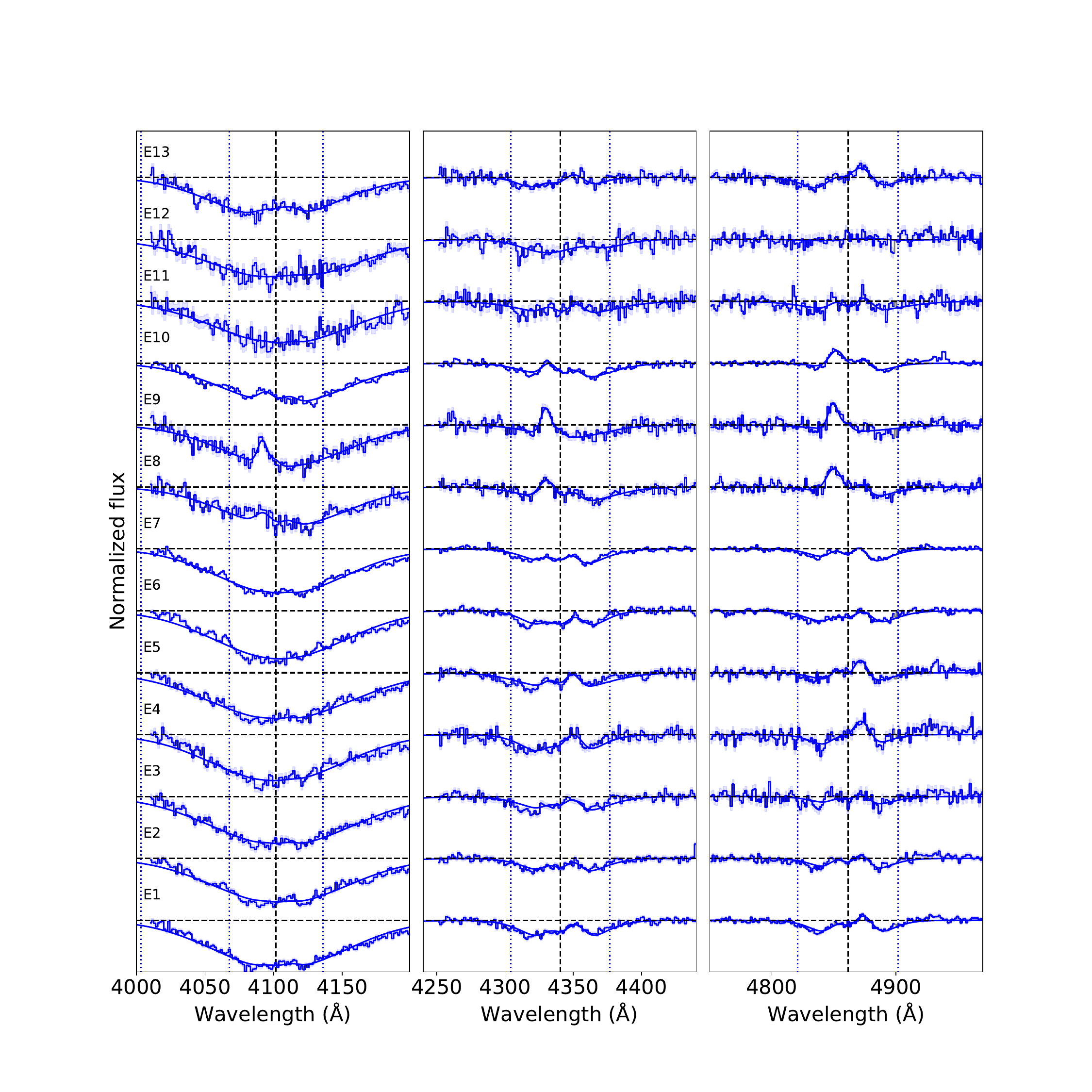}\includegraphics[keepaspectratio,trim=1cm 0cm 2.5cm 3cm, clip=true,width=0.5\textwidth]{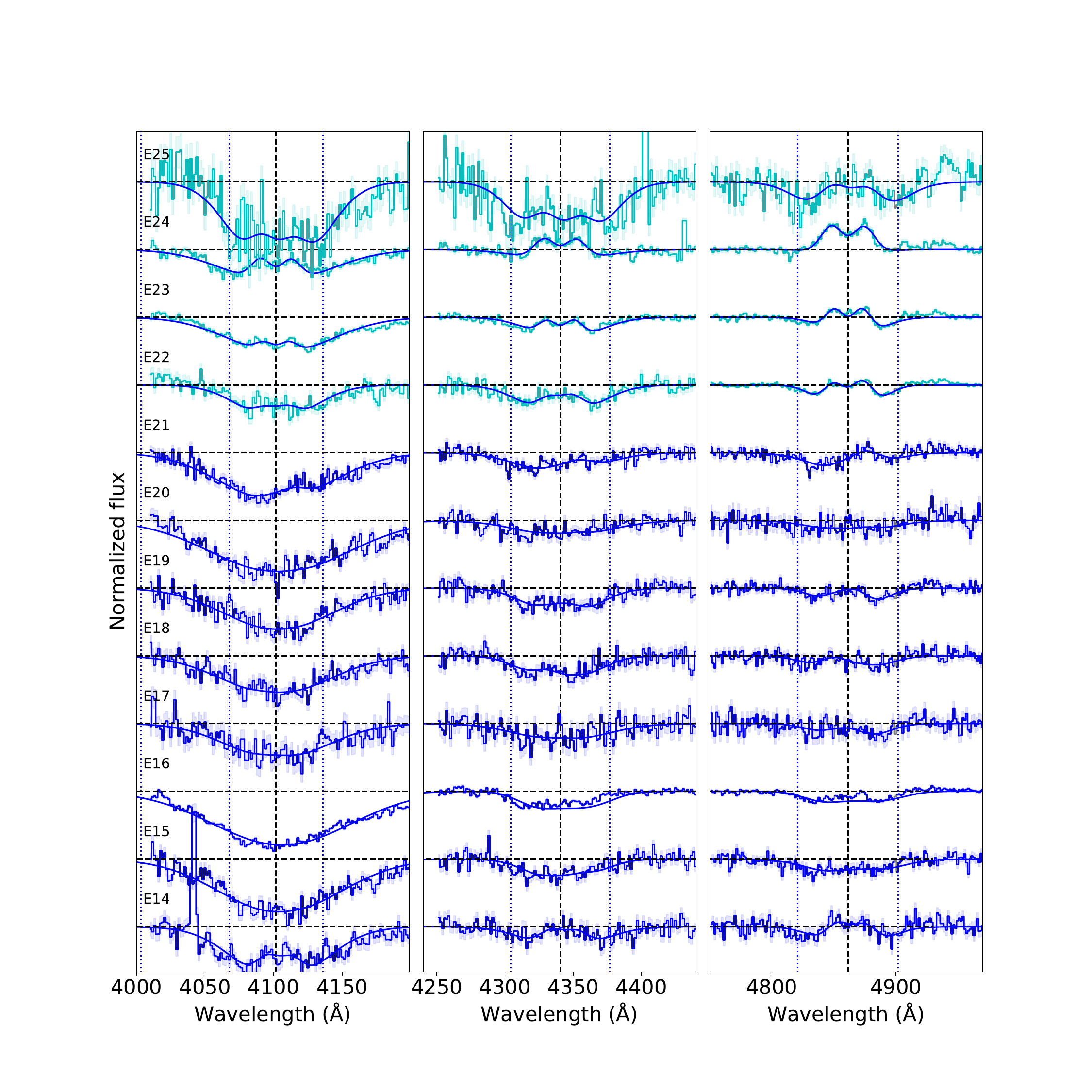}
\caption{J1118 spectra from the 2000 (blue) and the 2005 outburst events (cyan). The best fit corresponding with the parameters shown in Fig. \ref{fig:J1118fitparams} is shown as a solid blue line. Neither the DIB next to $\rm H\beta$, nor the nearby He I transition are present in the spectra, so they did not have to be masked out. The rest wavelength of the transition is marked with a black-dashed line, while blue-dotted lines mark velocity shifts of $\pm 2500\, \rm{km\,s^{-1}}$ as a visual reference.}
     \label{fig:J1118fitspec}%
 \end{figure*}

\begin{figure*}
\centering
\includegraphics[keepaspectratio, trim=0cm 0cm 1cm 1cm, clip=true,width=0.5\textwidth]{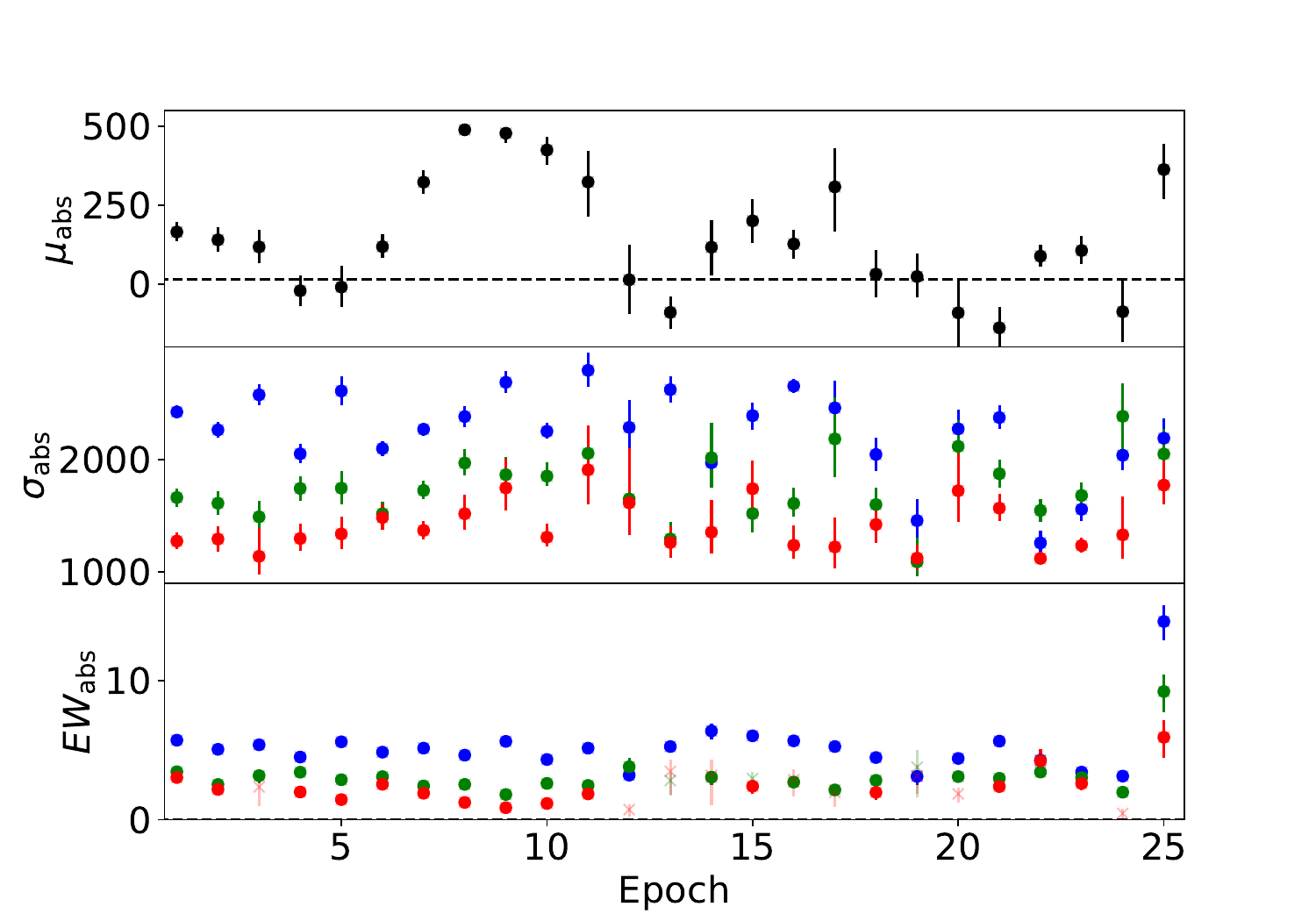}\includegraphics[keepaspectratio,trim=0cm 0cm 1cm 0cm, clip=true,width=0.5\textwidth]{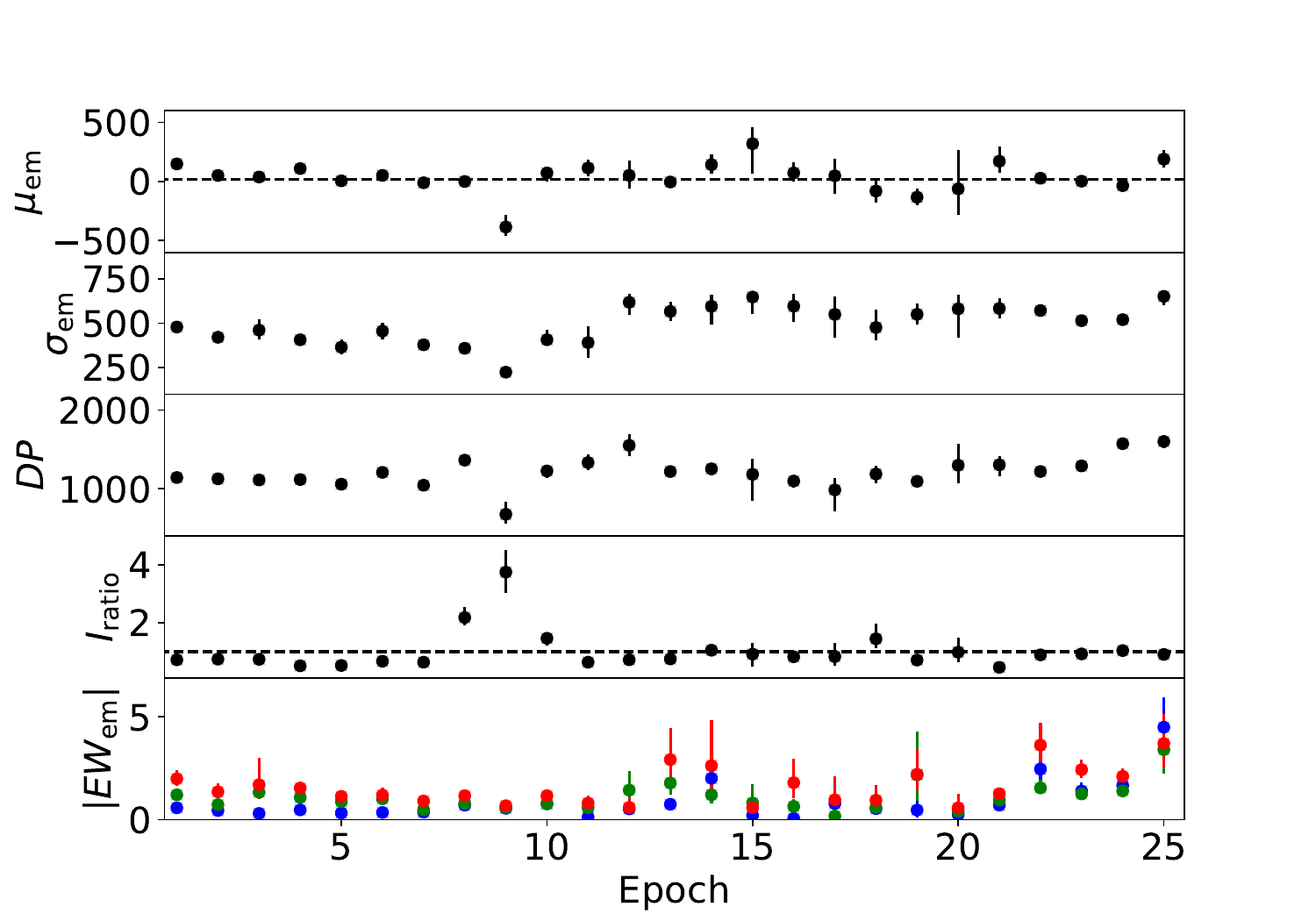}
\caption{Best fitting parameters for the complete spectroscopic sample of J1118 (see Fig. \ref{fig:J1118fitspec}), following the same description as in Fig. \ref{fig:J1727fitparams}.}
     \label{fig:J1118fitparams}%
 \end{figure*}

\begin{figure*}
\centering
\includegraphics[keepaspectratio, trim=1cm 0cm 2.5cm 3cm, clip=true,  width=0.4\textwidth]{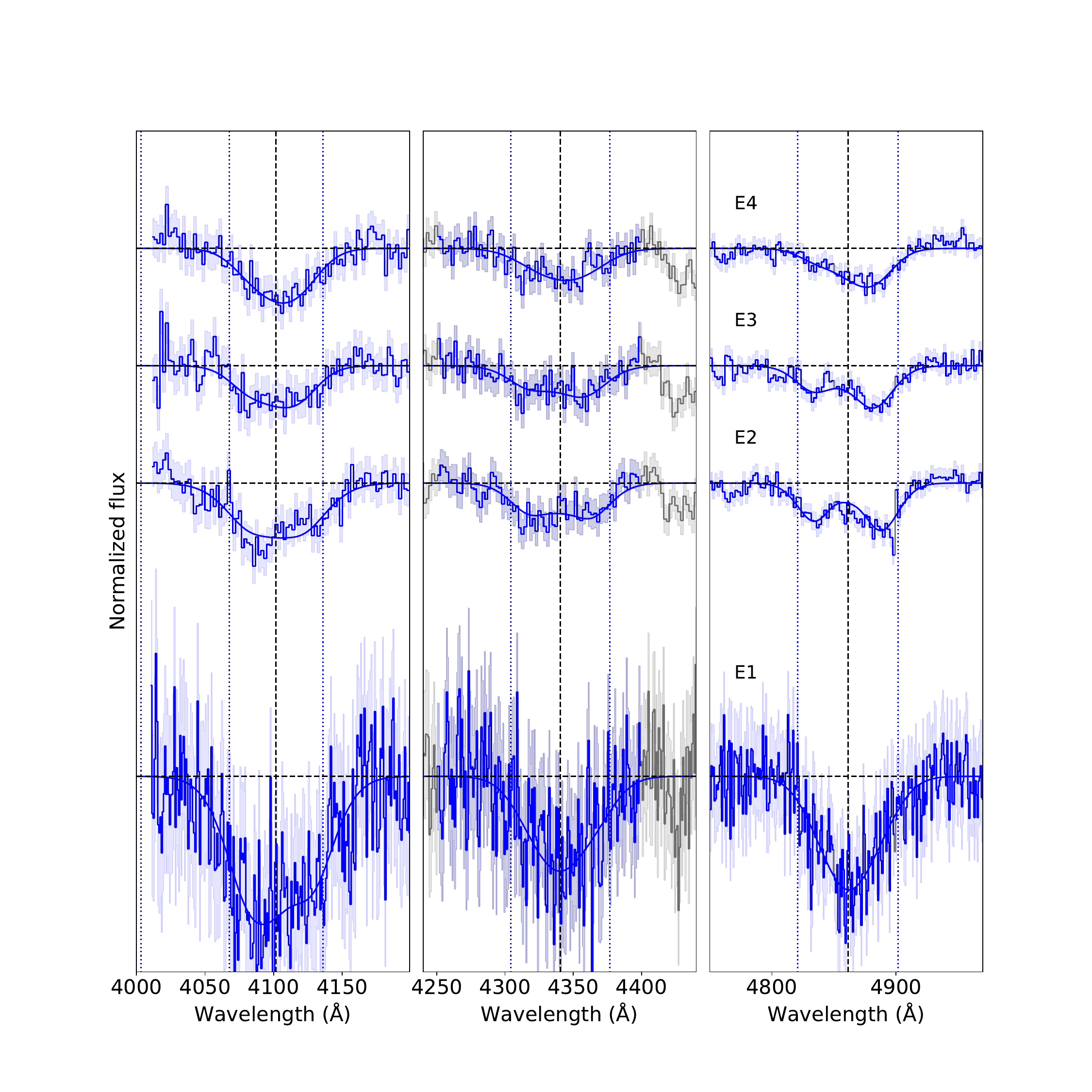}\includegraphics[keepaspectratio,trim=0cm 0cm 1cm 0cm, clip=true,width=0.6\textwidth]{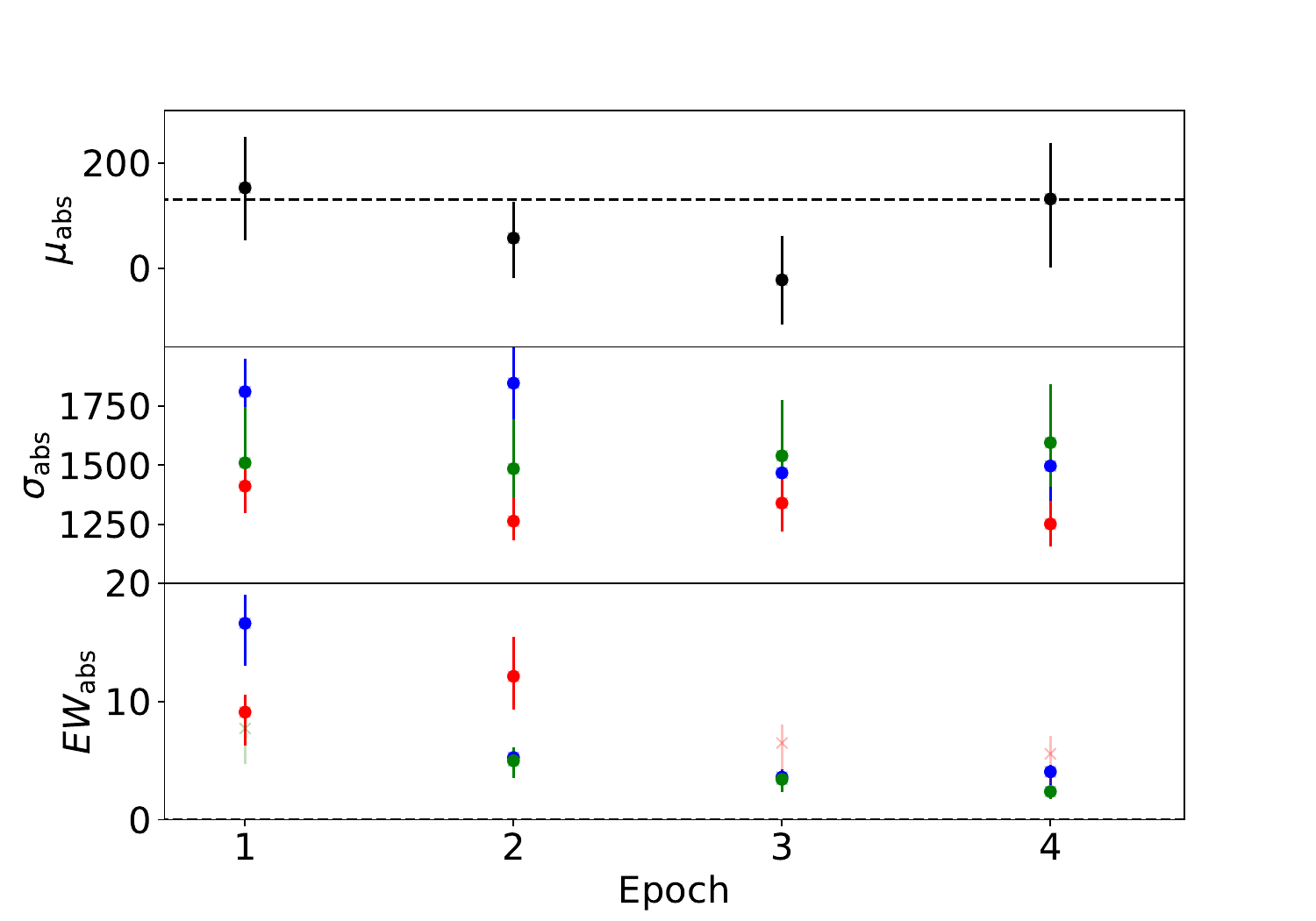}
\caption{Left panel: J1753 spectra showing the BAs evolution during its 2023 outburst, zoomed into the Balmer $\rm H\delta$, $\rm H\gamma$ and $\rm H\beta$ transitions. The colour of the spectra corresponds to the X-ray spectral state, following the same convention as in Fig. \ref{fig:J1727fitspec}. The best fit corresponding with the parameters shown in the right panel is shown as a solid blue line. The rest wavelength of the transition is marked with a black-dashed line, while blue-dotted lines mark velocity shifts of $\pm 2500\, \rm{km\,s^{-1}}$ as a visual reference. Black-coloured regions of the spectra were masked during the fit due to the presence of a telluric band. Right panel: Best fitting parameters for the spectroscopic sample of J1357, following the same description as in Fig. \ref{fig:J1727fitparams}.}
     \label{fig:J753fitspec}%
 \end{figure*}

\begin{figure*}
\centering
\includegraphics[keepaspectratio, trim=1cm 0cm 2.5cm 3cm, clip=true,  width=0.4\textwidth]{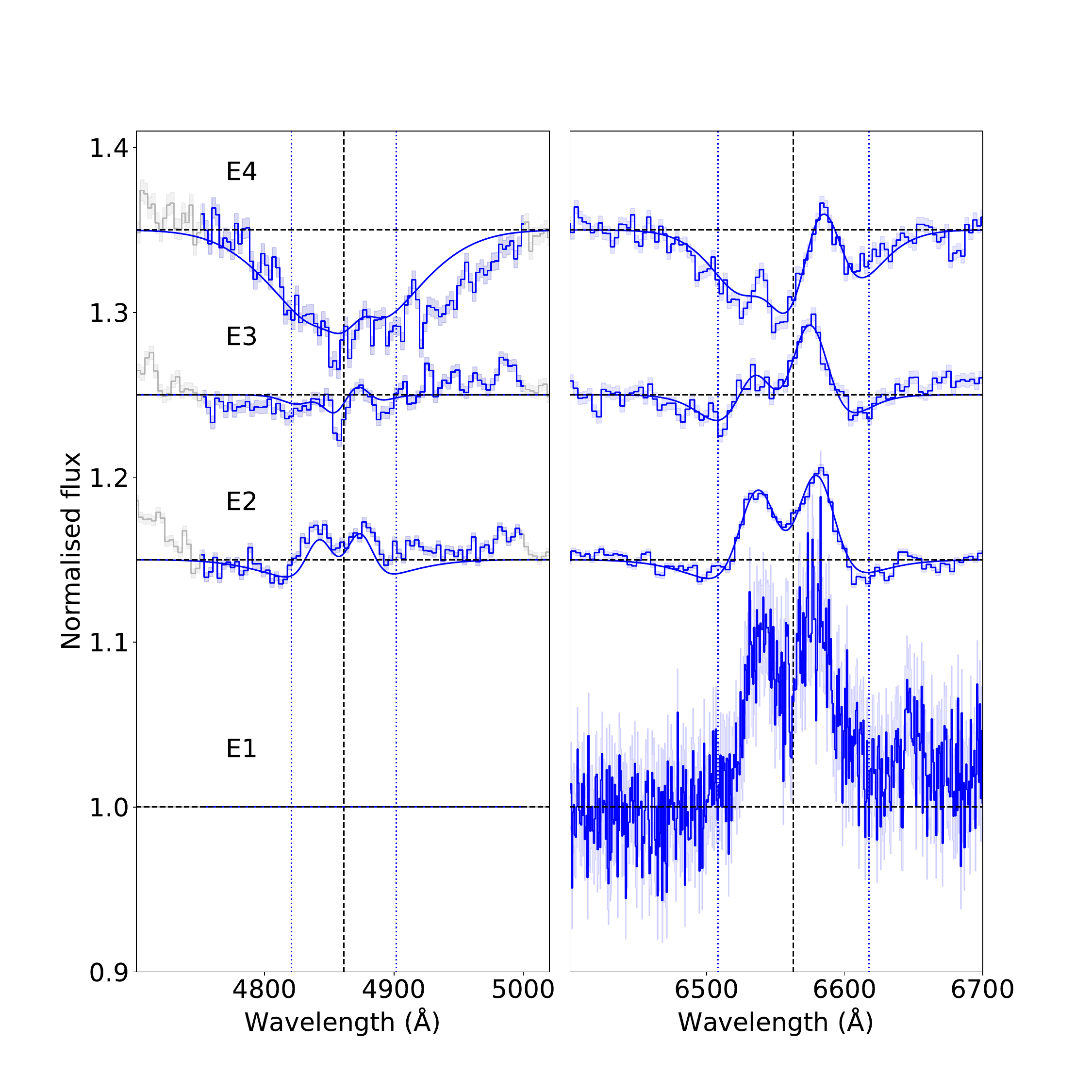}\includegraphics[keepaspectratio,trim=0cm 0cm 1cm 0cm, clip=true,width=0.6\textwidth]{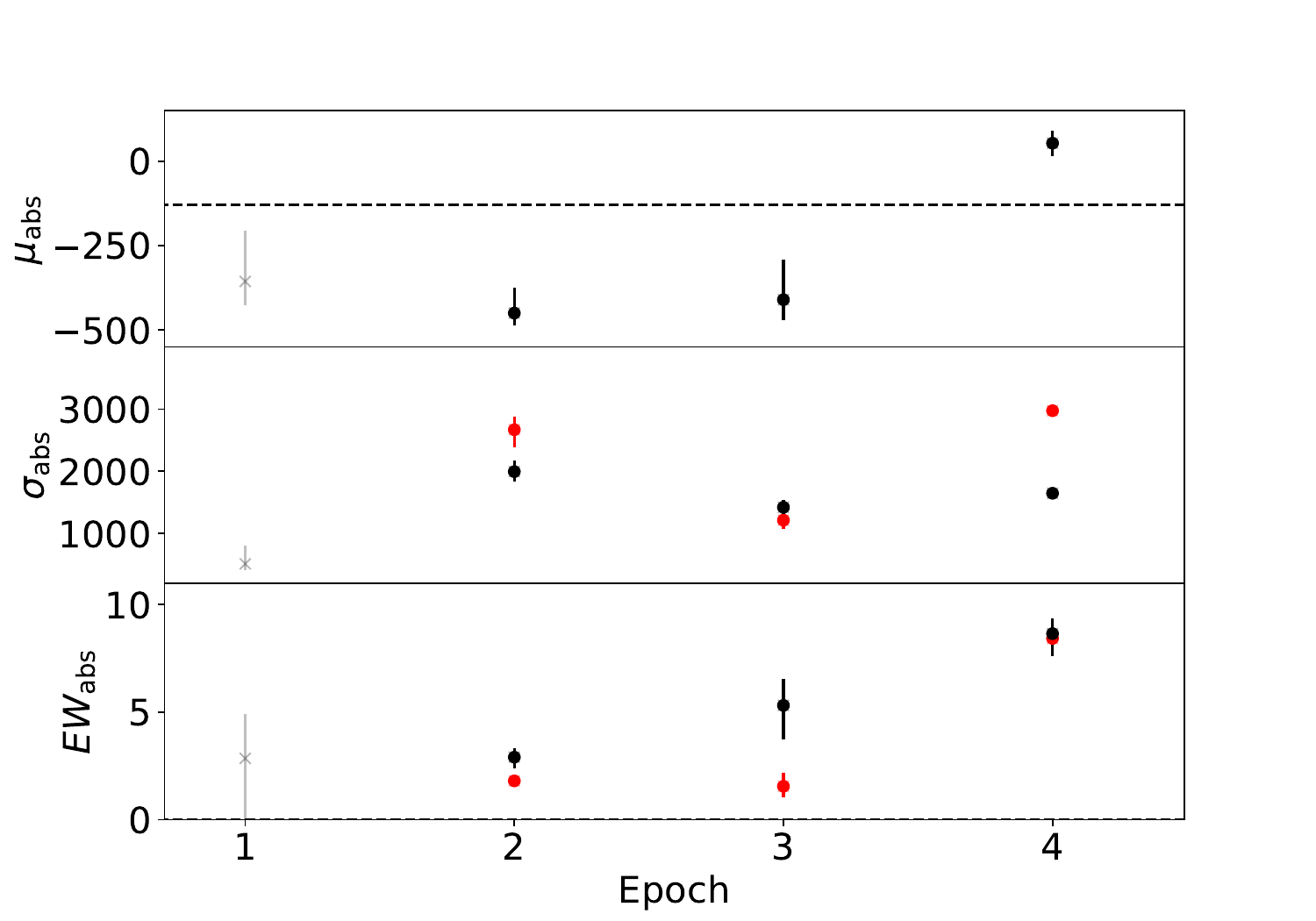}
\caption{Left panel: J1357 spectra showing the BAs evolution during its 2011 outburst, zoomed into the Balmer $\rm H\beta$ and $\rm H\alpha$ transitions. The colour of the spectra corresponds to the X-ray spectral state, following the same convention as in Fig. \ref{fig:J1727fitspec}. The best fit corresponding with the parameters shown in the right panel is shown as a solid blue line. The rest wavelength of the transition is marked with a black-dashed line, while blue-dotted lines mark velocity shifts of $\pm 2500\, \rm{km\,s^{-1}}$ as a visual reference. Right panel: Best fitting parameters for the spectroscopic sample of J1357, following the same description as in Fig. \ref{fig:J1727fitparams}, but using black dots for the $\rm H\alpha$ transition.}
     \label{fig:J1357fitspec}%
 \end{figure*}

\begin{figure*}
\centering
\includegraphics[keepaspectratio, trim=1cm 0cm 2.5cm 3cm, clip=true,  width=0.5\textwidth]{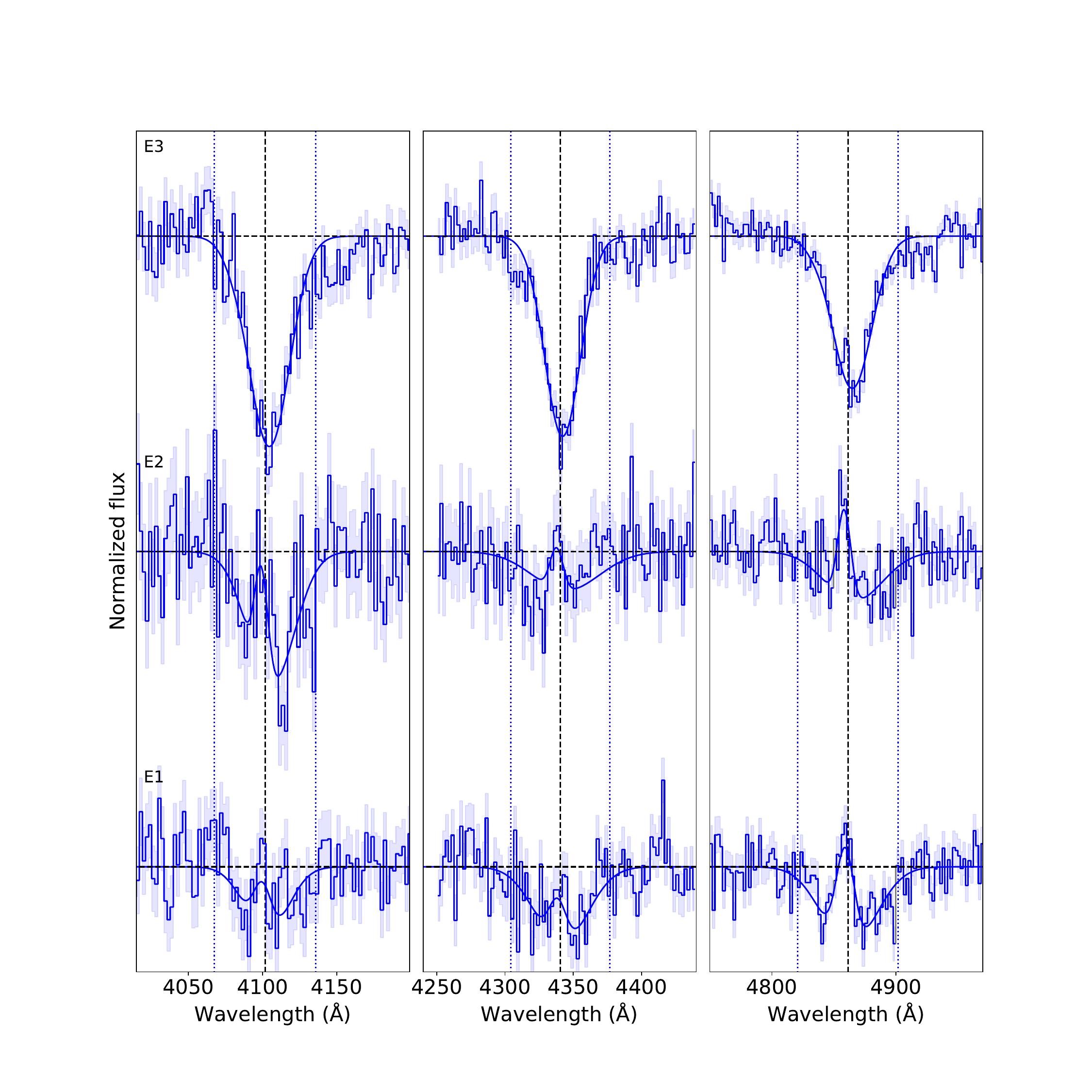}\includegraphics[keepaspectratio, trim=1cm 0cm 2.5cm 3cm, clip=true, width=0.5\textwidth]{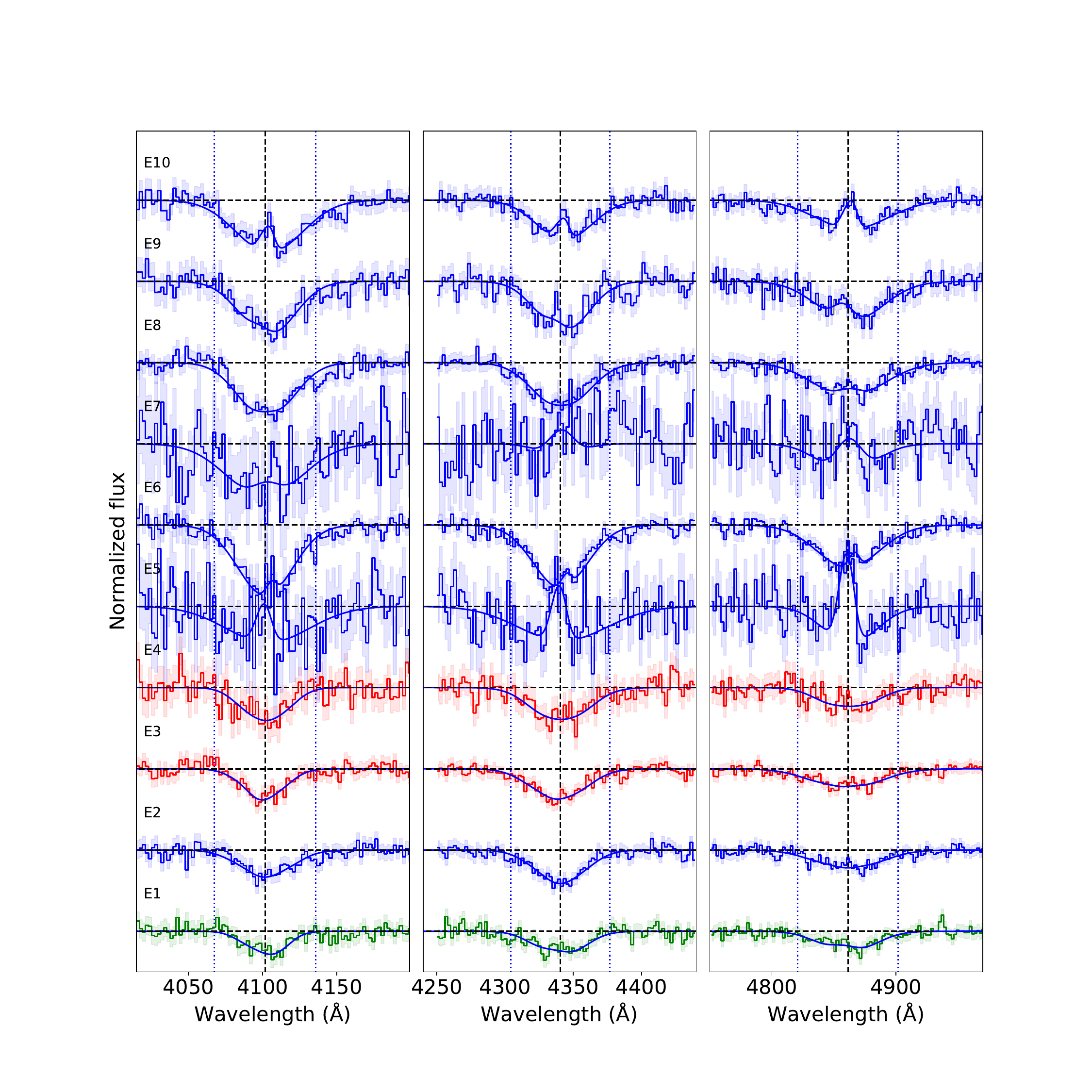}
\caption{J1807 spectra from the 2017 (left) and 2023 (right) outburst events. The colour code mark the X-ray state, similarly to Fig. \ref{fig:J1727fitparams}. The best fit corresponding with the parameters compiled in Fig. \ref{fig:J1807fitparams} is shown as a solid blue line. Neither the DIB next to $\rm H\beta$, nor the nearby He I transition are present in the spectra, so they did not have to be masked out. The rest wavelength of the transition is marked with a black-dashed line, while blue-dotted lines mark velocity shifts of $\pm 2500\, \rm{km\,s^{-1}}$ as a visual reference.}
     \label{fig:J1807fitspec}%
 \end{figure*}

\begin{figure*}
\centering
\includegraphics[keepaspectratio, trim=0cm 0cm 1cm 1cm, clip=true,width=0.5\textwidth]{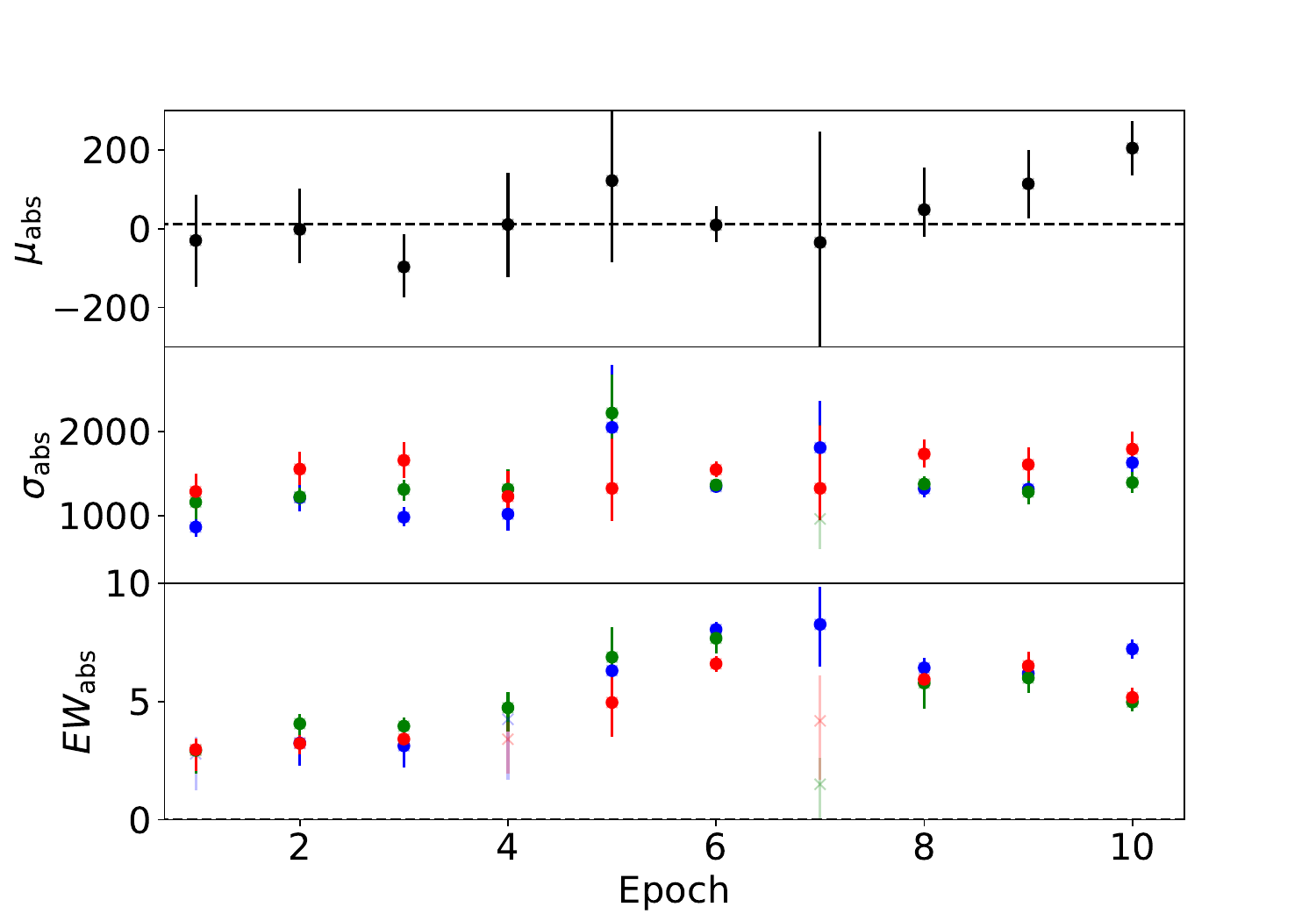}\includegraphics[keepaspectratio, trim=0cm 0cm 1cm 0cm, clip=true,width=0.5\textwidth]{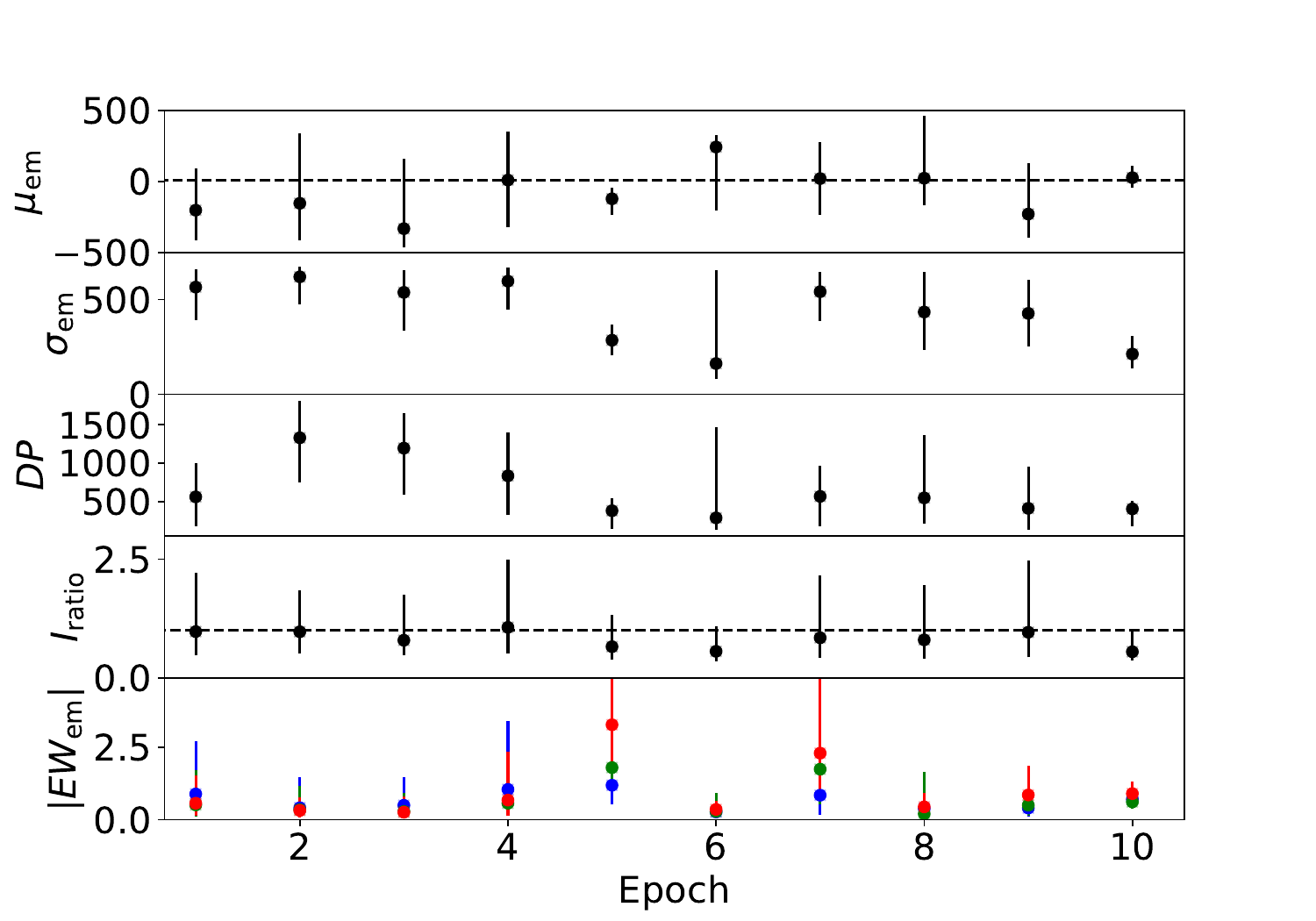}
\caption{Best fitting parameters for the 2023 spectroscopic sample of J1807 (Fig. \ref{fig:J1807fitspec}), following the same description as in Fig. \ref{fig:J1727fitparams}.}
     \label{fig:J1807fitparams}%
 \end{figure*}

\end{appendix}

\end{document}